\documentclass{article}

\usepackage[left=1.25in,top=1.25in,right=1.25in,bottom=1.25in,head=1.25in]{geometry}
\usepackage{amsfonts,amsmath,amssymb,amsthm}
\usepackage{verbatim,float,url,graphicx,psfrag}
\usepackage{natbib}
\usepackage{graphicx}
\usepackage{mathtools} 
\usepackage{mathabx}   
\usepackage{hyperref} 
\usepackage{bbm}      
\usepackage{cancel}   
\usepackage{mathtools} 
\usepackage{dsfont}
\usepackage{subfigure} 
\usepackage[export]{adjustbox} 
\usepackage{tabularx} 
\usepackage[section]{placeins} 

\usepackage{marginnote}
\usepackage{mdframed} 
\usepackage{comment}
\usepackage{framed}

\usepackage{xr}
\newtheorem{algorithm}{Algorithm}

\theoremstyle{remark}

\theoremstyle{definition}

\newcommand{\argmin}{\mathop{\mathrm{argmin}}}
\newcommand{\argmax}{\mathop{\mathrm{argmax}}}

\newcommand{\st}{\mathop{\mathrm{subject\,\,to}}}

\def\P{\mathbb{P}}

\def\half{\frac{1}{2}}

\def\col{\mathrm{col}}
\def\row{\mathrm{row}}
\def\nul{\mathrm{null}}
\def\rank{\mathrm{rank}}
\def\nuli{\mathrm{nullity}}
\def\sign{\mathrm{sign}}

\def\diag{\mathrm{diag}}

\def\hu{\hat{u}}
\def\hbeta{\hat{\beta}}

\def\R{\mathbb{R}}

\def\cB{\mathcal{B}}

\def\cN{\mathcal{N}}

\def\cV{\mathcal{V}}

\def\one{\mathds{1}}

\title{Exact Post-Selection Inference for Changepoint Detection and Other  
  Generalized Lasso Problems} 
\author{Sangwon Hyun \and Max G'Sell \and Ryan J. Tibshirani}
\date{}

\begin{document}
\maketitle

\begin{abstract}
We study tools for inference conditioned on model selection events that are defined by the generalized lasso regularization path. The generalized lasso estimate is given by the solution of a penalized least squares regression problem, where the penalty is the $\ell_1$ norm of a matrix $D$ times the coefficient vector. The generalized lasso path collects these estimates as the penalty parameter $\lambda$ varies (from $\infty$ down to 0).  Leveraging a (sequential) characterization of this path from \citet{genlasso}, and recent advances in post-selection inference from \citet{exactlasso,exactlar}, we develop exact hypothesis tests and confidence intervals for linear contrasts of the underlying mean vector, conditioned on any model selection event along the generalized lasso path (assuming Gaussian errors in the observations).  

Our construction of inference tools holds for any penalty matrix $D$. By inspecting specific choices of $D$, we obtain post-selection tests and confidence intervals for specific cases of generalized lasso estimates, such as the fused lasso, trend filtering, and the graph fused lasso.  In the fused lasso case, the underlying coordinates of the mean are assigned a linear ordering, and our framework allows us to test selectively chosen breakpoints or {\it changepoints} in these mean coordinates.  This is an interesting and well-studied problem with broad applications; our framework applied to the trend filtering and graph fused lasso cases serves several applications as well.  Aside from the development of selective inference tools, we describe several practical aspects of our methods such as (valid, i.e., fully-accounted-for) post-processing of generalized estimates before performing inference in order to improve power, and problem-specific visualization aids that may be given to the data analyst for he/she to choose linear contrasts to be tested.  Many examples, both from simulated and real data sources, are presented to examine the empirical properties of our inference methods. 

\end{abstract}

\section{Introduction}
\label{sec:introduction}

\begin{figure}[htb]
\centering
\includegraphics[width=0.475\textwidth]{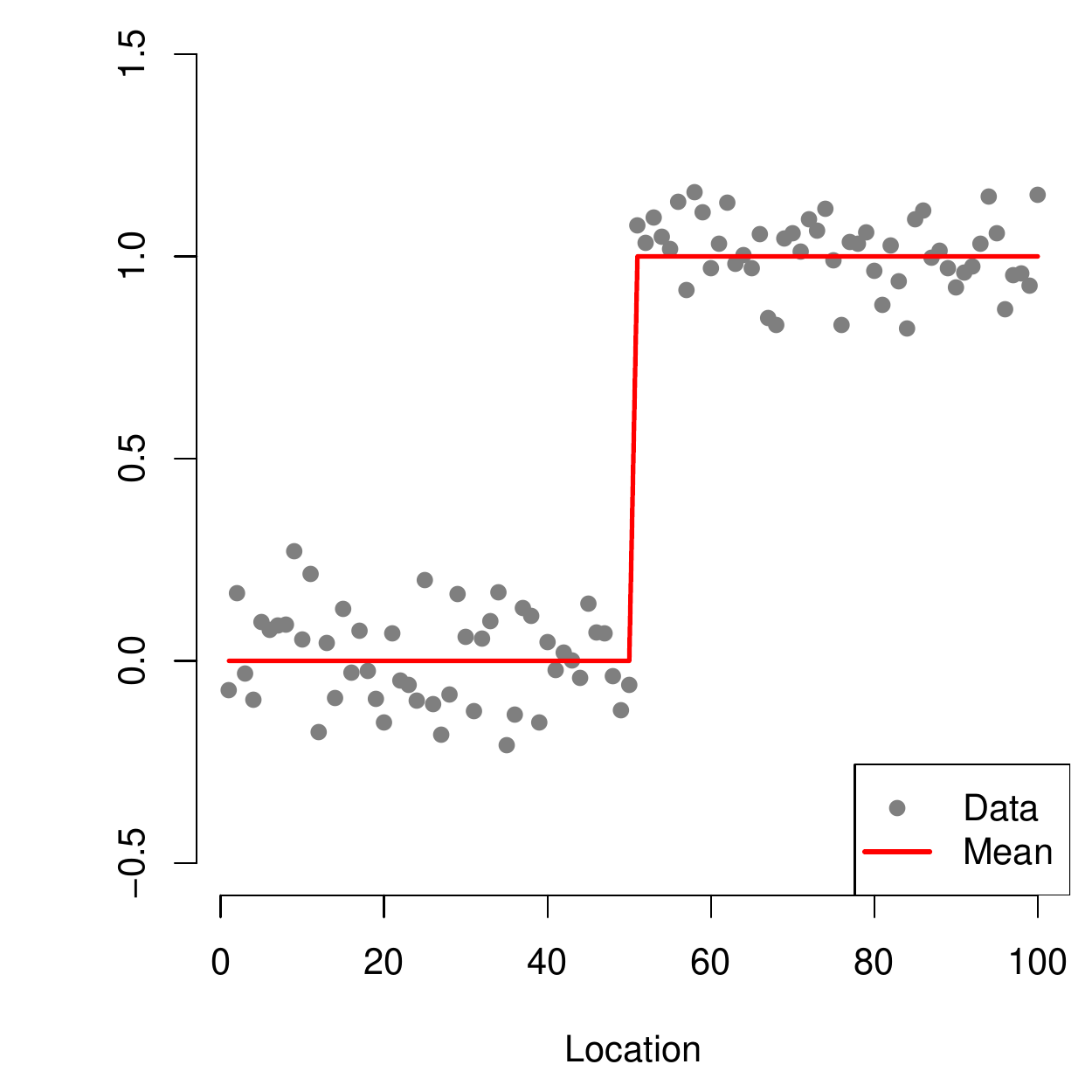}
\hspace{2pt}
\includegraphics[width=0.475\textwidth]{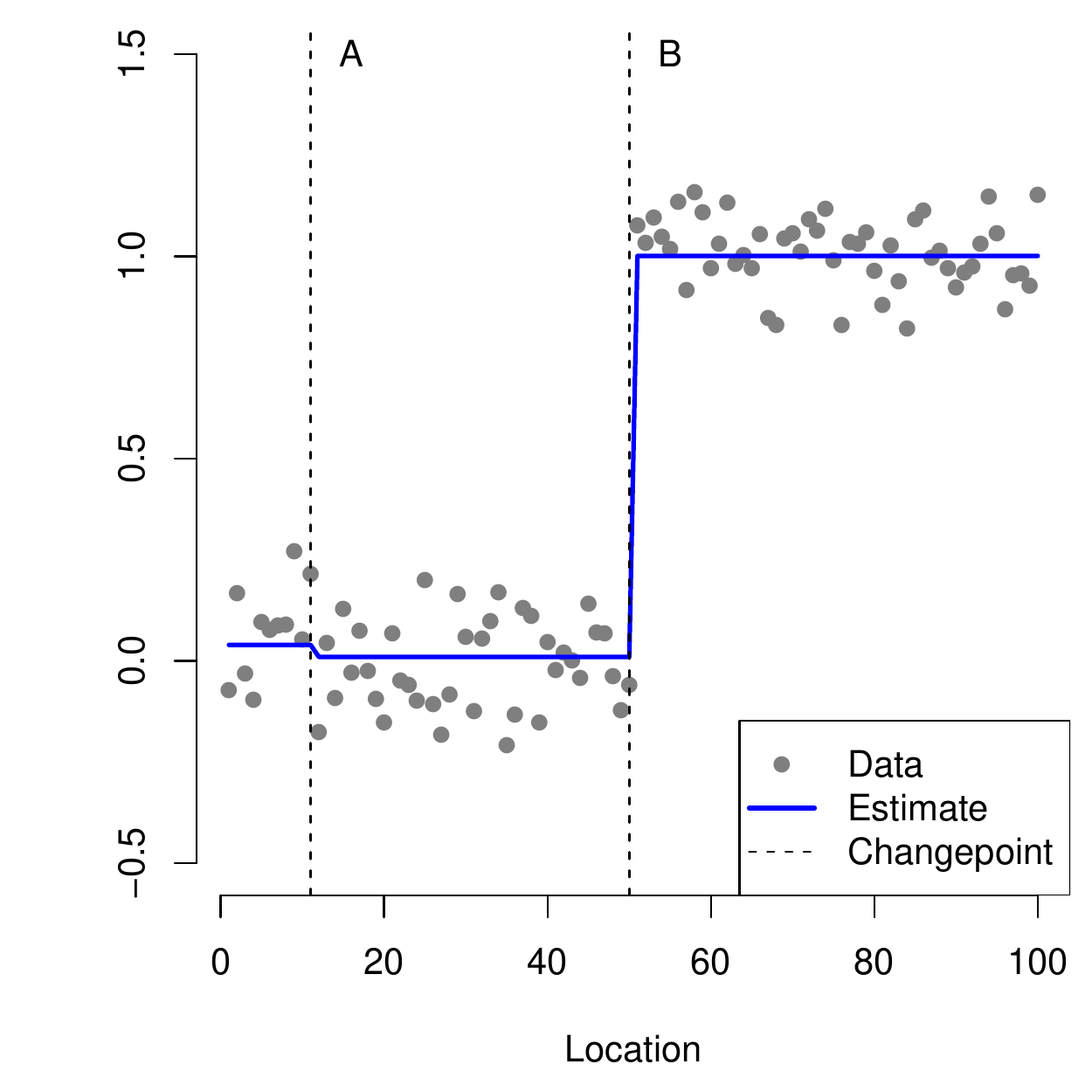} 

\smallskip\smallskip
\begin{tabular}{|r|r|r|r|}
  \hline
 & Location & Naive p-values & TG p-values \\  
  \hline
A & 11 & 0.057 & 0.359 \\ 
  B & 50 & 0.000 & 0.000 \\ 
   \hline
\end{tabular}

\caption{\it\small A simple example with $n=100$ points generated
  around a piecewise constant mean with one true changepoint at
  location 50, shown in the top left panel.  The 1d fused lasso path,
  stopped at the (end of the) second step, produces the estimate in
  the top right panel, with two detected changepoints at locations 11
  and 50, labeled $A$ and $B$ in the figure. The table reports
  p-values from the naive Z-test, which does not account for the
  data-dependent nature of the changepoints, and from our TG test 
  for the 1d fused lasso, which does.}   
\label{fig:intro}
\end{figure}

Consider a classic Gaussian model for observations $y \in \R^n$, with
known marginal variance $\sigma^2>0$, 
\begin{equation}
\label{eq:model}
y \sim \cN(\theta, \sigma^2 I),
\end{equation}
where the (unknown) mean $\theta \in \R^n$ is the parameter of
interest. In this paper, we examine problems in which $\theta$ is
believed to have some specific structure (at least approximately so),
in that it is sparse when parametrized with respect to a particular
basis.  A key example is the {\it changepoint detection} problem, in
which the components of the mean $\theta_1,\ldots,\theta_n$ correspond
to ordered underlying positions (or locations) $1,\ldots,n$, and many
adjacent components $\theta_i$ and $\theta_{i+1}$ are believed to be
equal, with the exception of a sparse number of breakpoints or {\it
  changepoints} to be determined.  See the left plot in Figure
\ref{fig:intro} for a simple example.   

Many methods are available for estimation and detection in the
changepoint problem. We focus on the {\it 1-dimensional fused lasso}
\citep{fuse}, also called {\it 1-dimensional total variation
  denoising} \citep{tv} in signal processing, for reasons that will
become clear shortly.  This method, which we call the 1d fused lasso
(or simply fused lasso) for short, is often used for piecewise
constant estimation of the mean, but it does not come with associated
inference tools after changepoints have been detected.  In the top
right panel of Figure \ref{fig:intro}, we inspect the 1d fused lasso
estimate that has been tuned to detect two changepoints, in a data
model where the mean $\theta$ only has one true changepoint.  Writing
the changepoint locations as $1 \leq I_1 <  I_2 \leq n$, we might
consider testing 
\begin{multline*}
H_{0,j} : \theta_{I_{j-1}} = \ldots = \theta_{I_j-1} = \theta_{I_j} =
\ldots = \theta_{I_{j+1}-1} 
\quad \text{versus} \\
H_{1,j} : \theta_{I_{j-1}} = \ldots = \theta_{I_j-1} \neq \theta_{I_j}
= \ldots = \theta_{I_{j+1}-1}, 
\quad j=1,2,
\end{multline*}
where we write $I_0=1$ and $I_3=n+1$ for notational convenience.  If
we were to naively ignore the data-dependent nature of $I_1,I_2$
(these are the estimated changepoints from the two-step fused lasso
procedure), i.e., treat them as fixed, then the natural tests for the
null hypothesess $H_{0,j}$, $j=1,2$ would be to reject for large
magnitudes of the statistics 
\begin{equation*}
T_j = \bar{y}_{(I_{j-1}): (I_j-1)} - \bar{y}_{(I_j) : (I_{j+1}-1)},
\quad j=1,2, 
\end{equation*}
respectively, where we use $\bar{y}_{a:b}=\sum_{i=a}^b y_i / (b-a)$ to
denote the average of components of $y$ between positions $a$ and $b$.
Indeed, these can be seen as likelihood ratio tests stemming from the
Gaussian model in \eqref{eq:model}.   

The table in Figure \ref{fig:intro} shows the results of running such naive
Z-tests.  At location $I_2$ (labeled location $B$ in the figure), which
corresponds to a true changepoint in the underlying mean, the test returns a
very small p-value, as expected.  But at location $I_1$ (labeled $A$ in the
figure), a spurious detected changepoint, the naive Z-test also produces a small
p-value.  This happens because the location $I_1$ has been selected by the 1d
fused lasso, which inexorably links it to an unusually large magnitude of $T_1$;
in other words, it is no longer appropriate to compare $T_1$ against its
supposed Gaussian null distribution, with mean zero and variance
$\sigma^2(1/(I_1-I_0)+1/(I_2-I_1))$.  Also shown in the table are the results of
running our new {\it truncated Gaussian} (TG) test for the 1d fused lasso, which
properly accounts for the data-dependent nature of the changepoints detected by
fused lasso, and produces p-values that are {\it exactly uniform} under the
null\footnote{Specifically, the TG test here tests the hypotheses
  \smash{$H_{0,j} : \bar\theta_{(I_{j-1}):(I_j-1)} =
    \bar\theta_{(I_j):(I_{j+1}-1)}$}, $j=1,2$; this is what we call the {\it
    segment test} in Section \ref{sec:1dfusedlasso}.}, conditional on $I_1,I_2$
having been selected by the fused lasso in the first place.  We now see that
only the changepoint at location $I_2$ has a small associated p-value.

\subsection{Summary}

In this paper, we make the following contributions.

\begin{itemize}
\item We introduce the usage of post-selection inference tools to selection
  events defined by a class of methods called {\it generalized lasso}
  estimators. The key mathematical task is to show that the model selection
  event defined by any (fixed) step the generalized lasso solution path can be
  expressed as a polyhedron in the observation vector $y$ (Section
  \ref{sec:kfixed}). The (conditionally valid) TG tests and confidence intervals
  of \citet{exactlasso,exactlar} can then be applied, to test or cover any
  linear contrast of the mean vector $\theta$.

\item We describe a stopping rule based on a generic information criterion (akin
  to AIC or BIC), to select a step along the generalized lasso path at which we
  are to perform conditional inference.  We give a polyhedral representation for
  the ultimate model selection event that encapsulates both the selected path
  step and the generalized lasso solution at this step (Section
  \ref{sec:kchosen}). Along with the TG tests and confidence intervals, this
  makes for a practical (nearly-automatic) and broadly applicable set
  of inference tools.

\item We study various special cases of the generalized lasso problem---namely,
  the 1d fused lasso, trend filtering, graph fused lasso, and regression
  problems---and for each, we develop specific forms for linear contrasts that
  can be used to test different population quantities of interest (Sections
  \ref{sec:1dfusedlasso} through \ref{sec:regression}). In each case, we believe
  that our tests provide new advancements in the set of currently available
  inferential tools. For example, in the 1d fused lasso, i.e., the changepoint
  detection problem, our tests are the first that we know of that are
  specifically designed to yield proper inferences {\it after} changepoint
  locations have been detected.

\item We present two of extensions of the basic tools described above for
  post-selection inference in generalized lasso problems:
  a post-processing tool, to improve the power of our
  methods, and a visualization aid, to improve practical useability. 

\item We conduct a comprehensive simulations across the various special problem
  cases, to investigate the (conditional) power of our methods, and verify their
  (conditional) type I error control (Sections \ref{sec:1dflexamples} through
  \ref{sec:regressionexample}). We also demonstrate a realistic
  application of our selective inference tools for changepoint detection to a
  data set of comparative genomic hybridization (CGH) measurements
  from two glioblas-toma multiforme (GBM) tumors (Section
  \ref{sec:cgh}). 
\end{itemize}

\subsection{Related work}
\label{sec:related}

Post-selection inference, also known as selective inference, is a new
but rapidly growing field.  Unlike other recent developments in
high-dimensional inference using a more classic full-population model,
the point of selective inference is to provide a means of testing
hypotheses derived from a {\it selected model}, the output of an
algorithm that has been applied to data at hand.  In a sequence of
papers, \citet{lp2003,lp2006,lp2008} prove impossibility results about
estimating the post-selection distribution of certain estimators in a
classical regression setting.  \citet{berk2013}, \citet{lockhart2014}
circumvent this by more directly conducting inference on
post-selection targets (rather than basing inference on the
distribution of the post-selection estimator itself).  The former work
is very broad and considers {\it all} selection mechanisms in
regression (hence yielding more conservative inference); the latter is
much more specific and considers the lasso estimator in particular.
\citet{exactlasso,exactlar} improve on the method in
\citet{lockhart2014}, and introduced a pivot-based framework for
post-selection inference. \citet{exactlasso} describe the application
to the lasso problem at a fixed tuning parameter $\lambda$;
\citet{exactlar} describe the application to the lasso path at a fixed
number of steps $k$ (and also, the least angle regression and forward
stepwise paths).  A number of extensions to different problem settings
are given in \citet{exactscreen,exactmeans,exactstep,exactpca}.
Asymptotics for non-Gaussian error distributions are presented in
\citet{tian2015asymptotics,tibshirani2015uniform}.  A broad treatment
of selective inference in exponential family models and selective
power is presented in \citet{optimalinf}.  An improvement based on
auxiliary randomization is considered in \citet{tian2015randomized}.
A study of selective sequential tests and stopping rules is given in
\citet{fithian2015selective}.  Ours is the first work to consider
selective inference in structural problems like the generalized
lasso. 

Changepoint detection carries a huge body of literature; reviews can
be found in, e.g., \citet{brodsky1993,chen2000,eckley2011}.   
Far sparser is the literature on changepoint inference, say, inference
for the location or size of changepoints, or segment lengths. 
\citet{Hinkley1970, Worsley1986, Bai1999} are some examples, and
\citet{Jandhyala2013,Horvath2014} provide nice surveys and
extensions. The main tools are built around likelihood ratio test
statistics comparing two nested changepoint models, but at {\it fixed}
locations. Since interesting locations to be tested are typically
estimated, these inferences can be clearly invalid (if estimation and
inference are both done on the same data samples). 

Probably most relevant to our goal of valid post-selection changepoint
inference is \citet{Frick2014}, who develop a simultaneous confidence
band for the mean in a changepoint model. Their Simultaneous
Multiscale Changepoint Estimator (SMUCE) seeks the most parsimonious
piecewise constant fit subject to an upper limit on a certain
multiscale statisic, 
and is solved via dynamic programming. Because the final confidence
band has simultaneous coverage (over all components of the mean), it
also has valid coverage for any (data-dependent) post-selection
target.  In contrast, our proposal does not give simultaneous coverage
of the mean, but rather, selective coverage of a particular
post-selection target.   An empirical comparison between the two
methods (SMUCE, and ours) is given in Section \ref{sec:smuce}.  While
this comparison is useful and informative, it is also worth
emphasizing that the framework in this paper applies far outside of
the changepoint detection problem, i.e., to trend filtering, graph
clustering, and regression problems with structured coefficients.

\subsection{Notation}
\label{sec:notation}

For a matrix $D$, we will denote by $D_S$ the submatrix whose rows are
in a set $S \subseteq \{1,\ldots,m\}$. We write $D_{-S}$ to mean
\smash{$D_{S^c} = D_{\{1,\ldots,m\} \setminus S}$}.  Similarly, for a
vector $x$, we write $x_S$ or $x_{-S}$ to extract the subvector whose
components are in $S$ or not in $S$, respectively.  
We use $A^+$ for the pseudoinverse of a matrix $A$, and $\row(A)$,
$\col(A)$, $\nul(A)$ for the column space, row space, and null space
of $A$, respectively. We write $P_L$ for the projection matrix onto a
linear subspace $L$.  Lastly, we will often abbreviate a sequence
$(x_1,\ldots,x_k)$ by $x_{1:k}$. 

\section{Preliminaries}
\label{sec:review}

\subsection{The generalized lasso regularization path}
\label{sec:genlasso}

Given a response $y \in \R^n$, the {\it generalized lasso} estimator is defined by the optimization problem 
\begin{equation}
\label{eq:genlasso}
\hbeta = \argmin_{\beta \in \R^n} \; \half \|y-X\beta\|_2^2 + \lambda
\|D\beta\|_1,
\end{equation}
where $X \in \R^{n\times p}$ is a matrix of predictors, $D \in \R^{m\times p}$ is a prespecified penalty matrix, and $\lambda \geq 0$ is  a regularization parameter.  This matrix $D$ is chosen so that sparsity of \smash{$D\hbeta$} induces some type of desired structure in the solution \smash{$\hbeta$} in \eqref{eq:genlasso}.  Important special cases, each corresponding to a specific class of matrices $D$, include the 1d fused lasso, trend filtering, and graph fused lasso problems.  More details on these problems is given in Section \ref{sec:applications}; see also Section 2 in \citet{genlasso}.  

We review the algorithm of \citet{genlasso} to compute the entire solution path in \eqref{eq:genlasso}, i.e., the continuum of solutions \smash{$\hbeta(\lambda)$} as the regularization parameter $\lambda$ desends from $\infty$ to 0.  We focus on the problem of {\it signal approximation}, where $X=I$: 
\begin{equation}
\label{eq:signalapprox}
\hbeta = \argmin_{\beta \in \R^n} \; \half \|y-\beta\|_2^2 + \lambda
\|D\beta\|_1.
\end{equation}
For a general $X$, a simple modification to the arguments used for \eqref{eq:signalapprox} will deliver the solution path for \eqref{eq:genlasso}, and we refrain from describing this until Section \ref{sec:regression}.  The path algorithm of \citet{genlasso} for \eqref{eq:signalapprox} is derived from the perspective of its equivalent Lagrange dual problem, namely
\begin{equation}
\label{eq:dual}
\hu = \argmin_{u \in \R^m} \;  \|y-D^T u\|_2^2 \;\; \st \;\;
\|u\|_\infty \leq \lambda.
\end{equation}
The primal and dual solutions, \smash{$\hbeta$} in \eqref{eq:signalapprox} and \smash{$\hu$} in \eqref{eq:dual}, are related by
\begin{equation}
\label{eq:primaldual}
\hbeta = y-D^T \hu,
\end{equation}
as well as 
\begin{equation}
\label{eq:dualsubg}
\hu_i \in \begin{cases}
\{+1\} & \text{if $(D\hbeta)_i > 0$} \\
\{-1\} & \text{if $(D\hbeta)_i < 0$} \\
[-1,1] & \text{if $(D\hbeta)_i = 0$}
\end{cases}, \quad i=1,\ldots,m.
\end{equation}
The strategy is now to compute a solution path \smash{$\hu(\lambda)$} in the dual problem, as $\lambda$ descends from $\infty$ to 0, and then use \eqref{eq:primaldual} to deliver the primal solution path.  Therefore it suffices to describe the path algorithm as it operates on the dual problem; this is given next.  

\begin{algorithm}[\textbf{Dual path algorithm for the generalized
    lasso, $X=I$}]
\label{alg:dualpath}
\hfill\par
\smallskip
\smallskip
Given $y \in \R^n$ and $D \in \R^{m\times n}$.
\begin{enumerate}
\item Compute \smash{$\hu = (DD^T)^+ Dy$}, and compute the first hitting time,
\begin{equation*}
\lambda_1 = \max_{i=1,\ldots,m} \; |\hu_i|.
\end{equation*}
Define the hitting coordinate $i_1$ to be the argmax of the above expression, and define the hitting sign \smash{$r_1=\sign(\hu_{i_1})$}.  Initialize the boundary set $\cB_1=\{i_1\}$ and the boundary sign list $s_{\cB_1}=(r_1)$. Record the solution as \smash{$\hu(\lambda)=\hu$} over $\lambda \in [\lambda_1, \infty)$, and set $k=1$. 

\item While $\lambda_k>0$:
\begin{enumerate}
\item Compute \smash{$a = (D_{-\cB_k}D_{-\cB_k}^T)^+ D_{-\cB_k} y$} and  
\smash{$b =  (D_{-\cB_k}D_{-\cB_k}^T)^+ D_{-\cB_k} D_{\cB_k}^T s_{\cB_k}$}.  Also define
\begin{align*}
c &=\diag\big(s_{\cB_k}\big)  D_{\cB_k} (y - D_{-\cB_k}^T a), \\
d &= \diag\big(s_{\cB_k}\big) D_{\cB_k} ( D_{\cB_k}^T s - D_{-\cB_k}^T b).
\end{align*}

\item Compute the next hitting time,
\begin{equation}
\label{eq:hittingtime}
\lambda_{k+1}^{\mathrm{hit}} = 
\max_{i \notin \cB_k, \; r \in \{-1,1\}} \; 
\frac{a_i}{r+b_i} \cdot 
\one \Bigg\{ \frac{a_i}{r+b_i} \geq 0 \Bigg\}.
\end{equation}
Define the hitting coordinate \smash{$i^{\mathrm{hit}}_{k+1}$} and
hitting sign \smash{$r^{\mathrm{hit}}_{k+1}$} to be the pair achieving
the maximum in the above expression.  

\item Compute the next leaving time,
\begin{equation}
\label{eq:leavingtime}
\lambda_{k+1}^{(\mathrm{leave})} = \argmax_{i \in \cB_k} \;
\frac{c_i}{d_i} \cdot 
\one \Big\{ c_i \leq 0, \; d_i < 0 \Big\},
\end{equation}
Define the leaving coordinate \smash{$i^{\mathrm{leave}}_{k+1}$} to be the argmax of the above expression, and define the leaving sign 
\smash{$r^{\mathrm{leave}}_{k+1} = r_{i^{\mathrm{leave}}_{k+1}}$}.

\item Define the next knot according to
\begin{equation}
\label{eq:nextknot}
\lambda_{k+1}=\max\Big\{ 
\lambda_{k+1}^{(\mathrm{hit})}, \; \lambda_{k+1}^{(\mathrm{leave})} \Big\}.
\end{equation}
If the next hitting time is larger, \smash{$\lambda_{k+1}^{(\mathrm{hit})} \geq \lambda_{k+1}^{(\mathrm{leave})}$}, then define the new boundary set $\cB_{k+1}$ by appending the hitting coordinate \smash{$i^{\mathrm{hit}}_{k+1}$} to $\cB_k$, and define the new boundary sign list \smash{$s_{\cB_{k+1}}$} by appending the hitting sign \smash{$r^{\mathrm{hit}}_{k+1}$} to \smash{$s_{\cB_k}$}. Otherwise, define $\cB_{k+1}$ by removing the leaving coordinate from \smash{$i^{\mathrm{leave}}_{k+1}$} from $\cB_k$ and define \smash{$s_{\cB_{k+1}}$} by removing the leaving sign \smash{$r^{\mathrm{leave}}_{k+1}$} from \smash{$s_{\cB_k}$}.  Record the solution as \smash{$\hu(\lambda) = a-\lambda b$} over $\lambda \in [\lambda_{k+1},\lambda_k]$, and update $k=k+1$.
\end{enumerate}
\end{enumerate}
\end{algorithm}

Explained in words, the dual path algorithm in Algorithm \ref{alg:dualpath} tracks the coordinates of the computed dual solution $\hu(\lambda)$ that are equal to $\pm \lambda$, i.e., that lie on the boundary of the constraint region $[-\lambda,\lambda]^m$.  The collection of such coordinates, at any given step $k$ in the path, is called the {\it boundary set}, and is denoted $\cB_k$.  Critical values of the regularization parameter at which the boundary set changes (i.e., at which coordinates join or leave the boundary set) are called {\it knots}, and are denoted $\lambda_1 \geq \lambda_2 \geq \ldots \geq 0$.  From the form of the dual solution \smash{$\hat{u}(\lambda)$} as presented in Algorithm \ref{alg:dualpath}, and the primal-dual relationship \eqref{eq:primaldual}, the primal solution path may be expressed in terms of the current boundary set $\cB_k$ and boundary sign list \smash{$s_{\cB_k}$}, as in
\begin{equation}
\label{eq:primalsol}
\hbeta(\lambda) = P_{\nul(D_{-\cB_k})} (y - \lambda D_{\cB_k}^T s_{\cB_k})
\quad \text{for $\lambda \in [\lambda_{k+1},\lambda_k]$},
\end{equation}
As we can see, the primal solution lies in the subspace \smash{$\nul(D_{-\cB_k})$}, which implies it expresses a certain type of structure. This will become more concrete as we look at specific cases for $D$ in Section \ref{sec:applications}, but for now, the important point is that the structure of the generalized lasso solution \eqref{eq:primalsol} is determined by the boundary set $\cB_k$.  Therefore, by conditioning on the observed boundary set $\cB_k$ after a certain number of steps $k$ of the path algorithm, we are effectively conditioning of the observed {\it model structure} in the generalized lasso solution at this step.  This is essentially what is done in Section \ref{sec:inference}.

Lastly, we note the following important point.  In some generalized lasso problems, Step 2(c) in Algorithm \ref{alg:dualpath} does not need to be performed, i.e., we can formally replace this step by \smash{$\lambda_{k+1}^{\mathrm{leave}}=0$}, and accordingly, the boundary set $\cB_k$ will only grow over iterations $k$.  This is true, e.g., for all 1d fused lasso problems; more generally, it is true for any generalized lasso signal approximator problem in which $DD^T$ is diagonally dominant.

\subsection{Exact inference after polyhedral conditioning}
\label{sec:polyhedral}

Under the Gaussian observation model in \eqref{eq:model}, \citet{exactlasso,exactlar} build a framework for inference on an arbitrary linear constrast $v^T \theta$ of the mean $\theta$, conditional on $y \in G$, where $G \subseteq \R^n$ is an arbitrary polyhedron.  A core tool in these works is an exact pivotal statistic for $v^T \theta$, conditional on $y \in G$: they prove that there exists random variables \smash{$\cV^{\mathrm{lo}}, \cV^{\mathrm{up}}$} such that
\begin{equation}
\label{eq:pivot}
F^{[\cV^{\mathrm{lo}}, \cV^{\mathrm{up}}]}_{v^T \theta, \sigma^2 \|v\|_2^2} (v^T y) \; \Big| \; y \in G \; \sim \mathrm{Unif}[0,1],
\end{equation}
where \smash{$F^{[a,b]}_{\mu, \tau^2}$} denotes the cumulative distribution function of a univariate Gaussian random variable $Z \sim \cN(\mu,\tau^2)$ conditional on lying in the interval $[a,b]$.  The above is called the {\it truncated Gaussian} (TG) pivot.  The truncation limits \smash{$\cV^{\mathrm{lo}}, \cV^{\mathrm{up}}$} are easily computable, given a half-space representation for the polyhedron $G=\{ x : \Gamma x \geq w\}$, where $\Gamma \in \R^{q \times n}$ and $w \in \R^n$ and the inequality here is to be interpreted componentwise.  Specifically, we have 
\begin{align*}
\cV^{\mathrm{lo}} (y) &= v^T y - \min_{j : \rho_j > 0} \frac{(\Gamma y)_j - w_j}{\rho_j}, \\
\cV^{\mathrm{up}} (y) &= v^T y - \max_{j : \rho_j < 0} \frac{(\Gamma y)_j - w_j}{\rho_j},
\end{align*}
where $\rho = \Gamma v / \|v\|^2$. The TG pivotal statistic in \eqref{eq:pivot} enables us to test the null hypothesis $H_0 : v^T \theta = 0$ against the one-sided alternative $H_1 : v^T \theta > 0$.  Namely, it is clear that the TG test statistic 
\begin{equation}
\label{eq:tgstatistic}
T = 1 - F^{[\cV^{\mathrm{lo}}, \cV^{\mathrm{up}}]}_{0, \sigma^2 \|v\|_2^2} (v^T y) 
\end{equation}
is itself a p-value for $H_0$, with finite sample validity, conditional on $y \in G$.  (A two-sided test is also possible: we simply use $2 \min \{T, 1-T\}$ as our p-value; see \citet{exactlar} for a discussion of the merits of one-sided and two-sided selective tests.)  Confidence intervals follow directly from \eqref{eq:pivot} as well.  For an (equi-tailed) interval with exact finite sample coverage $1-\alpha$, conditional on the event $y \in G$, we take $[\delta_{\alpha/2}, \delta_{1-\alpha/2}]$, where $\delta_{\alpha/2}, \delta_{1-\alpha/2}$ are obtained by inverting the TG pivot, i.e., defined to satisfy
\begin{equation}
\label{eq:tginterval}
\begin{aligned}
1-F^{[\cV^\mathrm{lo}, \cV^\mathrm{up}]}_{\delta_{\alpha/2}, \sigma^2 \|v\|_2^2} (v^T y) &= \alpha/2, \\ 
1-F^{[\cV^\mathrm{lo}, \cV^\mathrm{up}]}_{\delta_{1-\alpha/2}, \sigma^2 \|v\|_2^2} (v^T y) &= 1-\alpha/2.
\end{aligned}
\end{equation}

At this point, it may seem unclear how this framework applies to
post-selection inference in generalized lasso problems.  The key
ingredients are, of course, the polyhedron $G$ and the contrast vector
$v$.  In the next section, we will show how to construct polyhedra
that correspond to model selection events of interest, at points along
the generalized lasso path.  In the following section, we will suggest
choices of contrast vectors that lead to interesting and useful tests
in specific settings, such as the 1d fused lasso, trend filtering, and
graph fused lasso problems. 

\subsection{Can we not just use lasso inference tools?}
\label{sec:whynotlasso}

When the penalty matrix $D$ is square and invertible, the generalized
lasso problem \eqref{eq:genlasso} is equivalent to a lasso problem, in
the variable $\alpha=D\beta$, with design matrix $XD^{-1}$.  More
generally, when $D$ has full row rank, problem \eqref{eq:genlasso} is
reducible to a lasso problem (see \citet{genlasso}).  In this case,
existing inference theory for the lasso path (from \citet{exactlar})
could be applied to the equivalent lasso problem, to perform
post-selection inference on generalized lasso models.  This covers
inference for the 1d fused lasso and trend filtering problems.  But
when $D$ is row rank deficient (when it has more rows than columns),
the generalized lasso is not equivalent to a lasso problem (see again
\citet{genlasso}), and we cannot simply resort to lasso inference
tools. This would hence rule out treating problems like the 2d fused
lasso, the graph fused lasso (for any graph with more edges than
nodes), the sparse 1d fused lasso, and sparse trend filtering from a
pure lasso perspective. Our paper presents a unified treatment of
post-selection inference across all generalized lasso problems,
regardless of the penalty matrix $D$. 


\section{Inference along the generalized lasso path}
\label{sec:inference}

\subsection{The selection event after a given number of steps $k$}
\label{sec:kfixed}

Here, we suppose that we have run a given (fixed) number of steps $k$ of the generalized lasso path algorithm, and we have a contrast vector $v$ in mind, such that $v^T \theta$ is a parameter of interest (to be tested or covered).  Define the {\it generalized lasso model} at step $\ell$ of the path to be
\begin{equation*}
M_\ell = (\cB_\ell, s_{\cB_\ell}, R^{\mathrm{hit}}_\ell, I^{\mathrm{leave}}_\ell),
\end{equation*}
where \smash{$\cB_\ell,s_{\cB_\ell}$} are the boundary set and signs at step $\ell$, and \smash{$R^{\mathrm{hit}}_\ell, I^{\mathrm{leave}}_\ell$} are quantities to be defined shortly.  We will show that the entire {\it model sequence} from steps $\ell=1,\ldots,k$, denoted $M_{1:k} = (M_1,\ldots,M_k)$, is a polyhedral set in $y$. By this we mean the following: if \smash{$\widehat{M}_{1:k}(y)$} denotes the model sequence as a function of $y$, and $M_{1:k}$ a given realization, then the set 
\begin{equation*}
G_k = \{ y : \widehat{M}_{1:k}(y) = M_{1:k} \}
\end{equation*} 
is a polyhedron, more specifically, a convex cone, and can therefore be expressed as $G_k = \{y : \Gamma y \geq 0\}$ for a matrix $\Gamma=\Gamma(M_{1:k})$ that we will show how to construct, based on $M_{1:k}$.

Our construction uses induction.  When $k=1$, and we write $\cB_1=\{i_1\}$ and \smash{$s_{\cB_1}=(r_1)$}, it is clear from the first step of Algorithm \ref{alg:dualpath} that $(i_1,r_1)$ is the hitting coordinate-sign pair if and only if
\begin{align*}
r_1 [(DD^T)^+D]_{i_1} \, y &\geq [(DD^T)^+ D]_i \, y, \quad i \neq i_1, \\
r_1 [(DD^T)^+D]_{i_1} \, y &\geq -[(DD^T)^+ D]_i \, y, \quad i \neq i_1.
\end{align*}
Hence we can construct $\Gamma(M_1)$ to have the corresponding
$2(m-1)$ rows---to be explicit, these are \smash{$r_1
  [(DD^T)^+D]_{i_1} \pm [(DD^T)^+ D]_i$}, $i \neq i_1$.  We note that
at the first step, there is no characterization needed for
\smash{$R^{\mathrm{hit}}_1$} and \smash{$I^{\mathrm{leave}}_1$} (for
simplicity, we may think of these as being empty sets). 

Now assume that, given a model sequence
$M_{1:(k+1)}=(M_1,\ldots,M_{k+1})$, we have constructed a polyhedral
representation for \smash{$G_k= \{ y : \widehat{M}_{1:k}(y) = M_{1:k}
  \}$}, i.e., we have constructed a matrix $\Gamma(M_{1:k})$ such that
$G_k = \{ y : \Gamma(M_{1:k}) \geq 0\}$.  To show that $G_{k+1} = \{ y
: \Gamma(M_{1:(k+1)}) \geq 0\}$ can also be written in the analogous
form, we will define $\Gamma(M_{1:{k+1}})$ by appending rows to
$\Gamma(M_{1:k})$ that capture the generalized lasso model at step
$k+1$ of Algorithm \ref{alg:dualpath}.  We will add rows to
characterize the hitting time \eqref{eq:hittingtime}, leaving time
\eqref{eq:leavingtime}, and the next action (either hitting or
leaving) \eqref{eq:nextknot}.  Keeping with the notation in
\eqref{eq:hittingtime}, a simple argument shows that the next hitting
time can be alternatively written as  
\begin{equation*}
\lambda_{k+1}^{\mathrm{hit}} = 
\max_{i \notin \cB_k} \;
\frac{a_i}{\sign(a_i)+b_i}.
\end{equation*}
Plugging in for
$a,b$, we may characterize the {\it viable hitting signs} 
at step $k+1$, 
\smash{$R^{\mathrm{hit}}_{k+1} = \{r_{k,i} : i \notin
  \cB_k\}$}, as well as the next hitting coordinate and
hitting sign, 
\smash{$i^{\mathrm{hit}}_{k+1}$} and \smash{$r^{\mathrm{hit}}_{k+1}$},
by the following inequalities: 
\begin{align*}
r_{k,i} \, [(D_{-\cB_k}D_{-\cB_k}^T)^+ D_{-\cB_k}]_i \, y &\geq 0, \quad i \notin \cB_k, \\
\frac{[(D_{-\cB_k}D_{-\cB_k}^T)^+ D_{-\cB_k}]_{i^{\mathrm{hit}}_{k+1}} \, y }{ r_k^\mathrm{hit} + 
[(D_{-\cB_k}D_{-\cB_k}^T)^+ D_{-\cB_k}]_{i^{\mathrm{hit}}_{k+1}} \, D_{\cB_k}^T s_{\cB_k}} &\geq  
\frac{[(D_{-\cB_k}D_{-\cB_k}^T)^+ D_{-\cB_k}]_i \, y }{ r_{k,i} + 
[(D_{-\cB_k}D_{-\cB_k}^T)^+ D_{-\cB_k}]_i \, D_{\cB_k}^T s_{\cB_k}}, \quad i \notin \cB_k.
\end{align*}
This corresponds to $2(m-|\cB_k|)$ rows to be appended to $\Gamma(M_{1:k})$.  

For \eqref{eq:leavingtime}, we first define the {\it viable leaving coordinates}, denoted \smash{$I^{\mathrm{leave}}_{k+1}$}, by the subset of $i \in \cB_k$ for which $c_i<0$ and $d_i<0$.  We may write \smash{$I^{\mathrm{leave}}_{k+1}=C^{\mathrm{leave}}_{k+1} \cap D^{\mathrm{leave}}_{k+1}$}, where \smash{$C^{\mathrm{leave}}_{k+1}$} is the set of $i$ for which $c_i<0$, and \smash{$D^{\mathrm{leave}}_{k+1}$} is the set of $i$ for which $d_i<0$.  Plugging in for $c,d$, we notice that only the former set \smash{$C^{\mathrm{leave}}_{k+1}$} depends on $y$, and \smash{$D^{\mathrm{leave}}_{k+1}$} is deterministic once we have characterized $M_{1:k}$.  This gives rise to the following inequalities determining \smash{$I^{\mathrm{leave}}_{k+1}=C^{\mathrm{leave}}_{k+1} \cap D^{\mathrm{leave}}_{k+1}$}:
\begin{align*}
\big[\diag\big(s_{\cB_k}\big)  D_{\cB_k} \big(I - D_{-\cB_k}^T (D_{-\cB_k}D_{-\cB_k}^T)^+ D_{-\cB_k}\big)\big]_i \,y &\leq 0, \quad i \in C^{\mathrm{leave}}_{k+1} \cap D^{\mathrm{leave}}_{k+1}, \\
\big[\diag\big(s_{\cB_k}\big)  D_{\cB_k} \big(I - D_{-\cB_k}^T (D_{-\cB_k}D_{-\cB_k}^T)^+ D_{-\cB_k}\big)\big]_i \,y &\geq 0, \quad i \in \big(C^{\mathrm{leave}}_{k+1}\big)^c \cap D^{\mathrm{leave}}_{k+1}, 
\end{align*}
which corresponds to \smash{$|D^{\mathrm{leave}}_{k+1}| \leq |\cB_k|$} rows to be appended to $\Gamma(M_{1:k})$.  Given this characterization for \smash{$I^{\mathrm{leave}}_{k+1}$}, we may now characterize the next leaving coordinate \smash{$i^{\mathrm{leave}}_{k+1}$} by:
\begin{multline*}
\frac{\big[\diag\big(s_{\cB_k}\big)  D_{\cB_k} \big(I - D_{-\cB_k}^T (D_{-\cB_k}D_{-\cB_k}^T)^+ D_{-\cB_k}\big)\big]_{i^{\mathrm{leave}}_{k+1}} \,y} 
{\big[\diag\big(s_{\cB_k}\big)  D_{\cB_k} \big(I - D_{-\cB_k}^T (D_{-\cB_k}D_{-\cB_k}^T)^+ D_{-\cB_k}\big)\big]_{i^{\mathrm{leave}}_{k+1}} \, D_{\cB_k}^T s_{\cB_k}} \geq \\
\frac{\big[\diag\big(s_{\cB_k}\big)  D_{\cB_k} \big(I - D_{-\cB_k}^T (D_{-\cB_k}D_{-\cB_k}^T)^+ D_{-\cB_k}\big)\big]_i \,y} 
{\big[\diag\big(s_{\cB_k}\big)  D_{\cB_k} \big(I - D_{-\cB_k}^T (D_{-\cB_k}D_{-\cB_k}^T)^+ D_{-\cB_k}\big)\big]_i \, D_{\cB_k}^T s_{\cB_k}}, \quad i \in I^{\mathrm{leave}}_{k+1}.
\end{multline*}
This corresponds to \smash{$|I^{\mathrm{leave}}_{k+1}| \leq |\cB_k|$} rows that must be appended to $\Gamma(M_{1:k})$.  Recall that the leaving coordinate is given by  \smash{$r^{\mathrm{leave}}_{k+1} = r_{i^{\mathrm{leave}}_{k+1}}$}.

Lastly, for \eqref{eq:nextknot}, we either use 
\begin{multline*}
\frac{[(D_{-\cB_k}D_{-\cB_k}^T)^+ D_{-\cB_k}]_{i^{\mathrm{hit}}_{k+1}} \, y }{ r_k^\mathrm{hit} + 
[(D_{-\cB_k}D_{-\cB_k}^T)^+ D_{-\cB_k}]_{i^{\mathrm{hit}}_{k+1}} \, D_{\cB_k}^T s_{\cB_k}} \geq \\
\frac{\big[\diag\big(s_{\cB_k}\big)  D_{\cB_k} \big(I - D_{-\cB_k}^T (D_{-\cB_k}D_{-\cB_k}^T)^+ D_{-\cB_k}\big)\big]_{i^{\mathrm{leave}}_{k+1}} \,y} 
{\big[\diag\big(s_{\cB_k}\big)  D_{\cB_k} \big(I - D_{-\cB_k}^T (D_{-\cB_k}D_{-\cB_k}^T)^+ D_{-\cB_k}\big)\big]_{i^{\mathrm{leave}}_{k+1}} \, D_{\cB_k}^T s_{\cB_k}} 
\end{multline*}
if \smash{$\lambda^{\mathrm{hit}}_{k+1} \geq \lambda^{\mathrm{leave}}_{k+1}$}, or the above with the inequality sign flipped, if \smash{$\lambda^{\mathrm{hit}}_{k+1} < \lambda^{\mathrm{leave}}_{k+1}$}.  In either case, only one more row is to be appended to $\Gamma(M_{1:k})$.  This completes the inductive proof.

It is worth noting that, in the inductive step that constructs $\Gamma(M_{1:(k+1)})$ by appending rows to $\Gamma(M_{1:k})$, we append a total of at most $2(m-|\cB_k|)+2|\cB_k|+1 = 2m+1$ rows.  Therefore after $k+1$ steps, the polyhedral representation for the model sequence $M_{1:(k+1)}$ uses a matrix $\Gamma(M_{1:(k+1)})$ with at most $(2m+1)(k+1)$ rows.

Combining the results of this subsection with the TG pivotal statistic from Section \ref{sec:polyhedral}, we are now equipped to perform conditional inference on the model that is selected at any fixed step $k$ of the generalized lasso path.  (Recall, we are assuming that a reasonable contrast vector $v$ has been determined such that $v^T \theta$ is a quantity of interest in the $k$-step generalized lasso model; in-depth discussion of reasonable choices of contrast vectors, for particular problems, is given in Section \ref{sec:applications}.)  Of course, the choice of which step $k$ to analyze is somewhat critical.  The high-level idea is to fix a step $k$ that is large enough for the selected model to be interesting, but not so large that our tests will be low-powered.  In some practical applications, choosing $k$ a priori may be natural; e.g., in the 1d fused lasso problem, where the selected model correponds to detected changepoints (as discussed in the introduction), we may choose (say) $k=10$ steps, if in our particular setting we are interested in detecting and performing inference on at most 10 changepoints.  But in most practical applications, fixing a step $k$ a priori is likely a difficult task.  Hence, we present a rigorous strategy that allows the choice of $k$ to be data-driven, next.

\subsection{The selection event after an IC-selected 
number of steps $k$} 
\label{sec:kchosen}

We develop approaches based on a generic information criterion (IC), like AIC or
BIC, for selecting a number of steps $k$ along the generalized path that admits
a ``reasonable'' model.  By ``reasonable'', our IC approach admits a $k$-step
generalized lasso solution balances training error and some notion of
complexity.  Importantly, we specifically design our IC-based approaches so that
the selection event determining $k$ is itself a polyhedral function of $y$.  We
establish this below.

Defined in terms of a generalized lasso model \smash{$M_k=(\cB_k,
  s_{\cB_k}, R^{\mathrm{hit}}_k, I^{\mathrm{leave}}_k)$} at step $k$,
we consider the general form IC: 
\begin{equation}
\label{eq:ic}
J(M_k) = \|y - P_{\nul(D_{-\cB_k})}y\|_2^2 + P_n\big(
\nuli(D_{-\cB_k}) \big). 
\end{equation}
The first term above is the squared loss between $y$ and its projection onto the
subspace $\nul(D_{-\cB_k})$; recall that the $k$-step generalized lasso solution
itself lies in this subspace, as written in \eqref{eq:primalsol}, and so here we
have replaced the squared loss between $y$ and \smash{$\hbeta(\lambda_k)$} with
the squared error loss between $y$ and the {\it unshrunken} estimate
\smash{$P_{\nul(D_{-\cB_k})} y$}.  (This is needed in order for our eventual
IC-based rule to be equivalent to a polyhedral constraint in $y$, as will be
seen shortly.)  The second term in \eqref{eq:ic} utilizes
\smash{$\nuli(D_{-\cB_k})$}, the dimension of $\nul(D_{-\cB_k})$, i.e., the
dimension of the solution subspace. It hence penalizes the complexity associated
with the $k$-step generalized lasso solution.  Indeed, from
\citet{genlasso,lassodf2}, the quantity \smash{$\nuli(D_{-\cB_k})$} is an
unbiased estimate of the {\it degrees of freedom} of
\smash{$\hbeta(\lambda_k)$}.
Further, $P_n$ is a penalty function that is allowed to depend on $n$ and
$\sigma^2$ (the marginal variance in the observation model
\eqref{eq:model}). Some common choices are: $P_n(d) = 2 \sigma^2 d$, which makes
\eqref{eq:ic} like AIC; $P_n(d) = \sigma^2 d \log{n}$, motivated by BIC; and
\smash{$P_n(d) = \sigma^2(d \log{n} + 2 \gamma \log {n \choose d})$}, where
$\gamma \in (0,1)$ is a parameter to be chosen (say, $\gamma=1/2$ for
simplicity), motivated by extended BIC (EBIC) of \citet{chenchen}.  Beyond
these, any choice of complexity penalty will do as long as $P_n(d)$ is an
increasing function of $d$.

\begin{figure}[htb]
\centering
\includegraphics[width=0.475\textwidth]{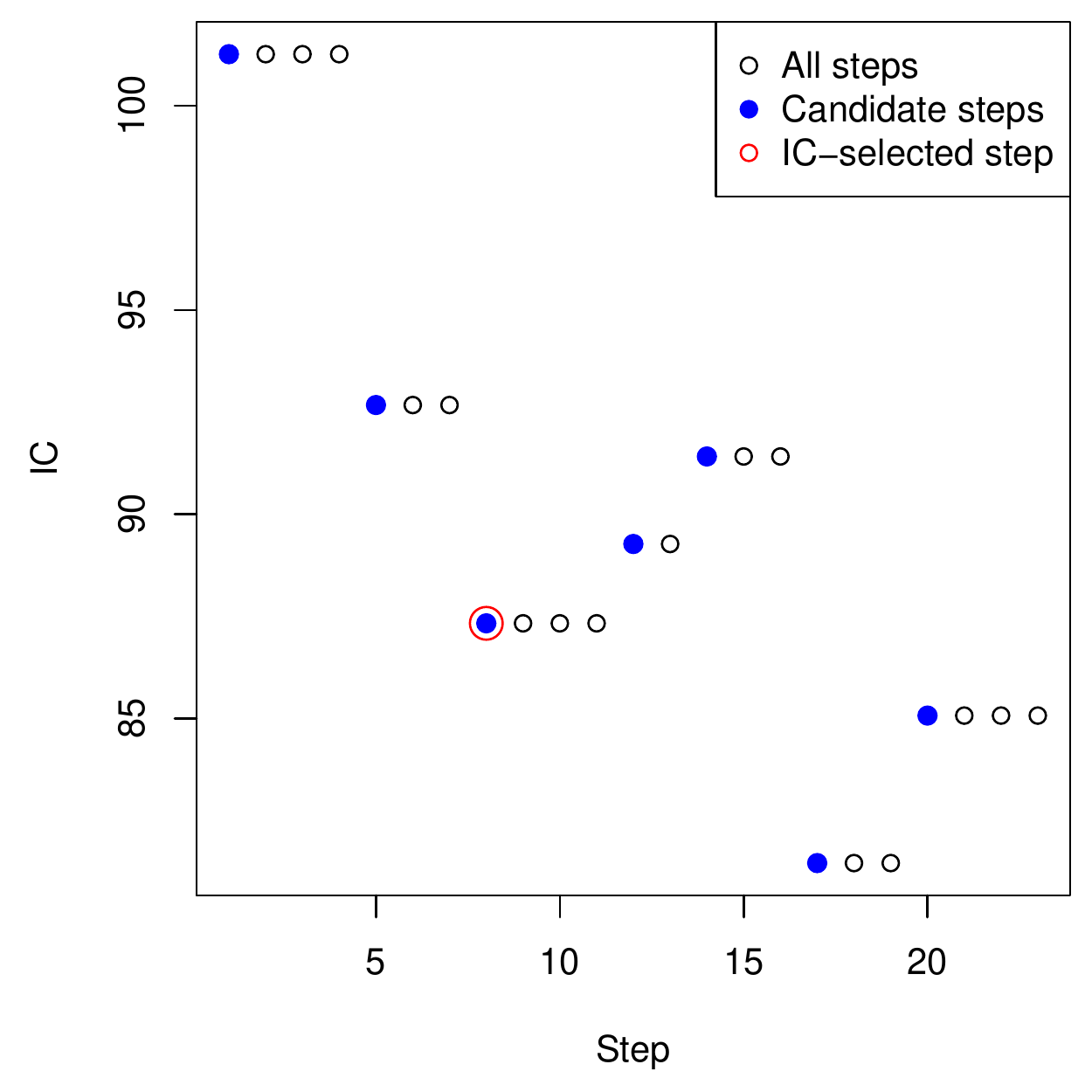}
\hspace{2pt}
\includegraphics[width=0.475\textwidth]{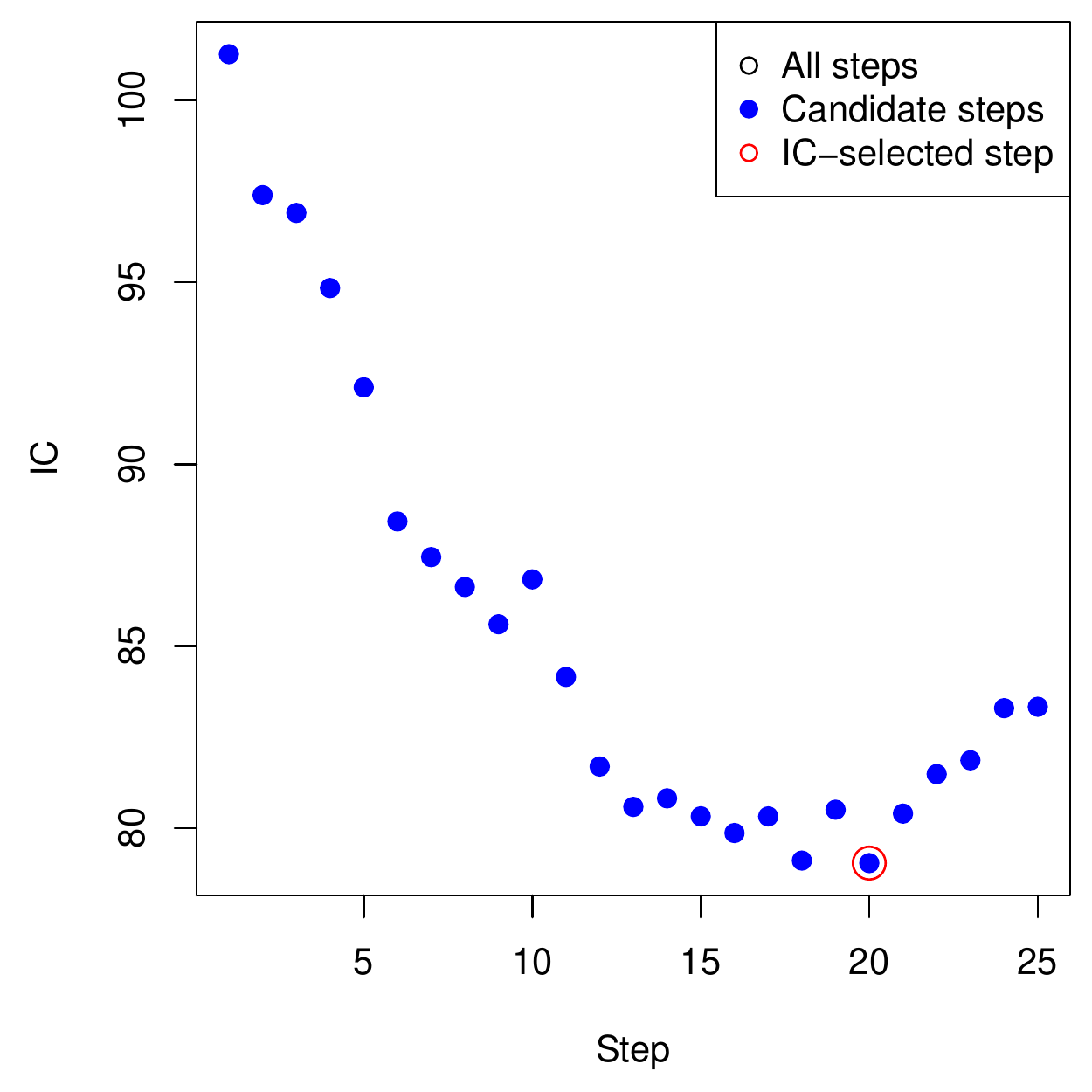}
\caption{\it\small Two illustrations of the IC selection rule; on the
  left, an example in which \smash{$\nul(D_{-\cB_k})$} changes at only
  6 steps (representing, e.g., the 2d fused lasso case); on the right,
  an example in which \smash{$\nul(D_{-\cB_k})$} changes at every step
  (representing, e.g., the 1d fused lasso case).  In both panels,
  solid blue circles mark the candidate set
  \smash{$\widehat{I}^{\mathrm{IC}}(y)$} in
  \eqref{eq:iccandidates}, and a large red circle is drawn around the
  IC-selected step in \eqref{eq:icrule}.}
\label{fig:ic}
\end{figure}

Unfortunately, choosing to stop the path at the step that minimizes the IC
defined in \eqref{eq:ic} does not define a polyhedron in $y$.  Therefore, we use
a modified IC-based rule.  We first define
\begin{equation}
  \label{eq:iccandidates}
  \widehat{I}^{\mathrm{IC}}(y) = \{1\} \cup \Big\{ k \in \{2,3,\ldots\}
  : \nul(D_{-\cB_k}) \neq \nul(D_{-\cB_{k-1}}) \Big\}, 
\end{equation}
the set of steps at which we see action (nonzero adjacent differences) in the
IC.\footnote{For generalized lasso problems in which $D$ is row rank deficient
  (e.g., the 2d fused lasso), it can happen at many path steps $k$ that
  \smash{$\nul(D_{-\cB_k})=\nul(D_{-\cB_{k-1}})$}; for others in which $D$ has
  full row rank (e.g., the 1d fused lasso) each path step $k$ marks a change in
  \smash{$\nul(D_{-\cB_k})$}.  For more, see \citet{genlasso}.}  For
\smash{$k \notin \widehat{I}^{\mathrm{IC}}(y)$}, we have
\smash{$\nul(D_{-\cB_k})=\nul(D_{-\cB_{k-1}})$}, meaning that the structure of
the primal solution is unchanged between steps $k-1$ and $k$, and the IC is
trivially constant as we move across these steps; we will hence restrict our
attention to candidate steps in \smash{$\widehat{I}^{\mathrm{IC}}(y)$} in
crafting our stopping rule. Denoting by $k_1 < k_2 < k_3 < \ldots$ the sorted
elements of \smash{$\widehat{I}^{\mathrm{IC}}(y)$}, we define for each
$j=1,2,3,\ldots$,
\begin{equation*}
\widehat{S}_j(y) =
\sign\big(J(M_{k_j})-J(M_{k_{j+1}})\big), 
\end{equation*}
the sign of the difference in IC values between steps $k_j$ and
$k_{j+1}$ (two adjacent elements in
\smash{$\widehat{I}^{\mathrm{IC}}(y)$} at which the IC values are
known to change nontrivially).  We are now ready to define our
stopping rule, which chooses to stop the path at the step 
\begin{equation}
\label{eq:icrule}
\hat{k}(y) = \min \Big\{ k_j \in \widehat{I}^{\mathrm{IC}}(y) :
\widehat{S}_j(y) = 1, \;
\widehat{S}_{j+1}(y) = 1, \; \ldots, \;
\widehat{S}_{j+q-1}(y) = 1 \Big\},  
\end{equation}
or in words, it chooses the smallest step $k$ such that the IC defined
in \eqref{eq:ic} experiences $q$ successive rises in a row, among the
elements of the candidate set \smash{$\widehat{I}^{\mathrm{IC}}(y)$}.
Here $q \geq 1$ is a prespecified integer; in practice, we have found
that $q=2$ works well in most scenarios.  It helps to see a visual
depiction of the rule, see Figure \ref{fig:ic}. 

We now show that the following set is a polyhedron in $y$,
\begin{equation*}
H = \Big\{ y : 
\widehat{M}_{1:(k_{j+q})}(y) = M_{1:(k_{j+q})}, \;
\hat{k}(y) = k_j, \;
\widehat{S}_{1:(j-1)}(y) = S_{1:(j-1)} \Big\}.
\end{equation*}
(By specifying
\smash{$\widehat{M}_{1:(k_{j+q})}(y) = M_{1:(k_{j+q})}$}, we have also
implicitly specified the first $j+q$ elements of 
\smash{$\widehat{I}^{\mathrm{IC}}(y)$}, and so we do not need to
explicitly include a realization of the latter set in the definition
of $H$.)  In particular, we show that we can express $H=\{ y : 
\Gamma y \geq 0, \; \Lambda y \geq w \}$, for the matrix
\smash{$\Gamma=\Gamma(M_{1:k_{j+q}})$} described in the previous   
subsection, and for 
\smash{$\Lambda=\Lambda(M_{1:(k_j+q)},k_j,S_{1:(j-1)})$} and 
\smash{$w=w(M_{1:(k_j+q)},k_j,S_{1:(j-1)})$}, whose construction is to
be described below.  Since the polyhedron (cone) $\{y : \Gamma y \geq
0\}$ characterizes the part \smash{$H_1 = \{ y :
  \widehat{M}_{1:(k_{j+q})}(y) = M_{1:(k_{j+q})}\}$}, it
suffices to study
\smash{$H_2 = \{y :
  \hat{k}(y) = k_j, \; \widehat{S}_{1:(j-1)}(y) = S_{1:(j-1)}\}$},
given \smash{$M_{1:(k_{j+q})}$}.  And as this part is defined entirely
by pairwise comparisons of IC values, it suffices to show that, for
any $\ell$, 
\begin{align}
\label{eq:iccompare}
J(M_{k_{\ell+1}}) \geq J(M_{k_\ell})
\end{align}
is equivalent to (a pair of) linear constraints on $y$.
By symmetry, if we reverse the inequality sign above, then this will
still be equivalent to linear constraints on $y$, and collecting these
constraints over steps $\ell=1,\ldots,j,\ldots,j+q-1$ gives the
polyhedron that determines $H_2$.  Simply recalling the IC definition in
\eqref{eq:ic}, and rearranging, we find that \eqref{eq:iccompare} is
equivalent to 
\begin{equation}
\label{eq:iccompare2}
y^T \big( P_{\nul(D_{-\cB_{k_\ell}})} - P_{\nul(D_{-\cB_{k_{\ell+1}}})}
\big) y \geq P_n\big(\nuli(D_{-\cB_{k_\ell}})\big) -
P_n\big(\nuli(D_{-\cB_{k_{\ell+1}}})\big). 
\end{equation}
Note that, by construction, the sets $\cB_{k_\ell}$ and
$\cB_{k_{\ell+1}}$ differ by at most one element. For concreteness,
suppose that $\cB_{k_\ell} \subset \cB_{k_{\ell+1}}$; the other
direction is similar.  Then \smash{$\nul(D_{-\cB_{k_\ell}}) \subset
  \nul(D_{-\cB_{k_{\ell+1}}})$}, and the two subspaces are of
codimension 1.  Further, it is not hard to see that the difference in
projection operators \smash{$P_{\nul(D_{-\cB_{k_{\ell+1}}})} -
  P_{\nul(D_{-\cB_{k_\ell}})}$} is itself the projection onto a
subspace of dimension 1.\footnote{This follows because, in 
  general, if $U,V$ are subspaces with $U \subset V$, then \smash{$P_V 
    - P_U = P_V - P_U P_V = P_U^\perp P_V$}, but also \smash{$P_V -
    P_U = P_V - P_V P_U = P_V P_U^\perp$}.  Since the product
  \smash{$P_U^\perp P_V = P_V P_U^\perp$} commutes, it is itself a
  projection matrix, onto the subspace $U^\perp V$.} 
Writing $a_\ell$ for the unit-norm basis vector for this subspace, and
$-b_\ell$ for the right hand side in \eqref{eq:iccompare2}, we see
that \eqref{eq:iccompare2} becomes 
\begin{equation*}
- (a_\ell^T y)^2 \geq -b_\ell,
\end{equation*}
or, since $b_\ell \geq 0$ (this is implied by
\smash{$\nuli(D_{-\cB_{k_{\ell+1}}}) > \nuli(D_{-\cB_{k_\ell}})$}, and
the complexity penalty $P_n$ being an increasing function), 
\begin{equation*}
-\sqrt{b_\ell} \leq a_\ell^T y \leq \sqrt{b_\ell}.
\end{equation*}
This is a pair of linear constraints on $y$, and we have verified the desired fact.

Altogether, with the final polyhedron $H$, we can use the TG pivot
from Section \ref{sec:polyhedral} to perform valid inference on linear
contrasts $v^T \theta$ of the mean $\theta$, conditional on having
chosen step $k$ with our IC-based stopping rule, and on having
observed a given model sequence over the first $k$ steps of the
generalized lasso path. 

\subsection{What is the conditioning set?}
\label{sec:conditioning}

For a fixed $k$, suppose that we have computed $k$ steps of the
generalized lasso path and observed a model sequence
\smash{$\widehat{M}_{1:k}(y)=M_{1:k}$}.  From Section
\ref{sec:kfixed}, we can form a matrix $\Gamma=\Gamma(M_{1:k})$ such
that \smash{$\{y:\widehat{M}_{1:k}(y)=M_{1:k}\} = \{y:\Gamma y \geq
  0\}$}. From Section \ref{sec:polyhedral}, for any vector $v$, we can
invert the TG pivot as in \eqref{eq:tginterval} to compute a
conditional confidence interval
$C_{1-\alpha}=[\delta_{\alpha},\delta_{1-\alpha/2}]$, with the
property 
\begin{equation}
\label{eq:fullconditioning}
\P \Big( v^T \theta \in C_{1-\alpha} \; \Big | \;
\widehat{M}_{1:k}(y)=M_{1:k} \Big) = 1-\alpha. 
\end{equation}
This holds for all possible realizations $M_{1:k}$ of model sequences,
and thus we can marginalize along any dimension to yield a valid
conditional coverage statement.  For example, by marginalizing over
all possible realizations $M_{1:(k-1)}$ of model sequences up to step
$k-1$, we obtain 
\begin{equation}
\label{eq:lessconditioning}
\P \Big( v^T \theta \in C_{1-\alpha} \; \Big | \; \widehat{\cB}_k(y) =
\cB_k, \; \hat{s}_{\cB_k}(y) = s_{\cB_k}, \;
\widehat{R}^{\mathrm{hit}}_k(y) = R^{\mathrm{hit}}_k, \;
\widehat{I}^{\mathrm{leave}}_k(y) = I^{\mathrm{leave}}_k\Big) =
1-\alpha. 
\end{equation}
Above, \smash{$\widehat{\cB}_k(y)$} is the boundary set at step $k$ as
a function of $y$, and likewise \smash{$\hat{s}_{\cB_k}(y),
  \widehat{R}^{\mathrm{hit}}_k(y), \widehat{I}^{\mathrm{leave}}_k(y)$}
are the boundary signs, viable hitting signs, and viable leaving
coordinates at step $k$, respectively, as functions of $y$.  Since a
data analyst typically never sees the viable hitting signs or viable
leaving coordinates at a generalized lasso solution (i.e., these are
``hidden'' details of the path computation, at least compared to the
boundary set and signs, which are reflected in the structure of
solution itself, recall \eqref{eq:primalsol} and \eqref{eq:dualsubg}),
the conditioning event in \eqref{eq:fullconditioning} may seem like it
includes ``unnecessary'' details.  Hence, we can again marginalize
over all possible realizations \smash{$R^{\mathrm{hit}}_k,
  I^{\mathrm{leave}}_k$} to yield  
\begin{equation}
\label{eq:evenlessconditioning}
\P \Big( v^T \theta \in C_{1-\alpha} \; \Big | \; \widehat{\cB}_k(y) =
\cB_k, \; \hat{s}_{\cB_k}(y) = s_{\cB_k} \Big) = 1-\alpha. 
\end{equation}
Among \eqref{eq:fullconditioning}, \eqref{eq:lessconditioning},
\eqref{eq:evenlessconditioning}, the latter is the cleanest statement
and offers the simplest interpretation. This is reiterated when we cover 
specific problem cases in Section \ref{sec:applications}.  

Similar statements hold when $k$ is chosen by our IC-based
rule, from Section \ref{sec:kchosen}. Applying the TG
framework 
from Section \ref{sec:polyhedral} to the full conditioning set,
in order to derive a confidence interval $C_{1-\alpha}$ for $v^T
\theta$, and following a reduction analogous to 
\eqref{eq:fullconditioning}, \eqref{eq:lessconditioning}, 
\eqref{eq:evenlessconditioning}, we arrive at the property
\begin{equation}
\label{eq:evenlessconditioning2}
\P \Big( v^T \theta \in C_{1-\alpha} \; \Big | \; \widehat{\cB}_k(y) = 
\cB_k, \; \hat{s}_{\cB_k}(y) = s_{\cB_k}, \; \hat{k}(y) = k \Big) =
1-\alpha. 
\end{equation}
Again this is a clean conditional coverage statement and offers a 
simple interpretation, for $k$ chosen in a data-driven manner.

\section{Special applications and extensions}
\label{sec:applications}

\subsection{Changepoint detection via the 1d fused lasso} 
\label{sec:1dfusedlasso}

Changepoint detection is an old topic with a vast
literature. It has applications in many areas, e.g., bioinformatics,
climate modeling, finance, and audio and video processing. Instead of
attempting to thoroughly review the changepoint detection literature,
we refer the reader to the comprehensive surveys and reviews in
\citet{brodsky1993,chen2000,eckley2011}.  Broadly speaking, a
changepoint detection problem is one in which the distribution of
observations along an ordered sequence potentially changes at some 
(unknown) locations. In a slight abuse of notation, we use the term 
changepoint detection to refer to the particular setting in which
there are changepoints in the underlying mean. Our focus is on 
conducting valid inference related to the selected changepoints.
The existing literature applicable to this goal is relatively small; 
is reviewed in Section \ref{sec:related} and compared to our 
methods in Section \ref{sec:smuce}.

Among various methods for changepoint detection,
the {\it 1d fused lasso} \citep{fuse}, also known as {\it 1d
  total variation denoising} in signal processing \citep{tv}, is of
particular interest because it is a special case of the generalized lasso.
Let $y=(y_1,\ldots,y_n)$ denote values observed at $1,\dots,n$.  Then the 1d
fused lasso estimator is defined as in
\eqref{eq:signalapprox}, with the penalty matrix being the discrete first
difference operator, $D=D^{(1)} \in
\R^{(n-1)\times n}$:
\begin{equation}
\label{eq:d1}
D^{(1)} = \left[\begin{array}{ccccc}
-1 & 1 & 0 & \ldots&  0\\
0 & -1 & 1 & \ldots&  0\\
\vdots &  & \ddots & \ddots & \\ 
0 & 0 & \ldots & -1 & 1 
\end{array}\right].
\end{equation}
In the 1d fused lasso problem, the dual boundary set tracked by
Algorithm \ref{alg:dualpath} has a natural interpretation: it provides 
the locations of changepoints in the primal solution, which we can
see more or less directly from \eqref{eq:dualsubg} (see also
\citet{genlasso, arnold2016}).
Therefore, we can rewrite \eqref{eq:primalsol} as 
\begin{equation}
\label{eq:1dflsol}
\hbeta(\lambda) = \sum_{j=1}^{k+1} \hat{b}_j(\lambda) \,
\one_{(I_{j-1}+1):I_j}, \quad \text{for $\lambda \in
  [\lambda_{k+1},\lambda_k]$}. 
\end{equation}
Here $I_1<\ldots<I_k$ denote the sorted elements of the boundary set
$\cB_k$, with $I_0=0$, $I_{k+1}=n$ for convenience, $\one_{p:q}$ denotes 
a vector with 1 in positions $p\,\dots q$ and 0 elsewhere, and 
\smash{$\hat{b}_1(\lambda),\ldots,\hat{b}_{k+1}(\lambda)$} denote
levels estimated by the fused lasso with parameter $\lambda$.
Note that in \eqref{eq:1dflsol}, we
have implicitly used the fact that the boundary set after $k$ steps of
the path algorithm has exactly $k$ elements; this is true since the
path algorithm never deletes coordinates from the boundary set in 1d
fused lasso problems (as mentioned following Algorithm
\ref{alg:dualpath}). 
The dual boundary signs also have a natural meaning: writing the elements
of \smash{$s_{\cB_k}$} as $s_{I_1},\ldots,s_{I_k}$, these record the signs of 
differences (or jumps) between adjacent levels, 
\begin{equation}
\label{eq:1dflsigns}
\sign\big(\hat{b}_{j+1}(\lambda) - \hat{b}_j(\lambda)\big) = s_{I_j},
\quad \text{for $j=1,\ldots,k$, $\lambda \in
  [\lambda_{k+1},\lambda_k]$}. 
\end{equation}

Below, we describe several aspects of selective inference with
1d fused lasso estimates. Similar discussions could be given
for the different special classes of generalized lasso problems, like trend 
filtering and the graph fused lasso, but for brevity we only go into such 
detail for the 1d fused lasso.

\paragraph{Contrasts for the fused lasso.}

The framework laid out in Section \ref{sec:inference} allows us to 
perform post-selection TG tests for hypotheses about $v^T\theta$, 
for any contrast vector $v$. We introduce two specific forms of interesting 
contrasts, which we call the segment and spike contrasts. 
From the $k$-step fused lasso solution, as portrayed in \eqref{eq:1dflsol}, 
\eqref{eq:1dflsigns}, there are two natural 
questions one could ask about the changepoint $I_j$, for some $j\in\{1,\dots,k\}$:
first, whether there is a difference in the underlying mean exactly at $I_j$,
\begin{equation}
\label{eq:hspike}
H_0 :  \theta_{I_j+1} = \theta_{I_j} \quad \text{versus} \quad
H_1 :  s_{I_j} (\theta_{I_j+1} - \theta_{I_j} ) >0.
\end{equation}
and second, whether there is an average difference in the mean between the regions
separated by $I_j$,
\begin{equation}
\label{eq:hsegment}
H_0 :  \bar\theta_{(I_j+1):I_{j+1}} = \bar\theta_{(I_{j-1}+1):I_j}
\quad \text{versus} \quad 
H_1 :  s_{I_j} (\bar\theta_{(I_j+1):I_{j+1}} -
\bar\theta_{(I_{j-1}+1):I_j}) > 0.
\end{equation}
These hypotheses are fundamentally different: that in
\eqref{eq:hspike} is sensitive to the exact location of the underlying mean
difference, whereas that in \eqref{eq:hsegment} can be non-null 
even if the change in mean is not exactly at $I_j$.  
To test \eqref{eq:hspike}, we use the so-called {\it spike contrast}
\begin{equation}
\label{eq:vspike}
v_{\mathrm{spike}} = s_{I_j} (\one_{I_j+1} - \one_{I_j}).
\end{equation}
The resulting TG test, as in \eqref{eq:tgstatistic} with
\smash{$v=v_{\mathrm{spike}}$}, is called the {\it spike test}, since it tests  
differences in the mean $\theta$ at exactly one location.
To test \eqref{eq:hsegment}, we use the so-called 
{\it segment contrast}
\begin{equation}
\label{eq:vsegment}
v_{\mathrm{seg}} = s_{I_j} \bigg(\frac{1}{I_{j+1}-I_j} \one_{(I_j+1):I_{j+1}} -
\frac{1}{I_j-I_{j-1}} \one_{(I_{j-1}+1):I_j}\bigg).
\end{equation}
The resulting TG test, as in \eqref{eq:tgstatistic} with
\smash{$v=v_{\mathrm{seg}}$}, is called the {\it segment test}, because 
it tests average differences across segments of the mean $\theta$.

In practice, the segment test often has more power than the spike test 
to detect a change in the underlying mean, since it averages over entire 
segments.  However, it is worth pointing out that the usefulness of the 
segment test at $I_j$ also depends on the quality of the {\it other} detected 
changepoints 1d fused lasso model (unlike the spike test, which does not), 
because these determine the
lengths of the segments drawn out on either side of $I_j$.
And, to emphasize what has already been said: unlike the 
spike test, the segment test does not test the precise location of a changepoint,
so a rejection of its null hypothesis must not be mistakenly interpreted
(also, refer to the corresponding coverage statement in \eqref{eq:isegment}). 

\begin{figure}[tb]
\centering
\includegraphics[width=0.475\textwidth]{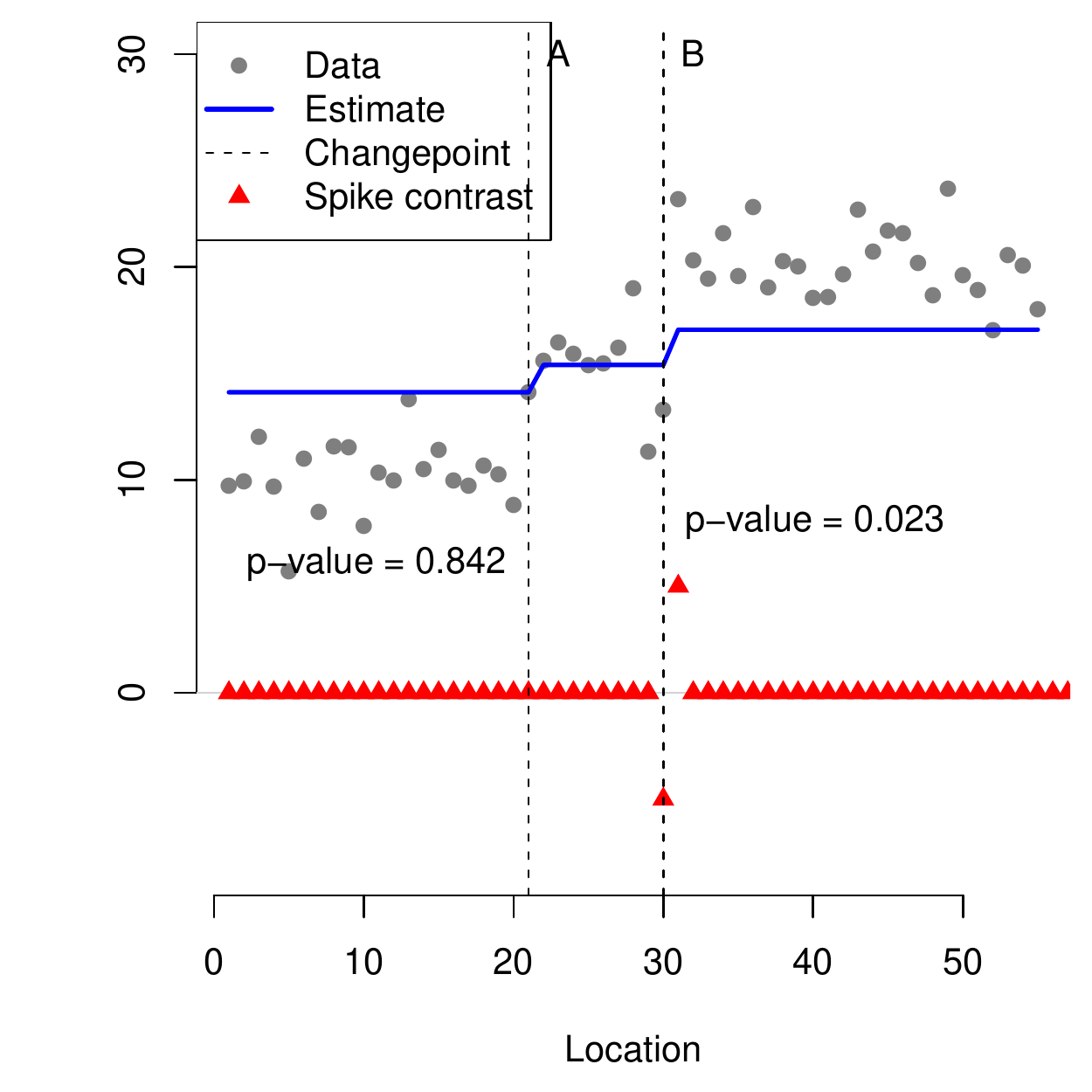}
\hspace{2pt}
\includegraphics[width=0.475\textwidth]{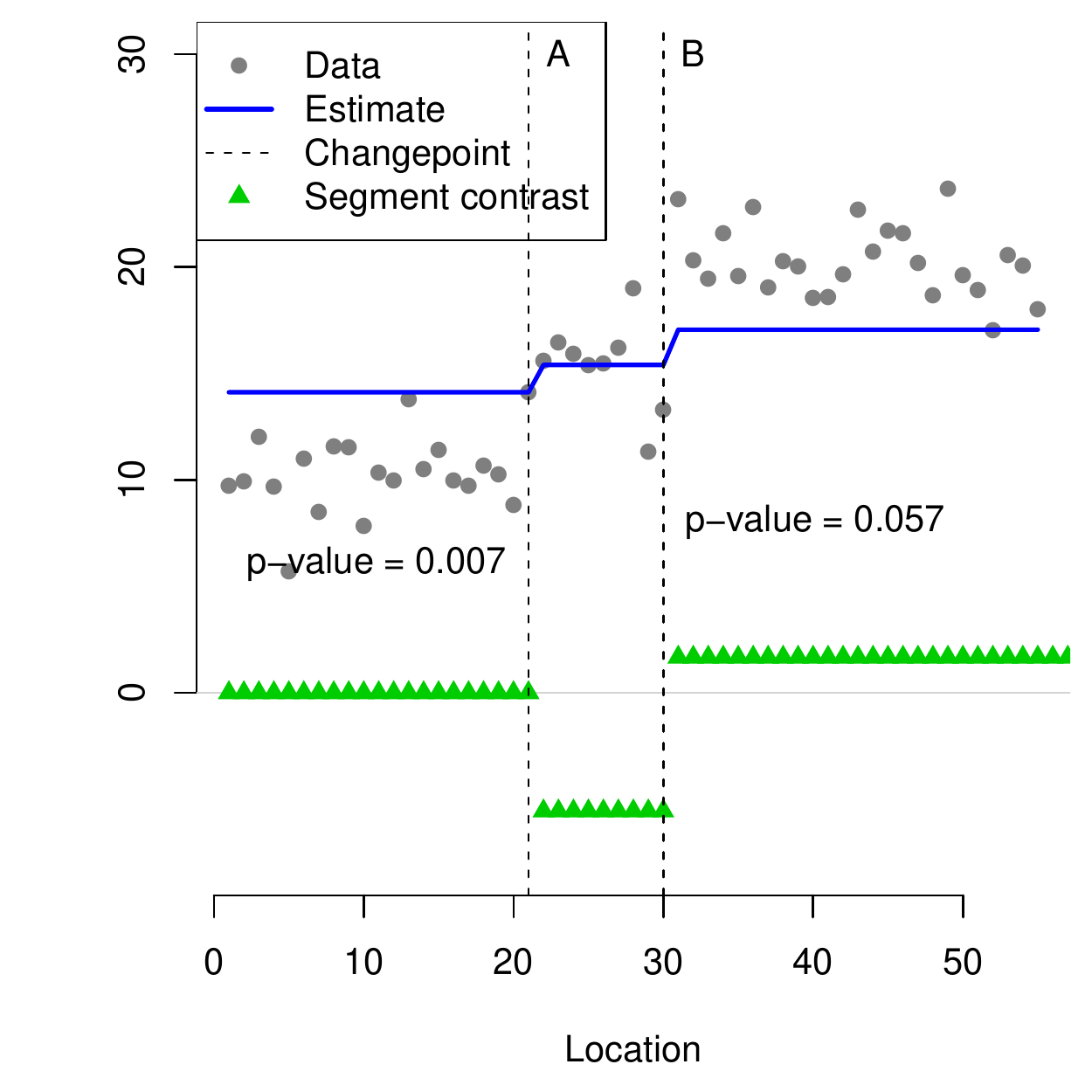}
\caption{\it\small An example with $n=60$ points, portraying the differences
  between the spike and segment tests for the fused lasso.  The underlying mean
  has true changepoints at locations 20 and 30; the 2-step fused lasso estimate,
  drawn in blue, detects changepoints at locations 21 and 30, labeled $A$ and
  $B$.  P-values from the segment test are reported in the left panel, and from
  the spike test in the right panel.  The segment and spike contrast vectors
  around changepoint $B$ are visualized on the panels (the entries of these
  vectors have been scaled up for visibility). We can see that both segment
  p-values are small, and both segment null hypotheses defined around locations
  $A$ and $B$ should be rejected; but only the spike p-value at location $B$ is
  small, and only the the spike null hypothesis around location $B$ should be
  rejected (as location $A$ does not correspond to a true changepoint in the
  underlying mean).}
\label{fig:1dfl}
\end{figure}

Which test is appropriate ultimately depends on the goals of
the data analyst. Figure \ref{fig:1dfl} shows a simple example of the spike 
and segment tests.  The behaviors of these two
tests will be explored more thoroughly in Section \ref{sec:1dflexamples}.   

\paragraph{Alternative motivation for the contrasts.}

It may be interesting to note that, for the segment contrast 
\smash{$v_{\mathrm{seg}}$} in \eqref{eq:vsegment}, the statistic 
\begin{equation*}
v_{\mathrm{seg}}^T y = \bar{y}_{(I_j+1):I_{j+1}} - \bar{y}_{(I_{j-1}+1):I_j}
\end{equation*} 
is the likelihood ratio test statistic for testing the null
\smash{$H_0:\theta_{I_{j-1}+1}=\ldots=\theta_{I_j}=\theta_{I_j+1}=\ldots=\theta_{I_{j+1}}$}   
versus the alternative
\smash{$H_1:\theta_{I_{j-1}+1}=\ldots=\theta_{I_j}\not=\theta_{I_j+1}=\ldots=\theta_{I_{j+1}}$},  
if the locations $I_{j-1},I_j,I_{j+1}$ were {\it fixed}.  An equivalent way
to write these hypotheses, which will be a helpful generalization going 
forward (as we consider other classes of generalized lasso problems), is
\begin{equation*}
H_0 :  \theta \in \nul(D_{-\cB_k})
\quad \text{versus} \quad 
H_1 :  \theta \in \nul(D_{-\cB_k\setminus\{I_j\}}).
\end{equation*}
In this notation, the segment contrast \smash{$v_{\mathrm{seg}}$} in
\eqref{eq:vsegment} is the unique (up to a scaling factor) basis vector 
for the rank 1 subspace 
\smash{$\nul(D_{-\cB_k})\setminus\nul(D_{-\cB_k\setminus\{I_j\}}) =  
\nul(D_{-\cB_k\setminus\{I_j\}})^\perp \nul(D_{-\cB_k})$}, and 
\smash{$v_{\mathrm{seg}}^T y$} is the likelihood ratio test statistic for
the above set of null and alternative hypotheses.

Lastly, both segment and spike tests can be viewed from an equivalent
regression perspective, after transforming the 1d fused lasso problem 
in \eqref{eq:signalapprox},
\eqref{eq:d1} into an equivalent lasso problem (recall Section
\ref{sec:whynotlasso}). In this context, it can be shown that the 
segment test corresponds to a test of a partial regression coefficient 
in the active model, whereas the spike test corresponds to a test of a 
marginal regression coefficient.

\paragraph{Inference with an interpretable conditioning event.}

As explained in Section \ref{sec:conditioning}, there are different levels of
conditioning that can be used to interpret the results of the TG tests for
model selection events along the generalized lasso path. Here we demonstrate
for the segment test in \eqref{eq:hsegment}, \eqref{eq:vsegment} what we see as 
the simplest interpretation of its conditional coverage property, with respect to
its parameter
\smash{$\bar\theta_{(I_j+1):I_{j+1}} -  \bar\theta_{(I_{j-1}+1):I_j}$}.  
The TG interval $C_{1-\alpha}=[\delta_\alpha,\delta_{1-\alpha/2}]$
in \eqref{eq:tginterval}, computed by inverting the TG pivot,
has the exact finite sample property
\begin{equation}
\label{eq:isegment}
\P \Big(\bar\theta_{(I_j+1):I_{j+1}} - \bar\theta_{(I_{j-1}+1):I_j}
\in C_{1-\alpha} \; \Big| \; I_1,\ldots,I_k,s_{I_1},\ldots,s_{I_k}
\Big) = 1-\alpha,
\end{equation} 
obtained by marginalizing over some dimensions of
the conditioning set, as done in Section \ref{sec:conditioning}.
In words, the coverage statement \eqref{eq:isegment} says that,
conditional on the estimated changepoints $I_1,\ldots,I_k$ and
estimated jump signs $s_{I_1},\ldots,s_{I_k}$ in the $k$-step 1d fused
lasso solution, the interval $C_{1-\alpha}$ traps the jump in segment 
averages \smash{$\bar\theta_{(I_j+1):I_{j+1}} - \bar\theta_{(I_{j-1}+1):I_j}$}
with probability $1-\alpha$. This all assumes that
the choice of step $k$ is fixed; for $k$ chosen by an IC-based rule as
described in Section \ref{sec:kchosen}, the interpretation is very similar 
and we only need to add $k$ to the right-hand side of the conditioning
bar in \eqref{eq:isegment}. 
A similar interpretation is also available for the spike test, which we omit
for brevity.


\paragraph{One-sided or two-sided inference?}

We note that both setups in \eqref{eq:hspike} and \eqref{eq:hsegment}
use a one-sided alternative hypothesis, and the contrast vectors 
in \eqref{eq:vspike} and \eqref{eq:vsegment} are defined accordingly.
To put it in words, we are testing for changepoint in the underlying mean
$\theta$ (either exactly at one location, or in an average sense across 
local segments) and are looking to reject when a jump in $\theta$
occurs {\it in the direction we have already observed in the 
fused lasso solution}, as dictated by the sign \smash{$s_{I_j}$}.  
On the other hand, for coverage statements as in \eqref{eq:isegment},
we are implicitly using a two-sided alternative, replacing the alternative
in \eqref{eq:hsegment} by
\smash{$H_1 : \bar\theta_{(I_j+1):I_{j+1}} \not= \bar\theta_{(I_{j-1}+1):I_j}$}
(since the coverage interval is the result of inverting a two-sided pivotal
statistic).
Two-sided tests and one-sided intervals are also possible in our inference
framework, however, we find them less natural, and our default is therefore
to consider the aforementioned versions.

\subsection{Knot detection via trend filtering} 
\label{sec:trendfiltering}

Trend filtering can be seen as an extension of the 1d fused lasso for
fitting higher-order piecewise polynomials \citep{hightv,l1tf,trendfilter}.  It can
be defined for any desired polynomial order, written as $r \geq 0$, with $r=0$ 
giving piecewise constant segments and reducing to the 1d fused lasso of the 
last subsection.  Here we focus on the case $r=1$, where piecewise linear 
segments are fitted.  The general case $r \geq 2$ is possible by following 
the exact same logic, though for simplicity, we do not cover it.

As before, we assume the data $y=(y_1,\ldots,y_n)$ has been measured at
ordered locations $1,\ldots,n$. The {\it linear trend filtering} estimate is 
defined as in \eqref{eq:signalapprox} with $D=D^{(2)} \in \R^{(n-2)\times n}$, the 
discrete second difference operator:
\begin{equation}
\label{eq:d2}
D^{(2)} = \left[\begin{array}{cccccc}
1 & -2 & -1 & 0 & \ldots&  0\\
0 & 1 & -2 & 1 & \ldots&  0\\
\vdots & & \ddots & \ddots & \ddots & \\ 
0 & 0 & \ldots & 1 & -2 & 1 
\end{array}\right].
\end{equation}
For the linear trend filtering problem, the elements of the dual boundary set
are in one-to-one correspondence with knots, i.e., changes in slope, in the
piecewise linear sequence \smash{$\hbeta=(\hbeta_1,\ldots,\hbeta_n)$}.  This
comes essentially from \eqref{eq:dualsubg} (for more, see \citet{genlasso,
arnold2016}).  Specifically, enumerating the elements of the boundary set 
$\cB_k$ as $I_1<\ldots<I_q$ (and using $I_0=0$ and $I_{q+1}=0$ for convenience),
each location $I_j+1$, $j=1,\ldots,q$ serves a knot in the trend filtering solution,
so that we may rewrite \eqref{eq:primalsol} as 
\begin{equation}
\label{eq:tfsol}
\hbeta(\lambda) = \sum_{j=1}^{q+1} \Big(\hat{b}_j(\lambda) + 
\hat{m}_j(\lambda) (j - I_{j-1} -1) \Big) 
\one_{(I_{j-1}+1):I_j}, \quad \text{for $\lambda \in
  [\lambda_{k+1},\lambda_k]$}. 
\end{equation}
Above, $q$ denotes the number of knots in the $k$-step linear trend filtering 
solution, which in general need not be equal to $k$, since (unlike the 1d fused
lasso) the path algorithm for linear trend filtering can both add to and 
delete from the boundary set at each step.  Also, for each $j=1,\ldots,q+1$,
the quantities
\smash{$\hat{b}_j(\lambda)$} and \smash{$\hat{m}_j(\lambda)$} denote the
``local'' intercept and slope parameters, respectively,
of the linear trend filtering solution, over the segment 
$\{I_{j-1}+1,\ldots,I_j\}$.\footnote{The parameters 
\smash{$\hat{b}_j(\lambda)$},
\smash{$\hat{m}_j(\lambda)$}, $j=1,\ldots,q+1$  
are not completely free to vary; the slopes are defined so that the
linear pieces in the trend filtering solution match at the knots,
\smash{$\hat{m}_j(\lambda) = (\hat{b}_{j+1}(\lambda) - 
\hat{b}_j(\lambda))/(I_j-I_{j-1})$}, $j=1,\ldots,q$.}
Denoting the dual boundary signs $s_{\cB_k}$ by $s_{I_1},\ldots,s_{I_q}$, we 
have
\begin{equation}
\label{eq:tfsigns}
\sign\big(\hat{m}_{j+1}(\lambda) - \hat{m}_j(\lambda)\big) = s_{I_j},
\quad \text{for $j=1,\ldots,q$, $\lambda \in
  [\lambda_{k+1},\lambda_k]$},
\end{equation}
i.e., these signs of changes in slopes between adjacent trend filtering segments.

\paragraph{Contrasts for linear trend filtering.}

We can construct both spike and segment tests for linear trend filtering
using similar motivations as in the 1d fused lasso. Given the trend filtering
solution in \eqref{eq:tfsol}, \eqref{eq:tfsigns}, we consider testing a particular
knot location $I_j+1$, for some $j=1,\ldots,q$. The spike contrast is defined by
\begin{equation}
\label{eq:vspike_tf}
v_{\mathrm{spike}} = s_{I_j} (\one_{I_j} - 2\one_{I_j+1} + \one_{I_j+2}),
\end{equation}
and the TG statistic in \eqref{eq:tgstatistic}
with \smash{$v=v_{\mathrm{spike}}$} provides us with a test for
\begin{equation}
\label{eq:hspike_tf}
H_0 :  \theta_{I_j+1} = \frac{\theta_{I_j}+\theta_{I_j+2}}{2}
\quad \text{versus} \quad
H_1 :  s_{I_j} (\theta_{I_j} - 2\theta_{I_j+1} + \theta_{I_j+2} ) >0.
\end{equation}
The segment contrast is harder to define explicitly from first principles, but
can be defined following one of the alternative motivations for the segment
contrast in the 1d fused lasso problem: consider the rank 1 subspace
\smash{$\nul(D_{-\cB_k})\setminus\nul(D_{-\cB_k\setminus\{I_j\}}) =  
\nul(D_{-\cB_k\setminus\{I_j\}})^\perp \nul(D_{-\cB_k})$}, and define 
$w$ to be a basis vector for this subspace (unique up to scaling).  The 
segment contrast is then
\begin{equation}
\label{eq:vsegment_tf}
v_{\mathrm{seg}} = \sign(w_{I_j} - 2w_{I_j+1} + w_{I_j+2}) s_{I_j} w,
\end{equation}
i.e., we align $w$ so that its second difference around the knot 
$I_j+1$ matches that in the trend filtering solution. To test
\smash{$v_{\mathrm{seg}}^T \theta=0$}, we can use the TG statistic  
in \eqref{eq:tgstatistic} with \smash{$v=v_{\mathrm{seg}}$}; however,
as $w$ is not easy to express in closed-form, this null   
hypothesis is also not easy to express in closed-form.  Still, we can
rewrite it in a slightly more explicit manner:
\begin{equation}
\label{eq:hsegment_tf}
H_0 :  h^T (\theta - \theta^{\mathrm{proj}}) = 0 
\quad \text{where} \quad
h = (\underbrace{0,\ldots,0}_{I_j+1},
\underbrace{1,2,3,\ldots,n-I_j-2}_{n-I_j-1}) \;\, \text{and} \;\, 
\theta^{\mathrm{proj}} = P_{\nul(D_{-\cB_k\setminus\{I_j\}})} \theta, 
\end{equation}
versus the appropriate one-sided alternative hypothesis. In words, 
\smash{$\theta^{\mathrm{proj}}$} is the projection of 
$\theta$ onto the space of piecewise linear vectors with knots at
locations $I_\ell+1$, $\ell \not= j$, and $h$ is a single piecewise
linear activation vector that rises from zero at location $I_j+1$. 

\begin{figure}[htb]
\centering
\includegraphics[width=0.475\textwidth]{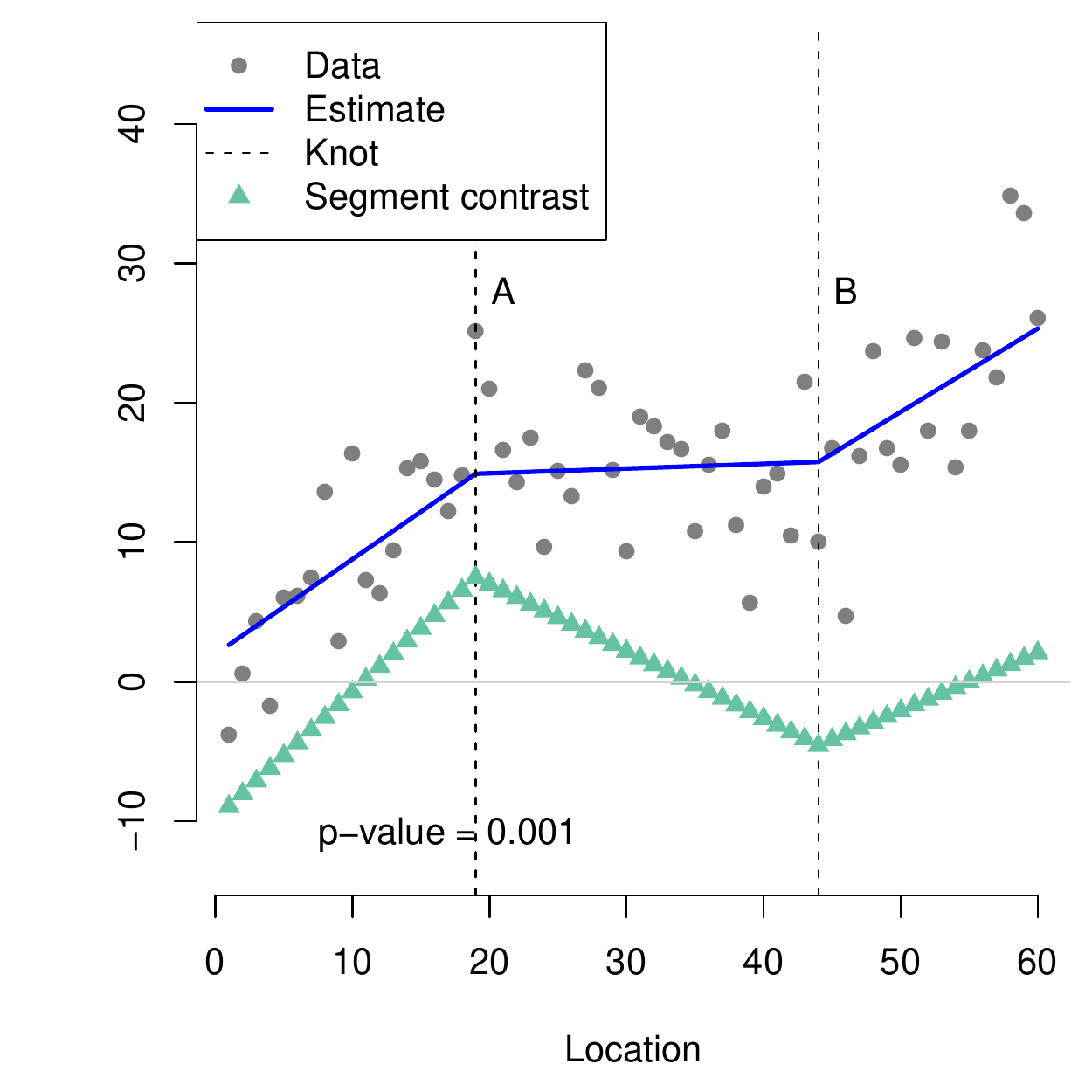}
\hspace{2pt}
\includegraphics[width=0.475\textwidth]{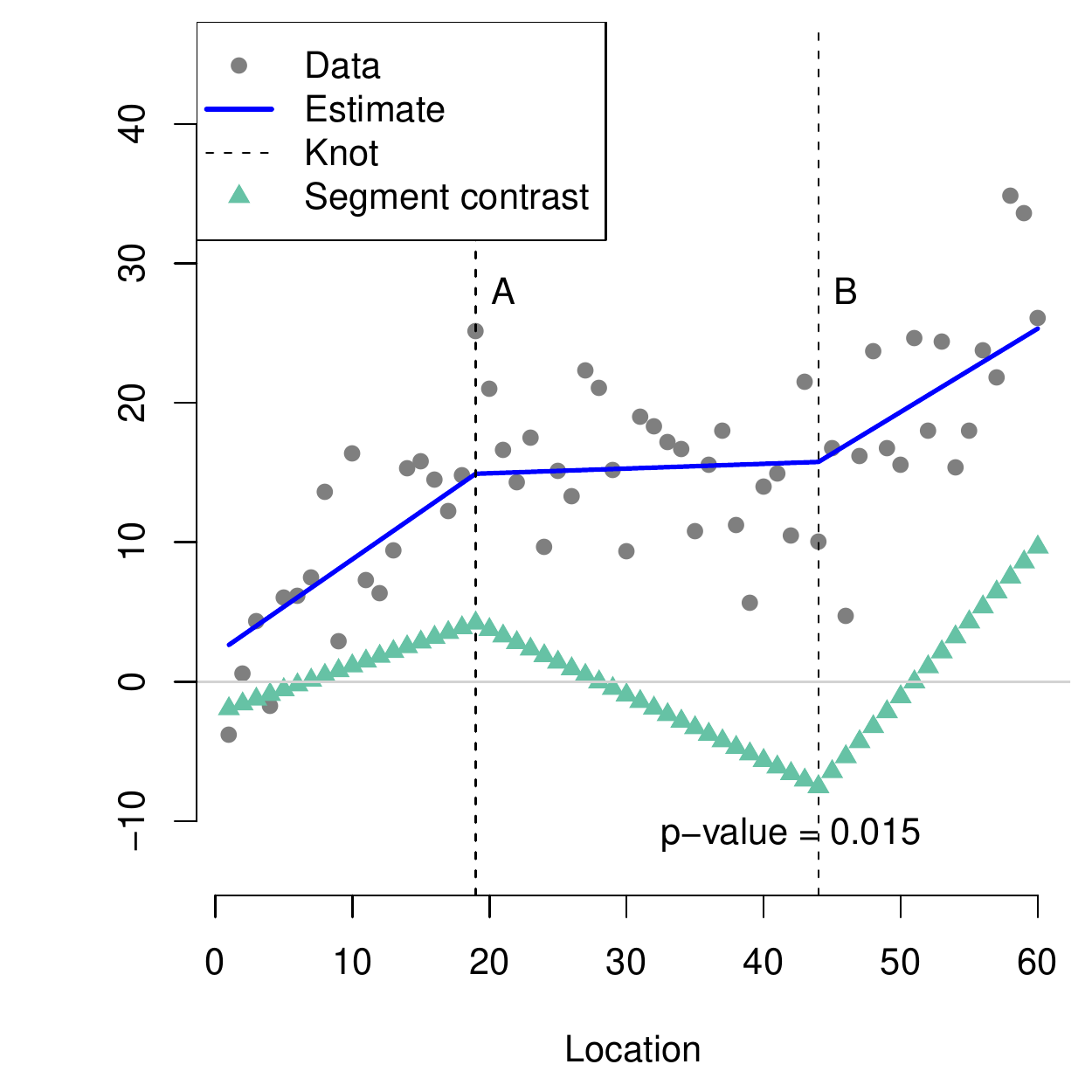}
\caption{\it\small An example with $n=60$ points, portraying two
  segment tests for trend filtering. The underlying piecewise linear
  mean has knots at locations 20 and 40; the 2-step linear trend
  filtering estimate, in blue, detects knots at locations 17 and 39,
  labeled A and B. The left plot shows the result the segment test at
  knot A, and the right plot at knot B. In each, the segment contrast
  is visualized. Both p-values are small.} 
\label{fig:tf}
\end{figure}

The same high-level points comparing the spike and segment tests for the
fused lasso also carry over to the linear trend filtering problem: the segment test
can often deliver more power, but at a given location $I_j+1$, the power
of the segment test will depend on the other knot locations in the estimated 
model.  The spike test at location $I_{j+1}$ does not depend on any other knot 
points in the trend filtering solution. Furthermore, the segment null does not 
specify a precise knot location, and one must be careful in interpreting 
a rejection here. Figure \ref{fig:tf} gives examples of the segment test for linear 
trend filtering. More examples are investigated in Section \ref{sec:tfexample}.  

\subsection{Cluster detection via the graph fused lasso}
\label{sec:graphfused}

The graph fused lasso is another generalization of the 1d fused lasso,
in which we depart from the 1-dimensional ordering of the components
of $y=(y_1,\ldots,y_n)$. 
Now we think of these components as being observed over nodes
$V=\{1,\ldots,n\}$ of a given (undirected) graph, with edges
$E=\{e_1,\ldots,e_m\}$, where say each $e_\ell=(i_\ell,j_\ell)$ joins 
some nodes $i_\ell$ and $j_\ell$, for $\ell=1,\ldots,m$.  
Note that the 1d fused lasso corresponds to the special case in which 
$E=\{(i,i+1): i=1,\ldots,n\}$, called the chain graph.  For a general
graph $G=(V,E)$, we define its edge incidence matrix $D_G \in
\R^{m\times n}$ by having 
rows of the form
\begin{equation}\label{eq:dgraph}
D_\ell = (0, \ldots 
\underset{\substack{\;\;\uparrow \\ \;\;i_\ell}}{-1}, 
\ldots \underset{\substack{\uparrow \\ j_\ell}}{1}, \ldots 0), 
\end{equation}
when the $\ell$th edge is $e_\ell=(i_\ell,j_\ell)$, with $i_\ell<j_\ell$,
for $\ell=1,\ldots,m$.  The {\it
graph fused lasso} problem, also called {\it graph total variation
denoising}, is given by \eqref{eq:signalapprox} with $D=D_G$.
This has been studied by many authors, particularly in the case when 
$G$ is a 2-dimensional grid, and the resulting program, called the
{\it 2d fused lasso}, is useful for image denoising (see, e.g., 
\citet{pco,fuseflow,hoefling2010path,genlasso,  
sharpnack2012sparsistency,arnold2016}).
Trend filtering can also be extended to graphs \citep{graphtf}; in 
principle our inferential treatment here extends to this problem as 
well, though we do not discuss it.

The boundary set constructed by the dual path algorithm, Algorithm
\ref{alg:dualpath}, has the following interpretation for the graph fused
lasso problem \citep{genlasso,arnold2016}. Denoting
$\cB_k=\{I_1,\ldots,I_q\}$, each element $I_\ell$
corresponds to an edge \smash{$e_{I_\ell}$} in the graph, $\ell=1,\ldots,q$.
The graph fused lasso solution is then piecewise constant over the sets
$C_1,\ldots,C_p$, which form partition of $\{1,\ldots,n\}$, 
and are defined by the connected components of 
\smash{$G=(V, E\setminus\{e_{I_\ell} : \ell,\ldots,q\})$}, 
i.e., the original graph with the edges \smash{$e_{I_\ell}$}, $\ell=1,\ldots,q$ 
removed.  That is, we may express \eqref{eq:primalsol} as
\begin{equation}
\label{eq:gflsol}
\hbeta(\lambda) = \sum_{j=1}^p \hat{b}_j (\lambda) \one_{C_j},
\quad \text{for $\lambda \in
  [\lambda_{k+1},\lambda_k]$},
\end{equation}
where $p$ denotes the number of connected components,
\smash{$\one_{C_j}$} denotes the indicator vector $C_j$, 
having $i$th entry 1 if $i \in C_j$ and 0 otherwise, and
\smash{$\hat{b}_j(\lambda)$} denotes an estimated level for component
$C_j$, for $j=1,\ldots,p$. The dual boundary signs \smash{$s_{\cB_k}=
\{s_{I_1},\ldots,s_{I_q}\}$}, capture the signs of differences between levels 
in the graph fused lasso solution,
\begin{multline}
\label{eq:gflsigns}
\sign\big(\hbeta_{j_\ell}(\lambda)-\hbeta_{i_\ell}(\lambda)\big) = s_{I_\ell},
\quad \text{when $e_{I_\ell}=(i_\ell,j_\ell)$, with $i_\ell < j_\ell$,} \\
\text{for $\ell=1,\ldots,q$, and $\lambda \in [\lambda_{k+1},\lambda_k]$}.
\end{multline}

\begin{figure}[htb]
\centering
\includegraphics[width=.32\textwidth]{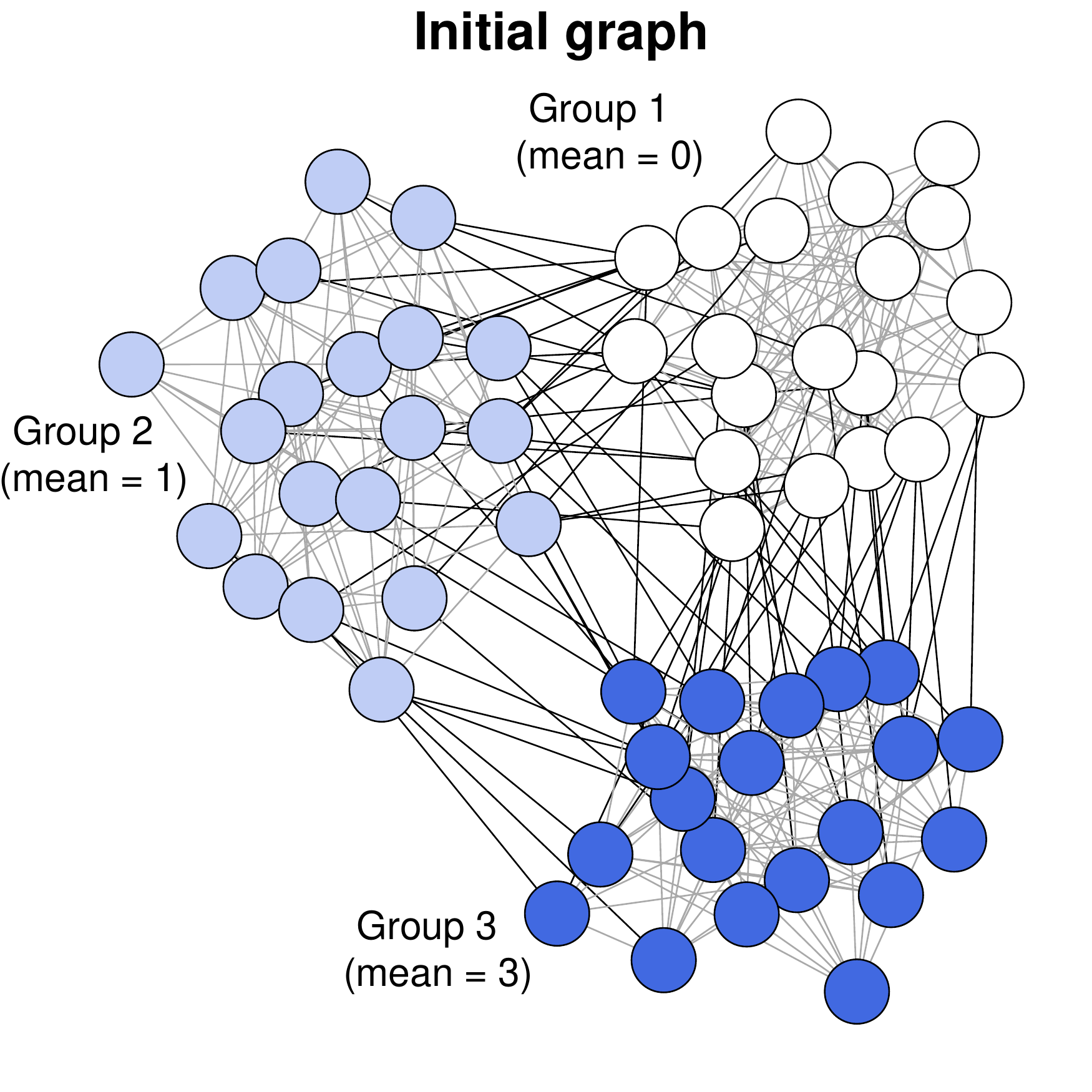}
\includegraphics[width=.32\textwidth]{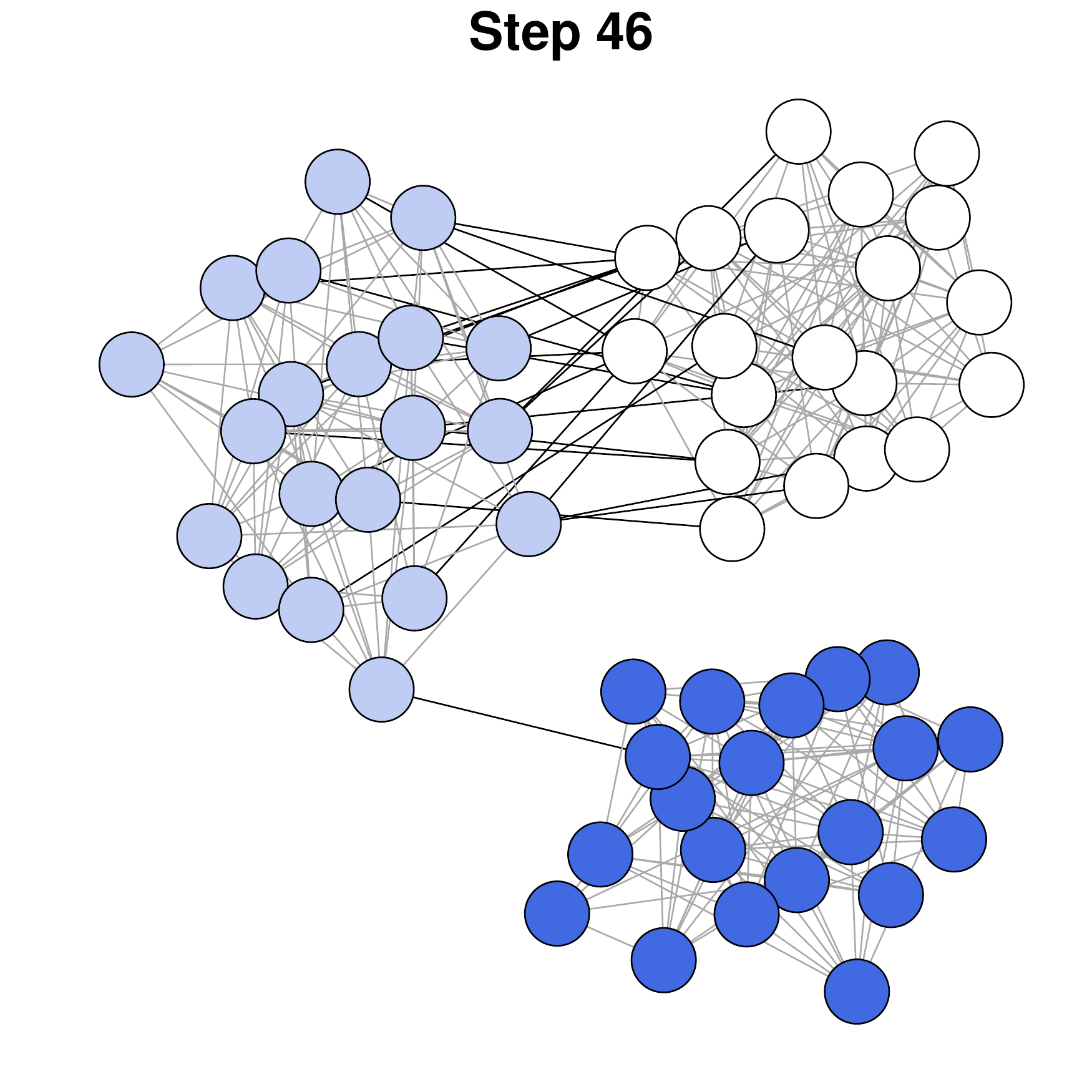}
\includegraphics[width=.32\textwidth]{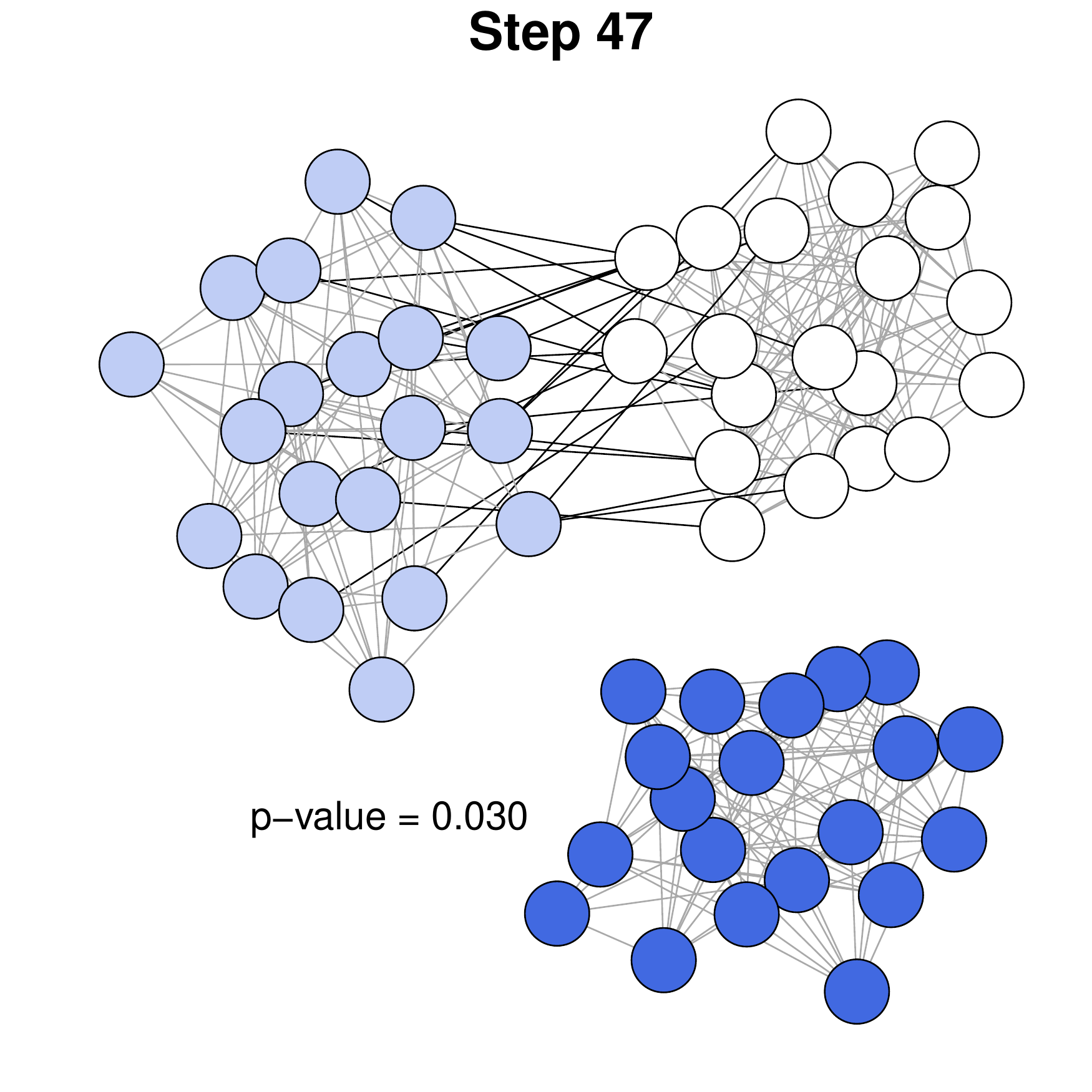}
\caption{\it\small An example with $n=60$ nodes illustrating the segment
test for the graph fused lasso. The graph was generated using a simple stochastic
block model with 3 groups of 20 nodes each.  The edge probabilities
were 0.5 for nodes in the same group and 0.05 for nodes in different 
groups.  This resulted in $m=369$ edges. 
The group means were defined to be 0, 1, and 3 (colored in white, light blue, and dark
blue, above). Data were generated by adding 
i.i.d.\ centered Gaussian noise, with standard deviation 0.15. The left plot displays
the initial graph, with 321 total edges. The middle plot displays the
graph fused lasso estimate after 46 path steps, where there is only
one edge left separating group 3 from groups 
1 and 2. At step 47, in the right plot, this last edge is removed
and the segment test \eqref{eq:hsegment_gfl}, \eqref{eq:vsegment_gfl} is applied,
with $C_a$ being the union of groups 1 and 2 (white and light blue) and
$C_b$ being group 3 (dark blue). The p-value is small, around $0.03$.}
\label{fig:gfl}
\end{figure}

\paragraph{Contrasts for the graph fused lasso.}  For the graph fused lasso problem, it 
is more natural to consider segment (rather than spike) type contrasts, conforming with the 
notation and concepts introduced for the 1d fused lasso problem.  Even restricting our
attention to segments tests, many possibilities are available to us, given the graph
fused lasso solution as in \eqref{eq:gflsol}, \eqref{eq:gflsigns}.  Say, we may choose 
any two ``neighboring'' connected components $C_a$ and $C_b$, for some
$a,b=1,\ldots,p$, meaning that there exists at least one edge (in the original graph)
between $C_a$ and $C_b$, and test
\begin{equation}
\label{eq:hsegment_gfl}
H_0 :  \bar\theta_{C_a} = \bar\theta_{C_b}
\quad \text{versus} \quad 
H_1 :  s_{ab} (\bar\theta_{C_b} - \bar\theta_{C_a}) > 0,
\end{equation}
where \smash{$s_{ab}=s_{I_\ell}$}, for some element $I_\ell \in \cB_k$ such that
\smash{$e_{I_\ell}=(i_\ell,j_\ell)$},
with $i_\ell<j_\ell$, and $i_\ell \in C_a$, $j_\ell \in C_b$.  Above, we use the notation
\smash{$\bar\theta_S=\sum_{i \in S} \theta_i/|S|$} for a subset $S$.  The hypothesis
in \eqref{eq:hsegment_gfl} tests whether the average of $\theta$ over components
$C_a$ and $C_b$ are equal, versus the alternative that they differ and their difference
matches the sign witnessed in the graph fused lasso solution.  To test 
\eqref{eq:hsegment_gfl}, we can use the TG statistic in \eqref{eq:tgstatistic} with 
$v=v_{\mathrm{seg}}$, where
\begin{equation}
\label{eq:vsegment_gfl}
v_{\mathrm{seg}} = s_{ab} \bigg( \frac{1}{|C_b|} \one_{C_b} - 
\frac{1}{|C_a|} \one_{C_a} \bigg).
\end{equation}
As in the 1d fused lasso problem, the above contrast can also be
motivated by the fact that \smash{$v_{\mathrm{seg}}^T y$} is the
likelihood ratio test for an appropriate
pair of null and alternative hypotheses.
More advanced segment tests are also possible, say, by testing whether averages 
of $\theta$ are equal over two subsets, each given by a union of connected 
components among $C_1,\ldots,C_q$.\footnote{However, here it is unclear how to
perform a one-sided test, since the preferred sign for rejection is not generally
specified by the graph fused lasso model selection event.}  Figure \ref{fig:gfl}
shows a simple example of a segment test of the form \eqref{eq:hsegment_gfl},
\eqref{eq:vsegment_gfl} for the graph fused lasso.  Section
\ref{sec:gflexample} gives another example.

\subsection{Problems with additional sparsity}
\label{sec:addedsparsity}

The generalized lasso signal approximator problem in \eqref{eq:signalapprox}
can be modified to impose {\it pure sparsity} regularization on $\beta$
itself, as in 
\begin{equation}
\label{eq:signalapprox_sp}
\hbeta = \argmin_{\beta \in \R^n} \; \half \|y-\beta\|_2^2 + \lambda
\|D\beta\|_1 + \alpha \lambda \|\beta\|_1,
\end{equation}
where $\alpha \geq 0$ is an another tuning parameter.  The above may be
called the {\it sparse generalized lasso} signal approximation
problem. In fused lasso settings, both 1d and graph-based, the estimate
\smash{$\hbeta$} in \eqref{eq:signalapprox_sp} will now be piecewise constant
across its components, with many attained levels being equal to zero
exactly (for a large enough value of $\alpha>0$).  In fact, the fused
lasso as  originally defined by \citet{fuse} was just as in
\eqref{eq:signalapprox_sp}, with  
both fusion and sparsity penalties.  In trend filtering settings, the estimate 
\smash{$\hbeta$} in \eqref{eq:signalapprox_sp} will be similar, except that
it will now have a piecewise polynomial structure whenever it is nonzero.  
There are many examples in which pure sparsity regularization is
a useful addition, see Section \ref{sec:cgh}, and also,
e.g., \citet{fuse,pco,fusedLassoCGH,trendfilter}.

Of course, problem \eqref{eq:signalapprox_sp} is still a generalized lasso
problem, since the two penalty terms in the criterion can be represented by
\smash{$\lambda \|\tilde{D} \beta\|_1$}, where 
\smash{$\tilde{D} \in \R^{(m+n) \times n}$} is given by row-binding 
$D \in \R^{m\times n}$ and $\alpha I \in \R^{n \times n}$.  This means that
all the tools presented so far in this paper are applicable, and post-selection
inference can be performed for problems like the sparse fused lasso and sparse
trend filtering.

\subsection{Generalized lasso regression problems}
\label{sec:regression}

Up until this point, our applications have focused on the signal approximation
problem in \eqref{eq:signalapprox}, but all of our methodology carries over
to the generalized lasso regression problem in \eqref{eq:genlasso}.  Allowing for a 
general regression matrix $X \in \R^{n\times p}$ greatly extends the scope of
applications; see Section \ref{sec:regressionexample}, and the discussions 
and examples in, e.g., \citet{fuse,pco,genlasso,arnold2016}.  

To tackle the regression problem in \eqref{eq:genlasso} with our framework, 
we must assume that $\rank(X)=p$ (which requires $n \geq p$).  We follow the 
transformation suggested by \citet{genlasso},
\begin{equation*}
\hbeta = \argmin_{\beta \in \R^p} \; \half \|y-X\beta\|_2^2 + \lambda
\|D\beta\|_1 \quad\iff\quad \hat\theta = \argmin_{\theta \in \R^n} \;
\half \|\tilde{y} - \theta\|_2^2 + \lambda \|\tilde{D} \theta\|_1,
\end{equation*}
where \smash{$\tilde{y}=XX^+ y$}, \smash{$\tilde{D} = DX^+$}, 
and the equivalence between solutions \smash{$\hbeta,\hat\theta$} 
is \smash{$\hat\theta = X\hbeta$}.  From what we can see
above, a generic generalized lasso regression problem can be transformed
into a generalized lasso signal approximation problem (just with a modified
response vector \smash{$\tilde{y}$} and penalty matrix \smash{$\tilde{D}$})
and so all of our tools can be applied to this transformed signal approximation 
problem in order to perform inference.  

When $\rank(X)<p$ (which always
happens in the high-dimensional case $n<p$), we can simply add a small
ridge penalty, which brings us back to the case in which the effective 
regression matrix is full column rank (see \citet{genlasso}). Then the
above transformation can be applied.

\subsection{Post-processing and visual aids}

We briefly discuss two extensions for the post-selection inference 
workflow. 

\paragraph{Post-processing.}  The choices of contrasts outlined in 
Sections \ref{sec:1dfusedlasso}--\ref{sec:regression} are defined
automatically from the generalized lasso selected model.  Given such a 
selected model, before we test a hypothesis or build a confidence 
interval, we can optionally choose to ignore or change some of the 
components of the selected model, in defining a contrast of interest. 
We refer to this as ``post-processing''; to be clear, it only 
affects the contrast vector being used, and not the conditioning set
in any way.  

\begin{figure}[htb]
\centering
\includegraphics[width=0.475\linewidth]{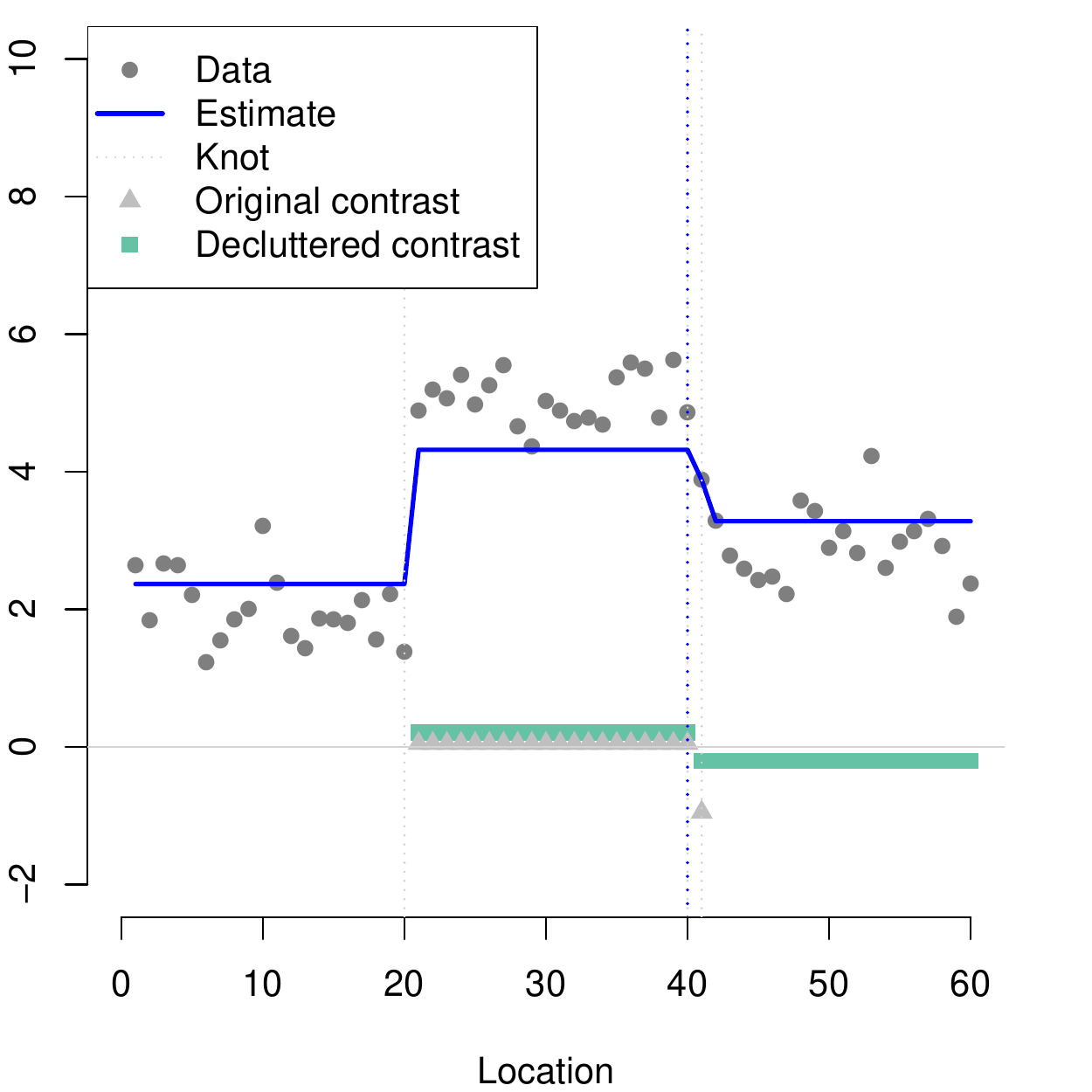}
\hspace{2pt}
\includegraphics[width=0.475\linewidth]{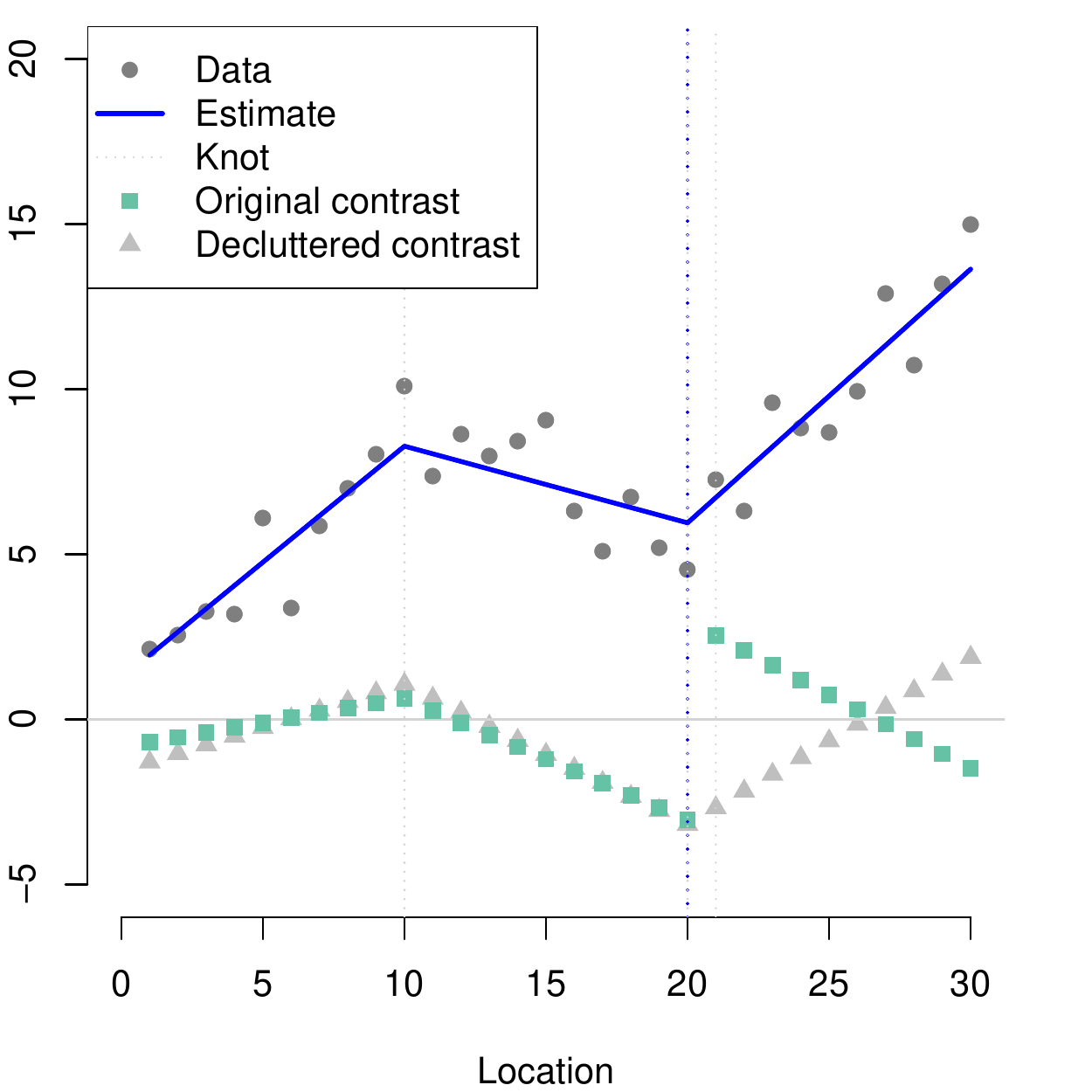}
\caption{\it\small Examples showing the segment test contrasts before 
  and after post-processing or ``decluttering'' for the 1d fused 
  lasso, in the left panel, and trend filtering, in the right panel.
  In both problems, the p-values for testing at locations marked
  by blue dashed vertical lines dropped considerably; on the left, the 
  p-value dropped from 0.236 to $<$0.001, and on the right, from 0.09 
  to 0.001.  For trend filtering, it can also be demonstrated that
  decluttering at one location helps the power for
  testing at another location that is farther away, but this
  phenomenon is absent in the fused lasso case (due to of the
  finite support of the segment test contrasts).} 
\label{fig:postprocess}
\end{figure}

It helps to give specific examples.  In the 1d fused lasso
problem, empirical examples reveal that the estimator sometimes places
several small jumps close to one larger jump.  The practitioner
could choose to merge nearby jumps before forming the segment contrast
of Section \ref{sec:1dfusedlasso}; we can see from
\eqref{eq:vsegment} that this would correspond to extending the
segment lengths on either side of the breakpoint in question, which
could result in greater power to detect a change in the underlying
mean. See the left panel of Figure \ref{fig:postprocess} for an
example.  In trend filtering, a practitioner could also choose
to merge nearby knots before forming the segment contrast in
\eqref{eq:vsegment_tf}, from Section
\ref{sec:trendfiltering}. See the right panel of
Figure \ref{fig:postprocess} for an example. Similar post-processing
ideas could be carried out for the graph fused lasso and generalized 
lasso regression problems. 

\begin{figure}[h!]
  \centering
  \includegraphics[width=0.45\linewidth]{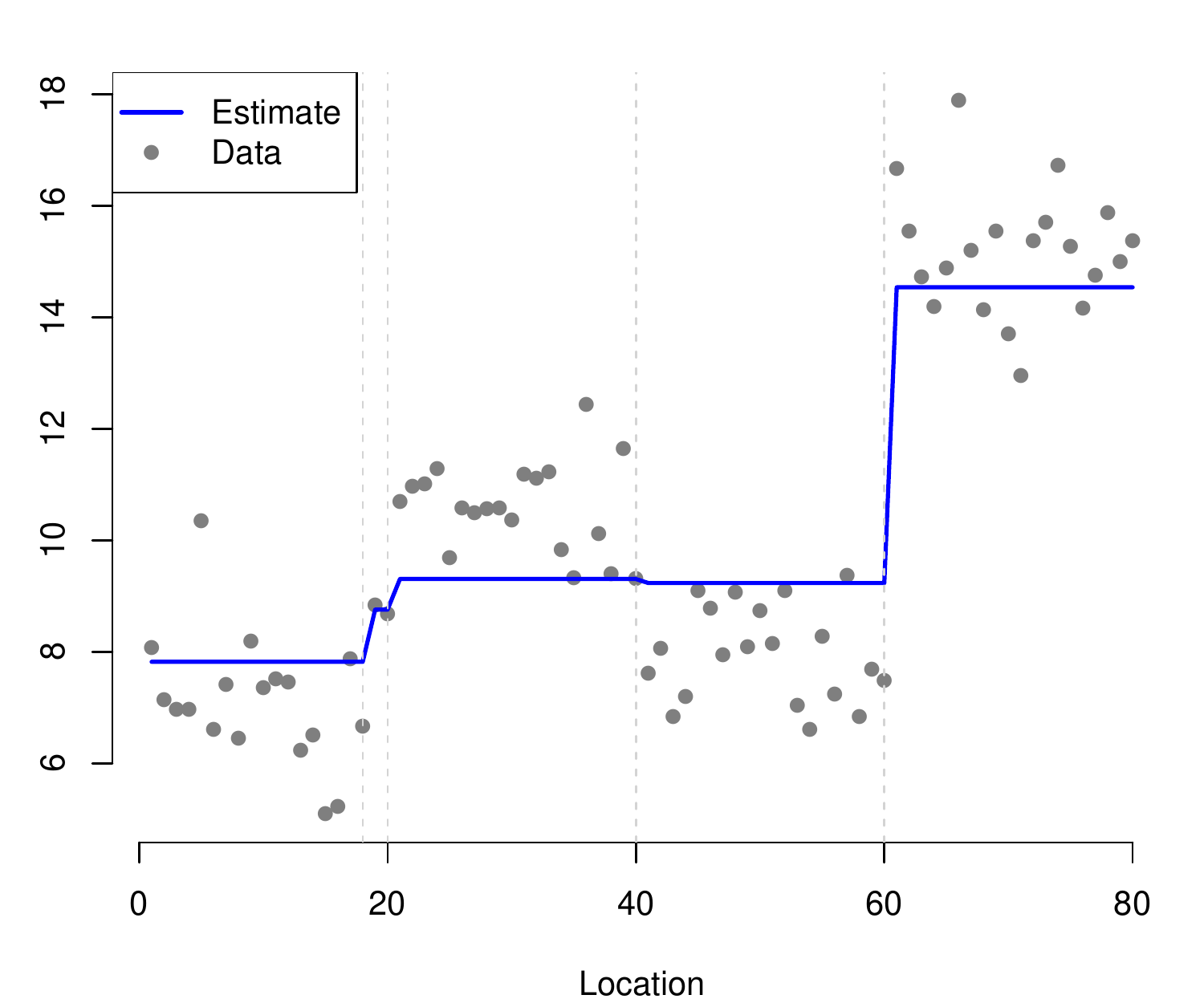}
  \hspace{2pt}
  \includegraphics[width=0.45\linewidth]{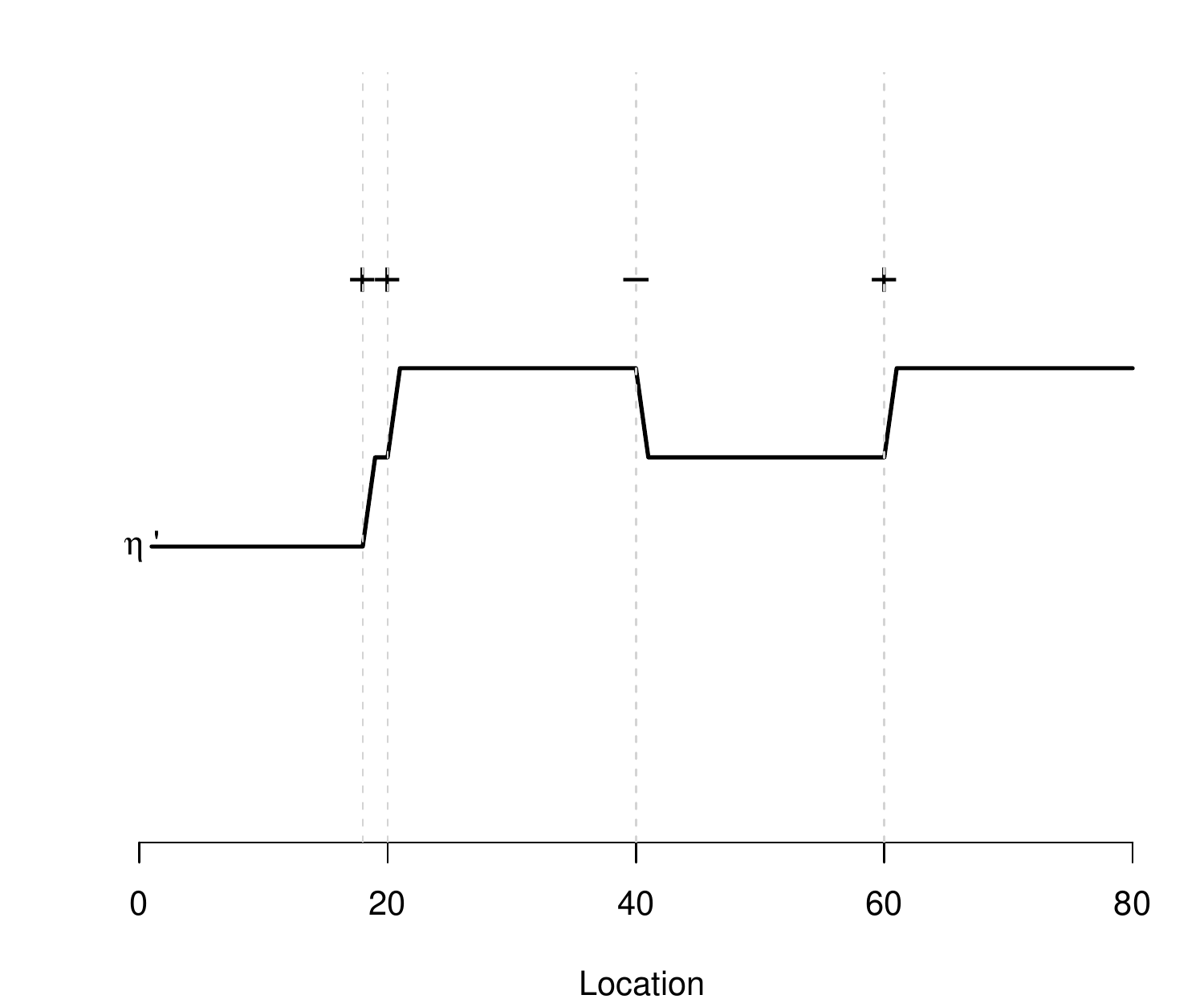}
  \caption{\it\small An example showing a 1d fused lasso solution 
    after 4 steps, in the left panel, and its corresponding step-sign
    plot, in the right panel. Based on the step-sign plot, the 
    data analyst may, e.g., deem locations 18 and 20 to be too close
    to be both interesting, and merge them before conducting segment
    tests.}   
  \label{fig:stepsignplot} 
\end{figure}


\paragraph{Visual aids.} In designing contrasts, the data analyst 
may also benefit from visualization of the generalized lasso selected
model. Such a ``visual aid'' has a similar goal to that of 
post-processing, namely, to improve the quality of the question asked,
i.e., the hypothesis 
tested, following a generalized lasso selection event.  For the
eventual inferences to be valid, the visual aid must not reveal
information about the data $y$ that is not contained in the selection
event, 
\smash{$\widehat{M}_{1:k}(y)=M_{1:k}$}, defined in Section  
\ref{sec:kfixed} (assuming a fixed step number $k$, for
simplicity).  Again, it helps to consider the fused lasso as a
specific use case. See Figure \ref{fig:stepsignplot} for an
example.  We cannot, e.g., reveal the 4-step fused lasso solution to
the analyst, ask him/her to hand-craft a contrast to be tested, and
then expect type I error control after applying our post-selection
inference tools.  This is because the solution itself contains
information about the data not contained in the selection
event---the magnitudes of the fitted jumps---and the decision
of which contrast to test could likely be affected by this
information.  This makes the conditioning set incomplete (said
differently, it means that the contrast vector no longer measurable
with respect to the conditioning event), and we should not expect our
previously established inference guarantees to apply, as a result.  We 
can, however, reveal a characature of the solution, as long as this
characature is based 
entirely on the selection event.  For the 1d fused lasso, this means
that the characature must be defined in terms of the changepoint
locations and signs of the fitted jumps, and we refer to it as a
``step-sign plot''. Examination of the jump locations and signs can
aid the analyst in designing interesting contrasts to test.

\section{Empirical examples}
\label{sec:examples}

\subsection{1d fused lasso examples}
\label{sec:1dflexamples}

\begin{figure}[htbp]
\centering
\makebox[\linewidth]{
\includegraphics[width=.35\linewidth]{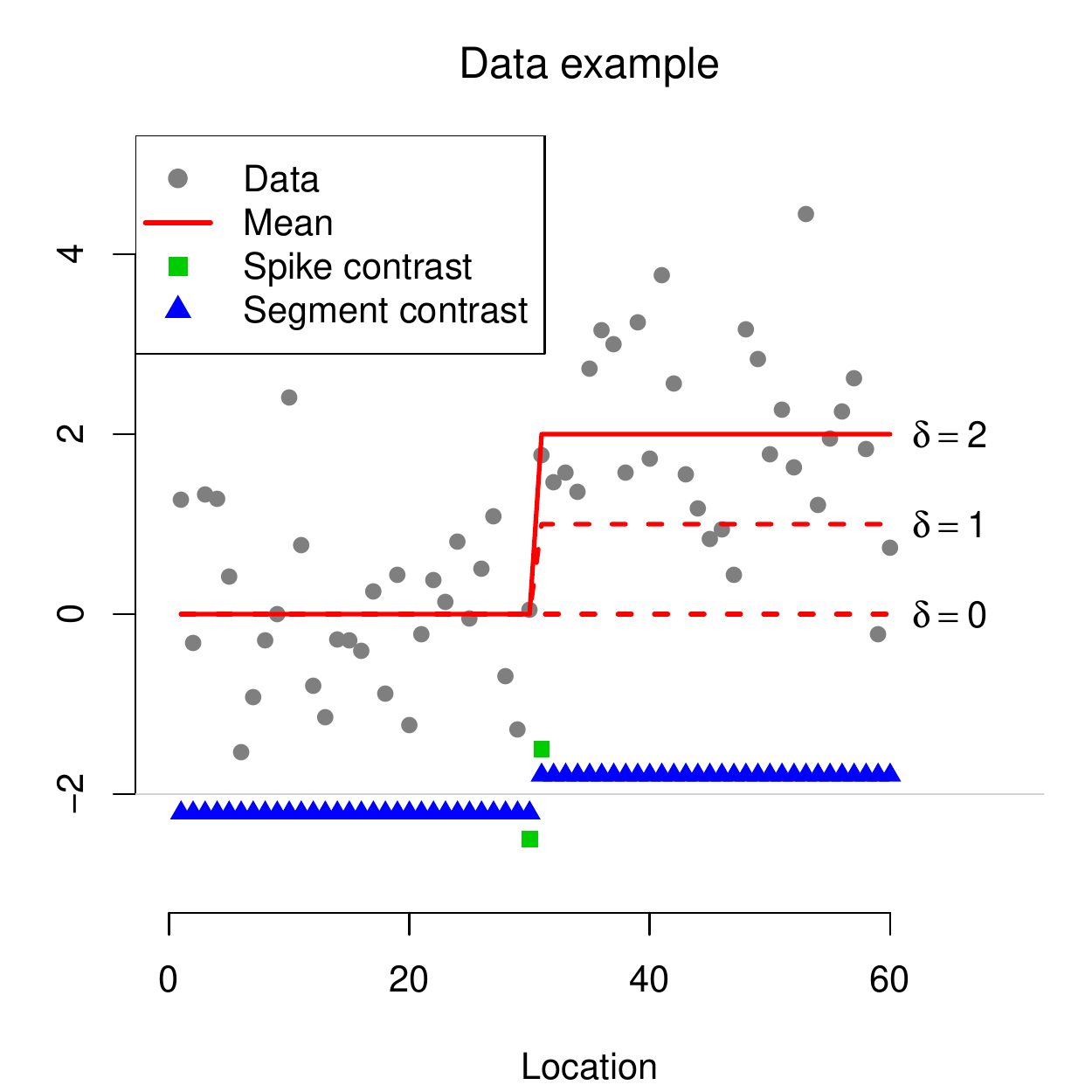}
\hspace{-5mm}
\includegraphics[width=.35\linewidth]{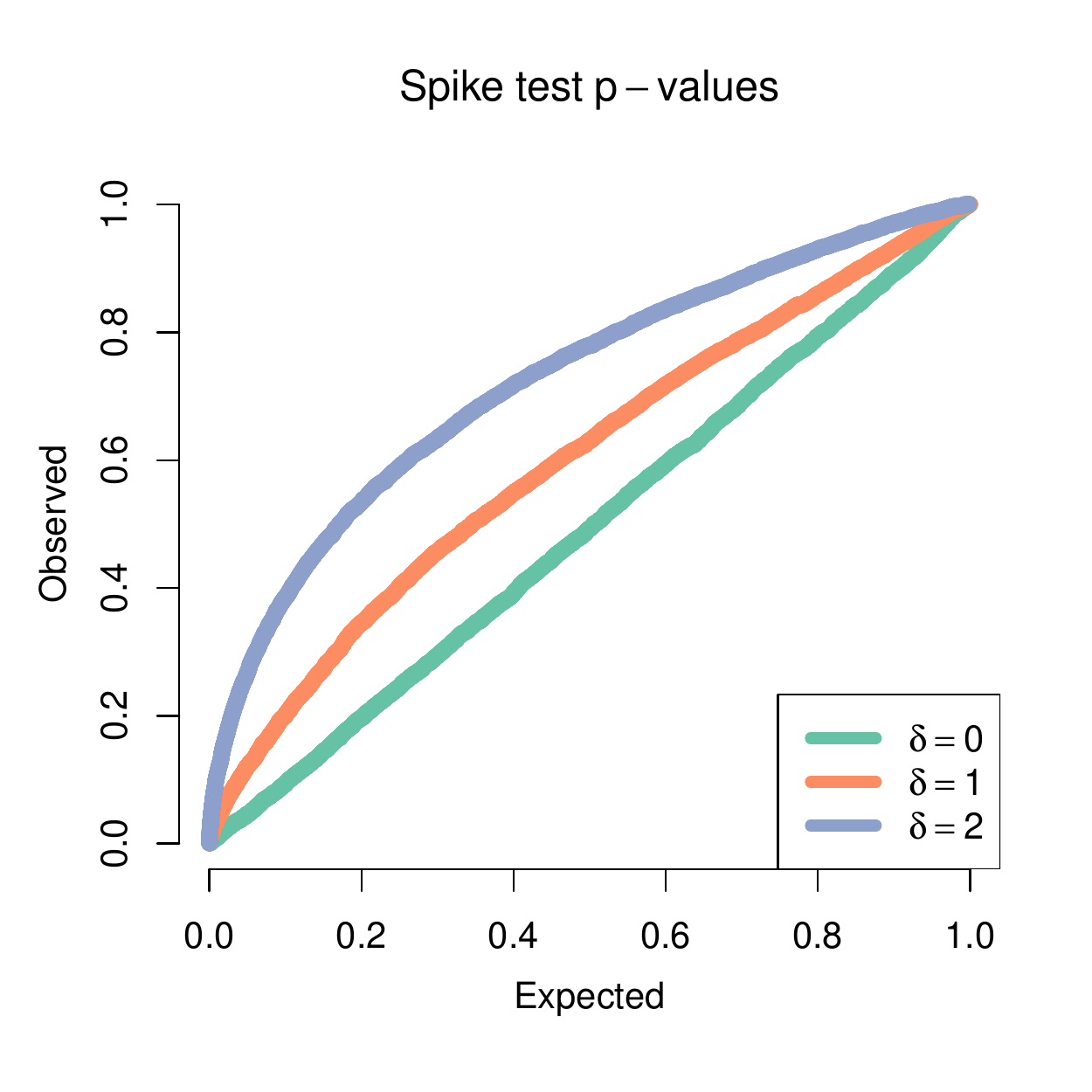}
\hspace{-5mm}
\includegraphics[width=.35\linewidth]{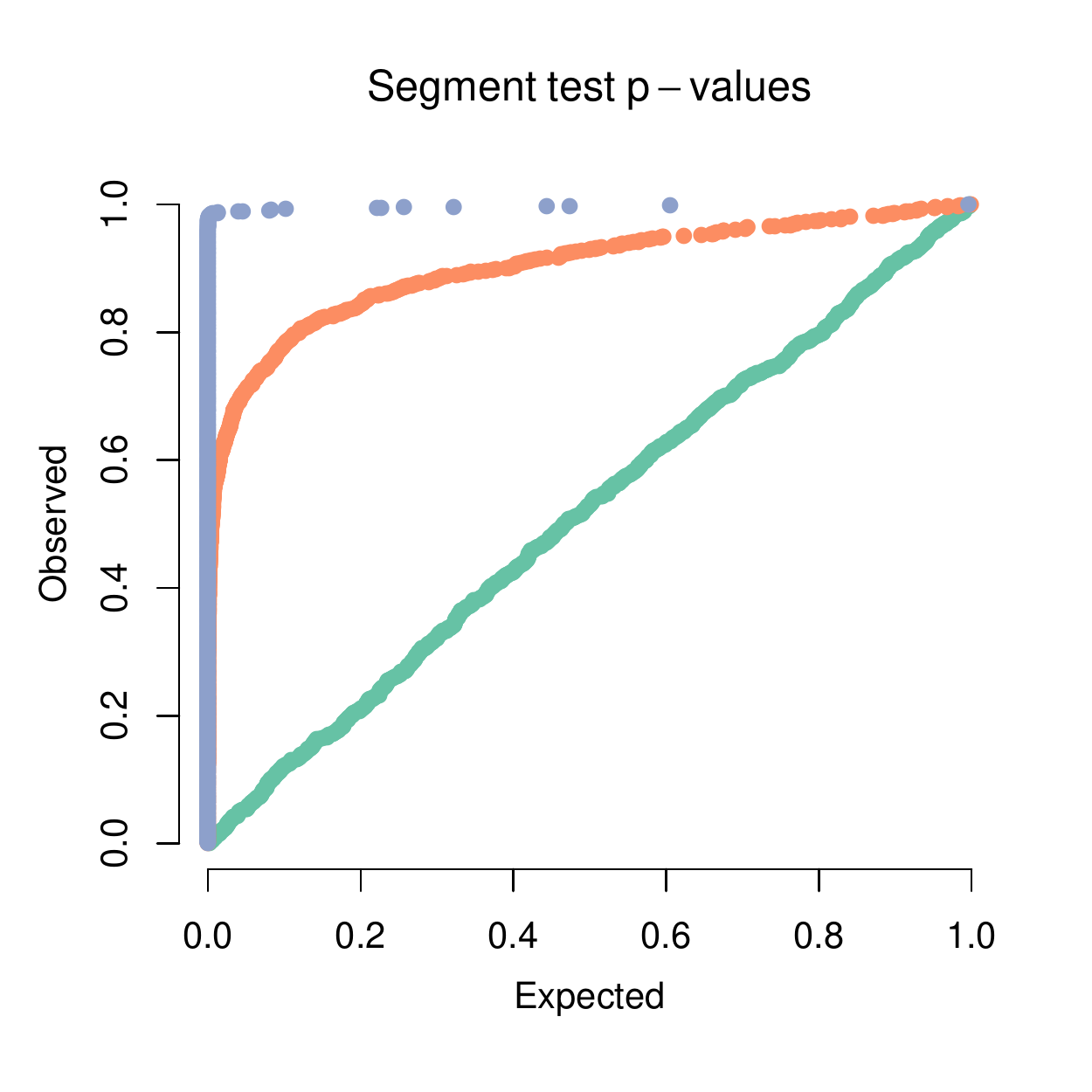} 
} \\
\bigskip
\makebox[\linewidth]{
\includegraphics[width=.35\linewidth]{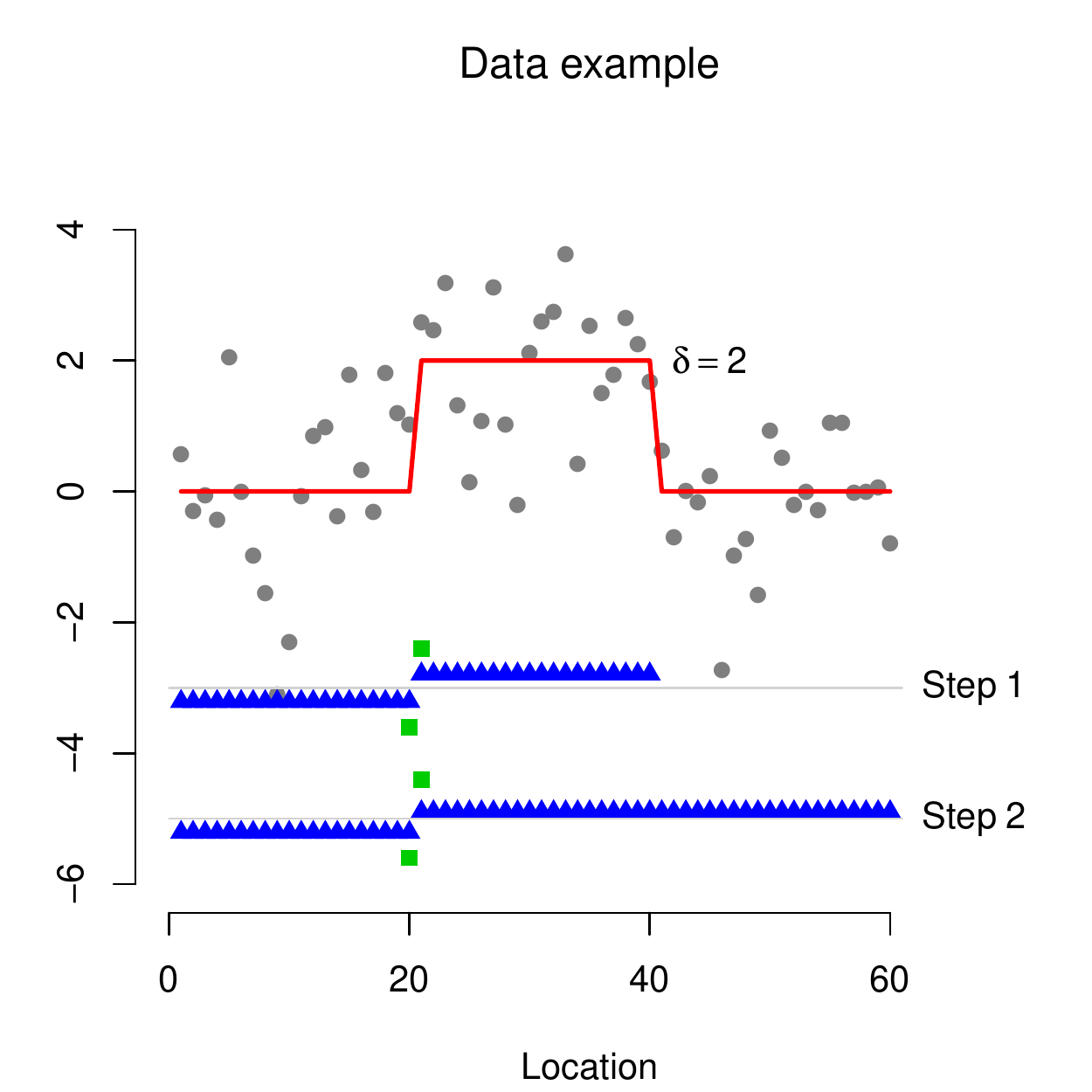}
\hspace{-5mm}
\includegraphics[width=.35\linewidth]{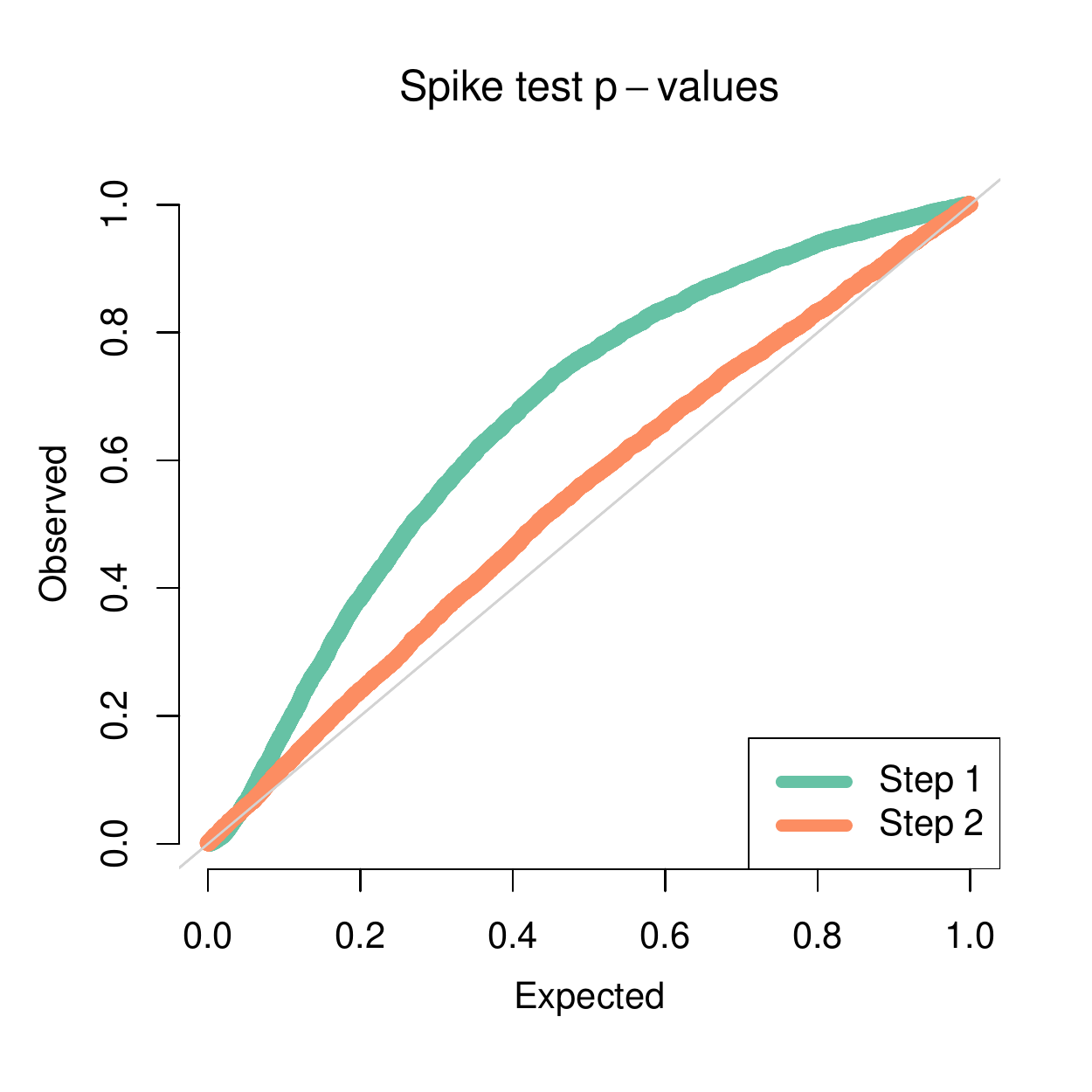}
\hspace{-5mm}
\includegraphics[width=.35\linewidth]{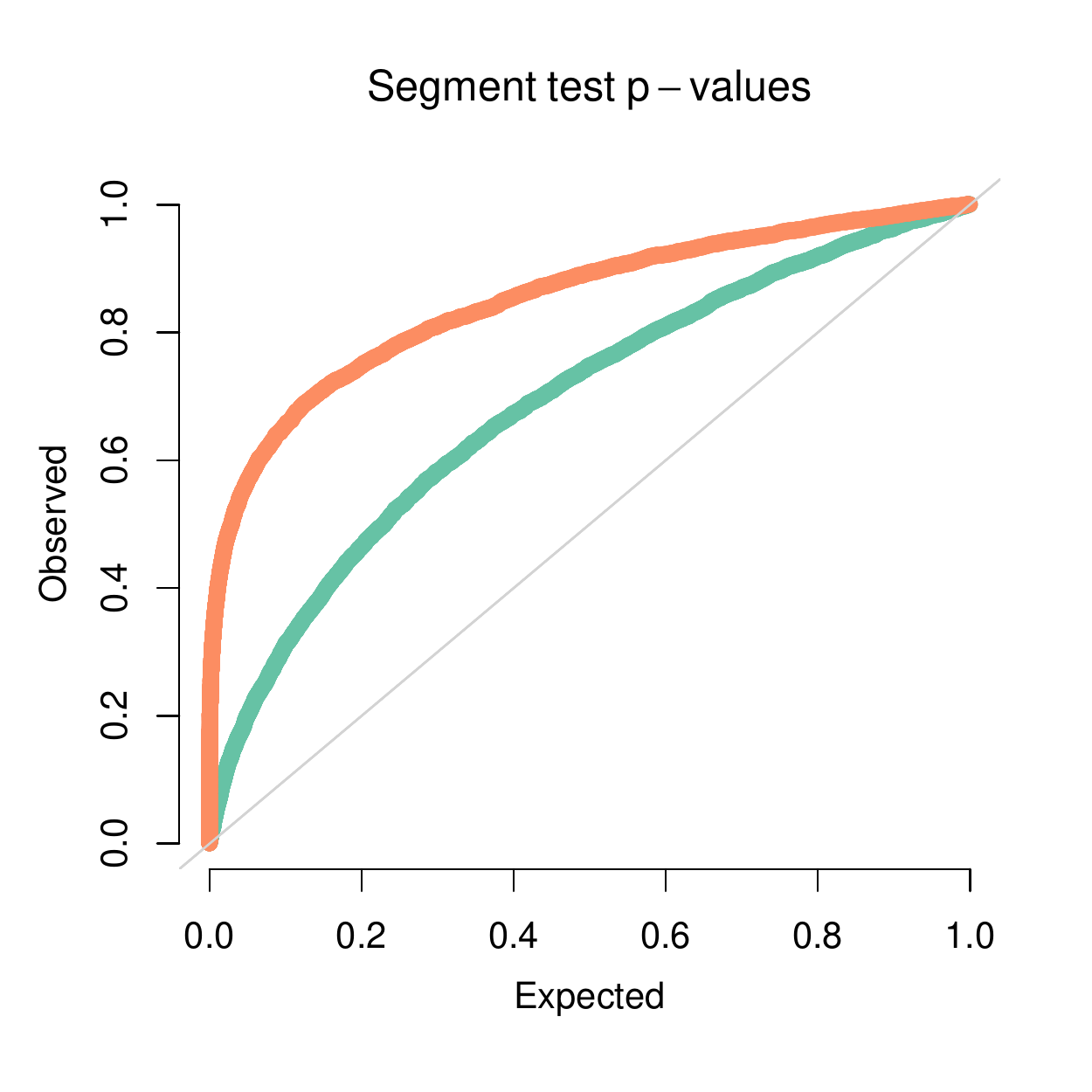}
}
\smallskip
\caption{\it\small Examination of p-values coming from the spike and 
  segment tests, in a setup with $n=60$ points and one or two true
  changepoints, shown in the top and bottom  
  rows, respectively.  In the one-jump setting, we considered three signal
  strengths: $\delta=0,1,2$.  The top left panel shows an example simulated data
  set from a one-jump signal with height $\delta=2$, and the middle and right
  panels show the p-values from the spike and segment tests, collected over
  simulations for which the 1-step fused lasso correctly detected a changepoint
  at location 30. We can see that the segment test generally has much better
  power. In the two-jump setting, we only considered the signal strength of
  $\delta=2$, and the bottom left panel shows an example simulated data set.
  The middle and right panels show p-values coming from the spike and segment
  tests, after 1 or 2 steps of the fused lasso.  At both steps, the
  spike and segment tests were used to test the 
  significance of location 20.  The p-values at step 1 were collected
  over simulations in which location 20 was detected, and at step 2
  over simulations in which locations 20 and 40 were detected (in
  either order).  We can see that the power of the segment test
  improves after 2 steps, as it uses a more effective contrast, but
  that of the spike test degrades, as it uses the same contrast and
  simply incurs more conditioning.}
\label{fig:fl-example}
\end{figure}

\paragraph{One-jump signal.} 
First, we examine a problem setup with $n=60$, and where
$\theta \in \R^{60}$ has one changepoint at location 30, of
height $\delta$.  Data $y \in \R^{60}$ were generated by adding
i.i.d.\ $\cN(0,1)$ noise to $\theta$.  We considered three settings
for the signal strength: $\delta=0$
(no signal), $\delta=1$ (moderate signal), and $\delta=2$ (strong
signal).  See the top left panel of Figure 
\ref{fig:fl-example} for an example.  Over 10,000 repetitions of the
data generation process, we fit the 1-step fused lasso estimate, and
computed both  
the spike and segment tests at the detected changepoint location.
Their p-values are displayed via QQ plots, in the  
top middle and top right panels of Figure \ref{fig:fl-example},
restricted to repetitions for which the detected location was 30.  
(This corresponded to roughly
2.2\%, 30\%, and 65\% of the 10,000 total trials when $\delta=0$, 1, 
and 2, respectively.)  When $\delta=0$, we see that both the spike and 
segment tests deliver uniform p-values, as they should.  When  
$\delta=1$ and 2, we see that the segment test provides much better 
power than the segment test, and has essentially full power at the
strong signal level $\delta=2$. 

When the fused lasso detects a changepoint at location 29 or 31, i.e.,
a location that is off by one from the true changepoint at location 30,
the spike and segment tests again perform very differently.  The
spike test yields uniform p-values, as it should, while the segment
test offers nontrivial power.  See Appendix \ref{supp:oneoff}
for QQ plots of these results.  

\paragraph{Two-jump signal.} 
Next, we examine a problem with $n=60$ and where
$\theta \in \R^{60}$ has two changepoints, at locations 20 and 40, 
each of height $\delta=2$. Data $y \in \R^{60}$ were again generated  
around $\theta$ by adding  i.i.d.\ $\cN(0,1)$ noise.  See the bottom
left panel of Figure 
\ref{fig:fl-example} for an example.  Over 10,000 repetitions, we fit
2 steps of the fused lasso and recorded spike and segment p-values, at 
each step, for testing the significance of location 20.   The bottom
middle and bottom right panels of Figure \ref{fig:fl-example} display
QQ plots, restricted at step 1 to simulations in which location 20 was
detected (corresponding to about 32\% of the total number of
simulations), and restricted at step 2 to simulations in which
locations 20 and 
40 were detected (in either order, corresponding to again about 32\%
of the total simulations).  We see that the spike
test has better power at step 1 versus step 2, however, for the
segment test, the story is reversed.  The spike test contrast for
testing at location 20 does not change between steps
1 and 2; the extra conditioning incurred at step 2 only hurts its
power.  On the other hand, the segment test uses a different contrast 
between steps 1 and 2, and the contrast at step 2 provides better
power, because it leads to an average over a shorter segment (to the 
right of location 20) over which the mean is truly constant. 

\begin{figure}[htb]
  \centering 
    \includegraphics[width=0.45\textwidth]{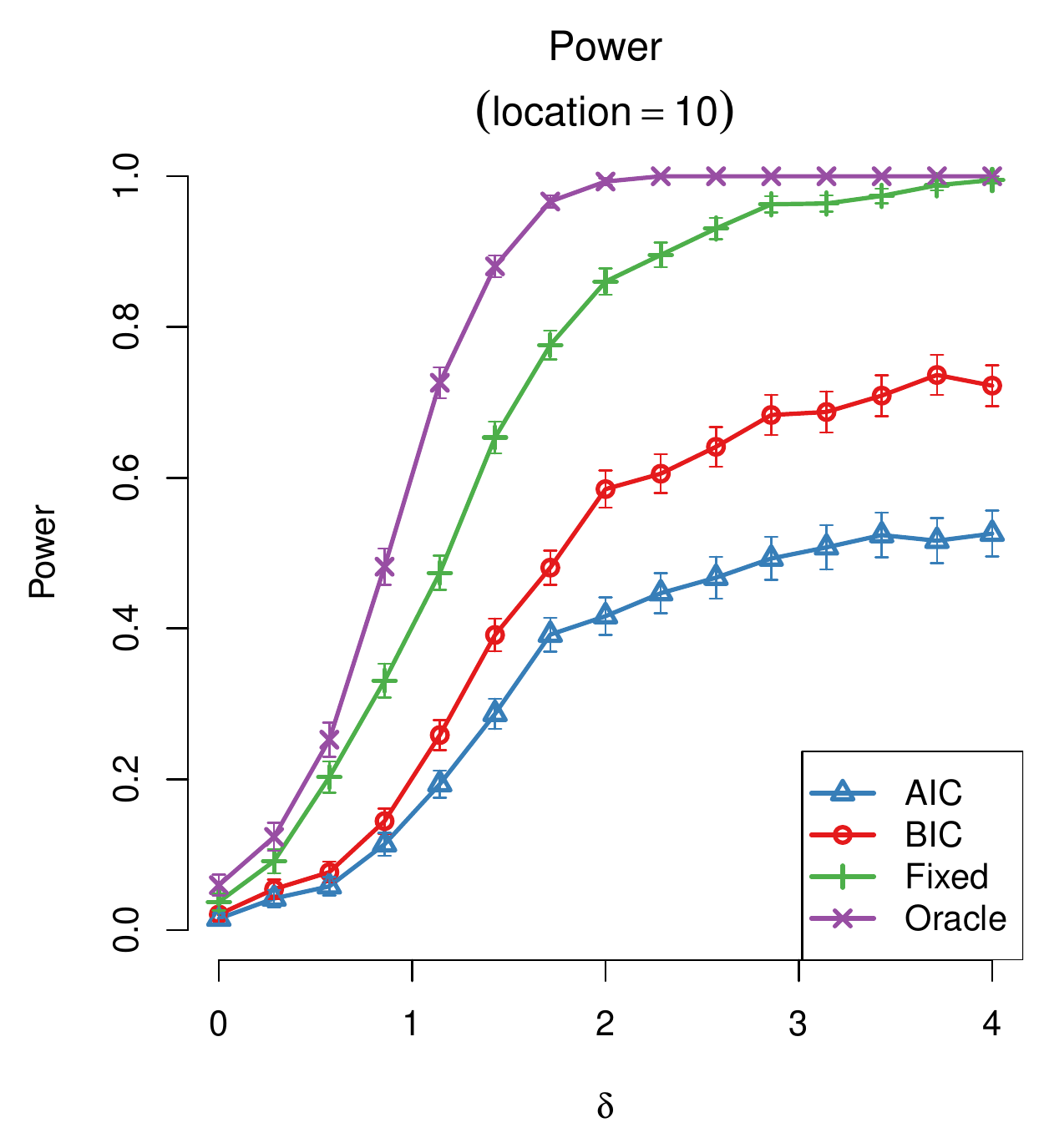}
    \hspace{2pt}
    \includegraphics[width=0.45\textwidth]{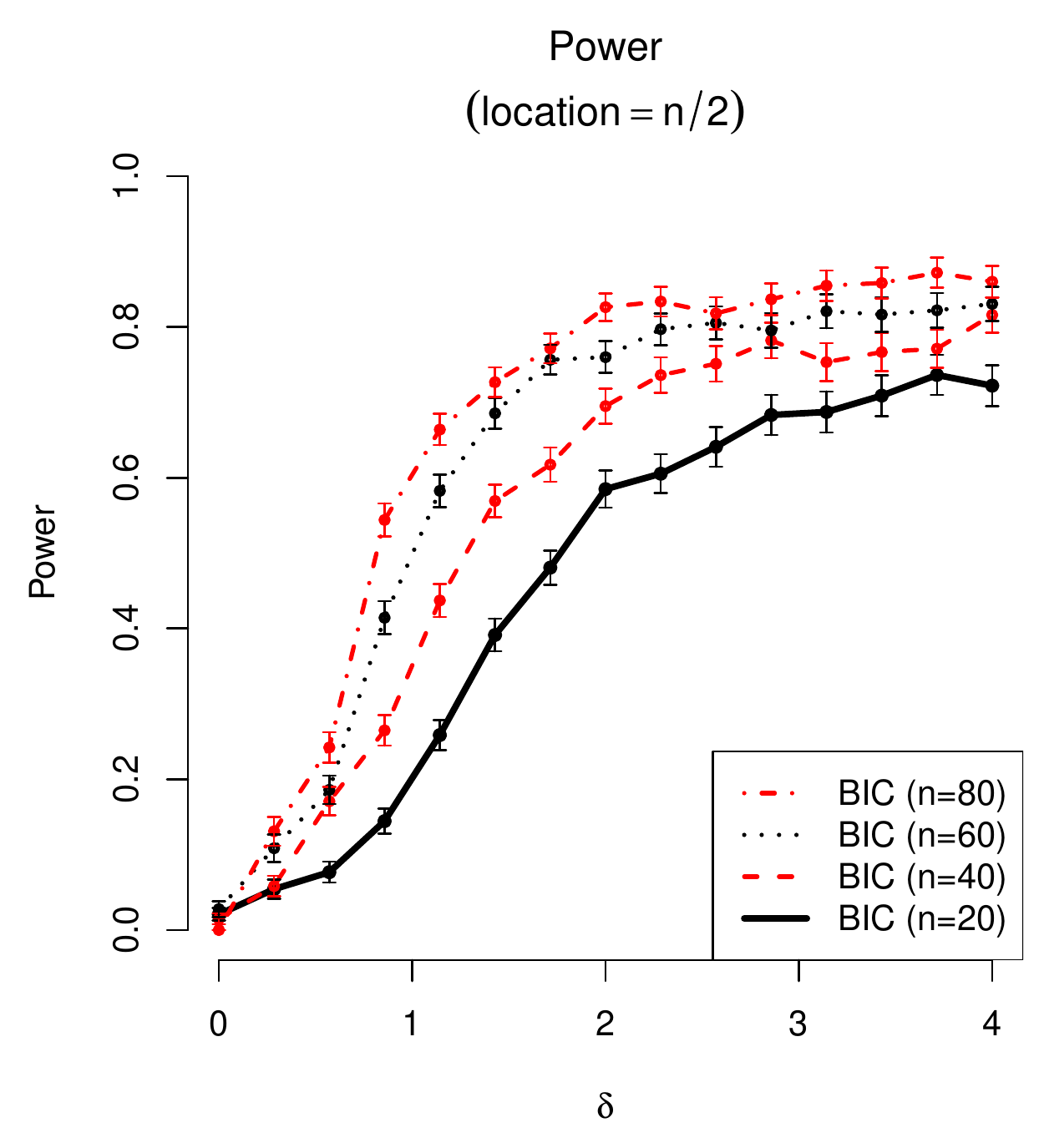}
\caption{\it\small Power curves for a one-jump signal with $n=20$
  points on the left, and $n=20,40,60,80$ on the 
  right, and in each case, having a jump at location $n/2$, of height   
  $\delta$.  The left panel shows the results from the segment
  test, either at step 1 (labeled ``Fixed'', in green), or at a step
  selected by the 2-rise AIC or BIC rule (labeled ``AIC'' and
  ``BIC'', in blue and red, respectively). The power curves were
  computed from p-values over 
  simulations in which location 10 appeared in the selected
  model (and the AIC- and BIC-based rules applied appropriate
  corrections for multiplicity).  The left panel also shows the
  results of applying an oracle test at location 10, for equality of 
  means. We can see a clear drop in power from the oracle to the fixed 
  rule to the IC-based rules.  The right panel shows the improvement
  in BIC power curves as $n$ increases.}
\label{fig:fl-power-example}
\end{figure}

\paragraph{IC-based stopping rules.}
The left panel of Figure \ref{fig:fl-power-example} shows
the segment test applied to a one-jump signal of length $n=20$, with a
jump at location 10 of height $\delta$, but this time incorporating
the IC-based stopping rules (to determine where along the fused lasso 
solution path to perform the test). This is a more practical
performance gauge because it requires minimal user input on model
selection.  Shown are power curves (fraction of rejections, at
the 0.05 level of type I error control) as functions of $\delta$,
computed over p-values from simulations in which location 10
was detected in the final model selected by the AIC- or BIC-type
rule described in Section \ref{sec:kchosen}, with $q=2$ (i.e.,
stopping after 2 rises in the criterion). Note that the p-values here
were all adjusted by the number of   
changepoints in the final AIC- or BIC-selected model, using a
Bonferroni  
correction (so that the familywise type I error is under control).
The BIC-type rule has better power than the AIC-type rule, as the
latter leads to larger models (AIC stops at 2.5 steps on average
versus 1.7 from BIC), resulting in further conditioning and also
misleading additional detected locations, both of which hurt its
performance. 

The results are compared to those from the segment test carried out at 
the 1-step fused lasso solution, over p-values from simulations  
in which location 10 was detected.  With less conditioning (and no
need for multiplicity correction), this method dominates the IC-based
rules in terms of power. The results are also compared to an oracle
rule who knows the correct  segments and carries out a test for
equality of means (with no conditioning); this serves as an upper
bound for what we can expect from our methods.
The right panel of Figure \ref{fig:fl-power-example} shows BIC
power curves as the sample size $n$ increases from 20 to 80, in
increments of 20.  We see a uniform improvement in power across all
signal strengths $\delta$, as $n$ increases.  However, at $n=80$, the
BIC-based test still delivers a power that is noticeably worse than
that of the oracle rule at $n=20$.

\subsection{Comparison to SMUCE-based inference}
\label{sec:smuce}

Here we compare our post-selection confidence intervals for the 1d
fused lasso to 
those based on the Simultaneous Multiscale Changepoint Estimator  
(SMUCE) of \citet{Frick2014}.  The SMUCE approach provides a
simultaneous confidence band for the components of the mean vector
$\theta$, from which confidence intervals for any linear
contrast of the mean can be obtained, and therefore, valid confidence 
intervals for post-selection  
targets can be obtained. Admittedly, a simultaneous band is a much
broader goal, and SMUCE was not designed for post-selection confidence
intervals, so we 
should expect such intervals to be wider than those from our
method. However, it is worthwhile to make empirical comparisons 
nonetheless. 

\begin{figure}[htb]
\centering
\makebox[\linewidth]{
\hspace{-5mm} 
\includegraphics[width=.35\linewidth]{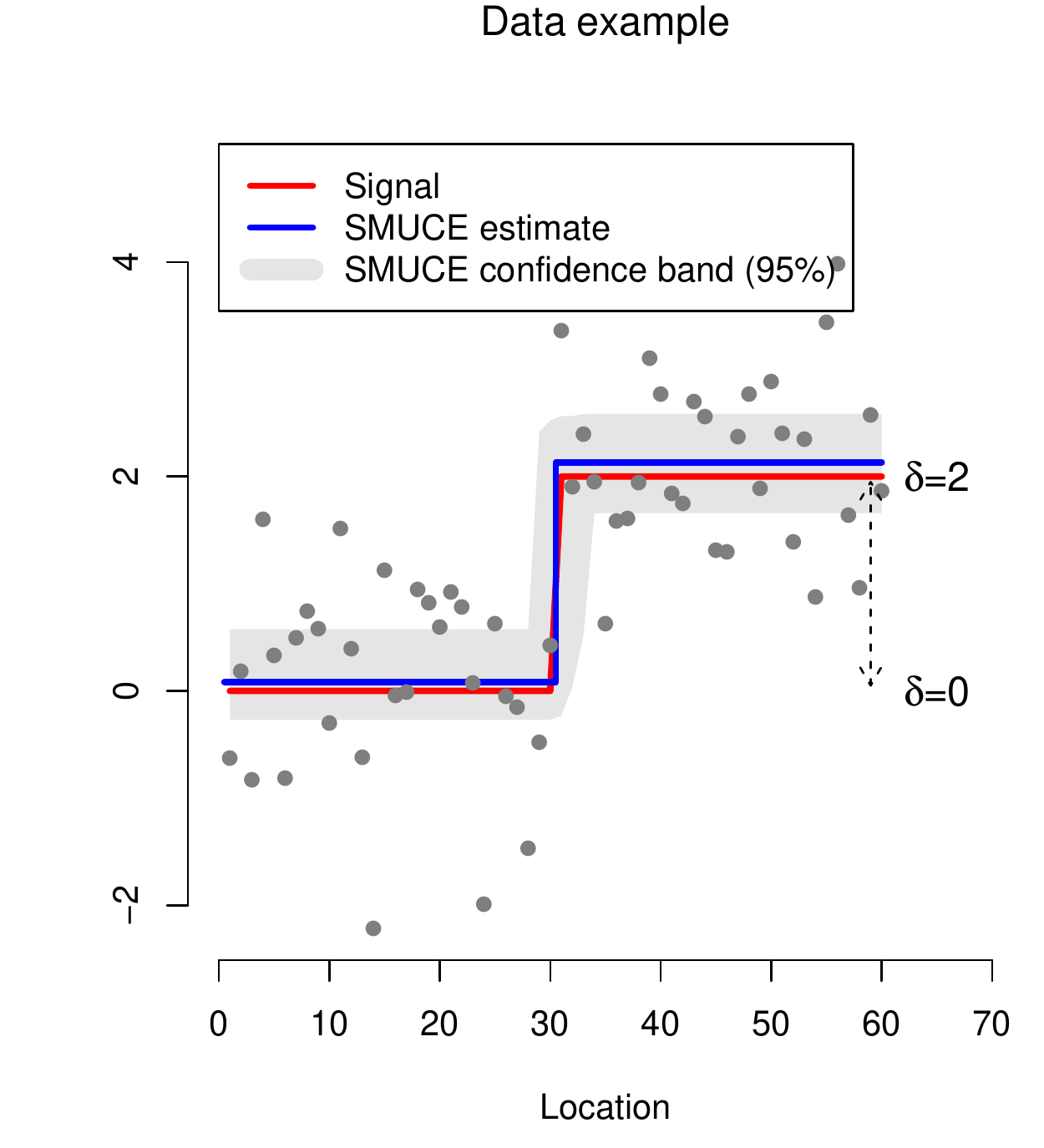}
\hspace{-5mm} 
\includegraphics[width=.35\linewidth]{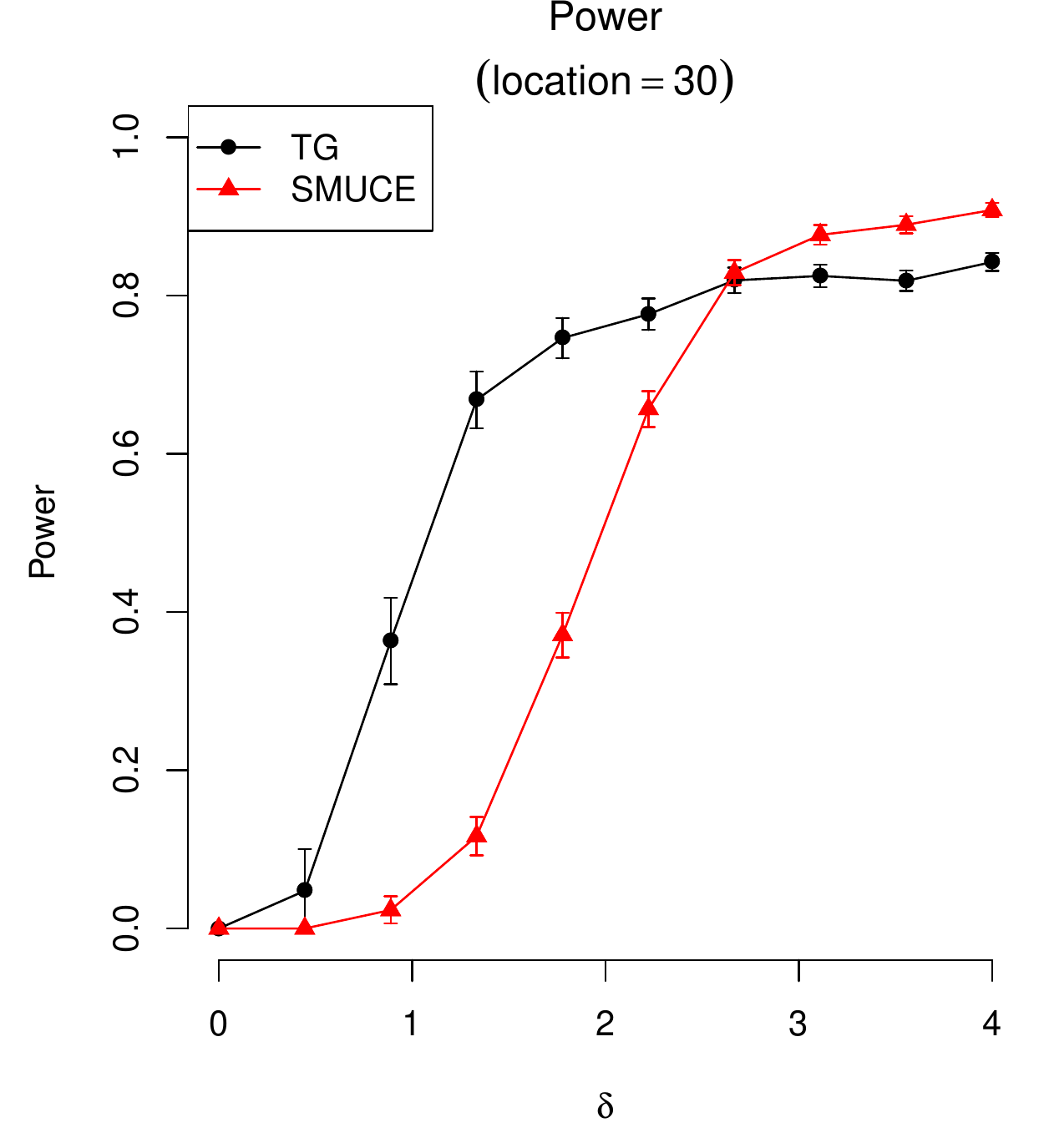}
\hspace{-5mm}
\includegraphics[width=0.35\linewidth]{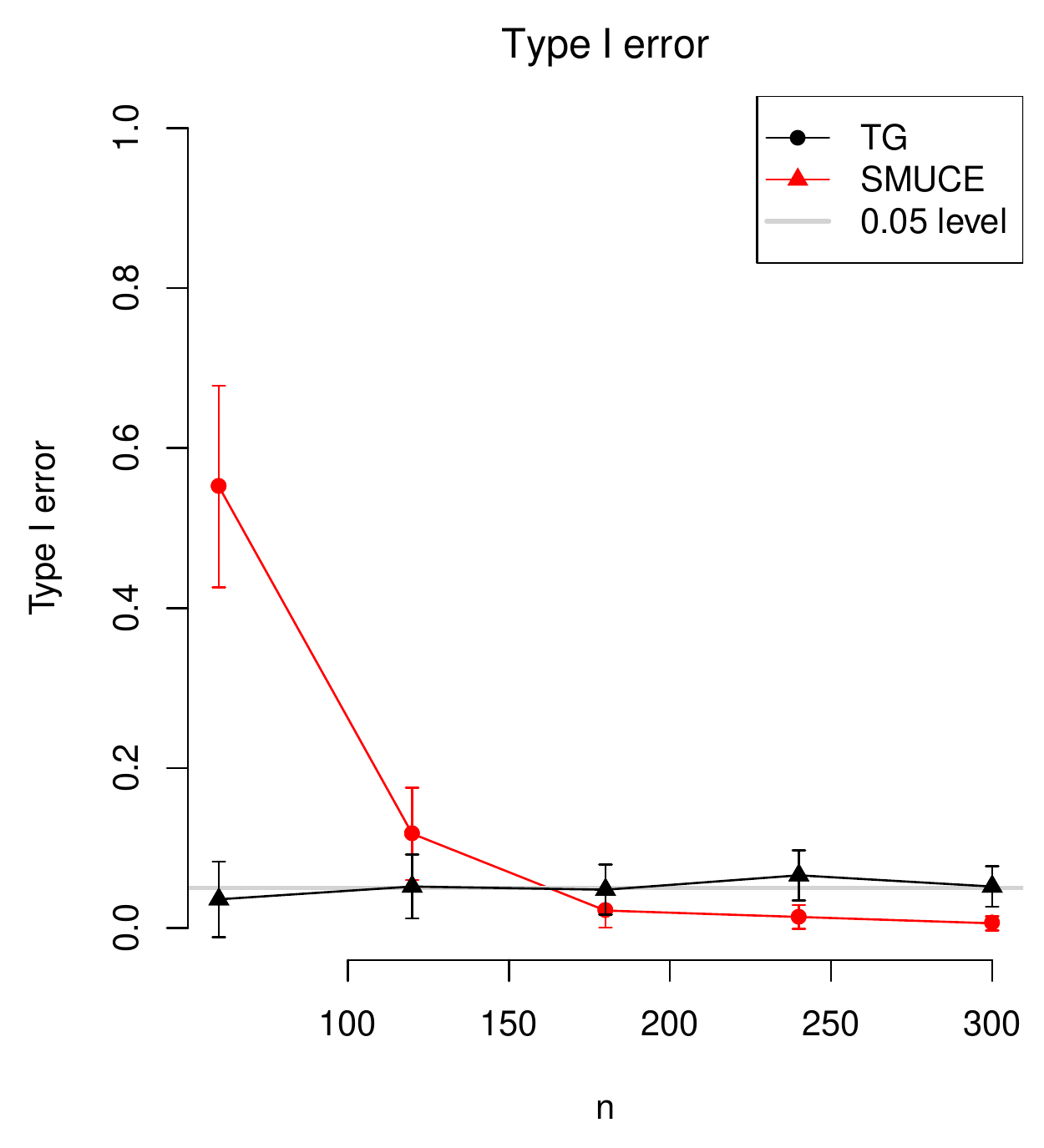}}
 \caption{\small\it Comparison of p-values from the TG test, and
   derived from the SMUCE simultaneous confidence band, for testing
   the same null hypothesis.  Data were generated under a problem
   setup that is the same as that in the top left panel of Figure
   \ref{fig:fl-example}, but with the signal strength $\delta$ varying
   between 0 and 4.  The top left panel of the 
   current figure shows an example with $\delta=2$.  In each
   simulation, the 1d fused lasso path was stopped using the 2-rise
   BIC rule, and segment test contrasts were formed around the
   detected changepoints.  The middle panel shows power curves,
   computed over simulations in which the location 30 appeared in the
   model selected by BIC.  These power curves were computed either
   from the SMUCE band having nominal confidence level 0.95, or the TG
   test with a type I error control of 0.05.  We can see that the
   latter method  has better power for smaller $\delta$, and both
   perform well for larger $\delta$, with the SMUCE-based method
   providing slightly more power.  The right panel displays the
   empirical type I error of the two testing methods, which emphasizes 
   that the SMUCE guarantees are only asymptotic, and this method can
   quite become conservative for large $n$, because in a way
   simultaneous coverage is a more abitious goal that
   post-selection coverage.} 
\label{fig:smuce}
\end{figure}

Data were generated as in the top left panel of Figure
\ref{fig:fl-example}, with the signal strength parameter $\delta$ varying
between 0 and 4.  We computed the 1d fused lasso path, and stopped
using the 2-rise BIC rule.  Over simulations in which the location 30
appeared in the eventual model selected by this rule, we computed the
segment test contrast \smash{$v_{\mathrm{seg}}$} around location 30,
and used the SMUCE band with a nominal confidence level of 0.95
to compute a post-selection interval for 
\smash{$v_{\mathrm{seg}}^T \theta$}.  A power curve was then computed,
as a function of $\delta$, by keeping track of the fraction of times
this interval did not contain 0. Again over simulations in which the
location 30 appeared in the model chosen by the BIC rule, we used
the TG test to compute p-values for the null hypothesis
\smash{$v_{\mathrm{seg}}^T \theta=0$}.  These p-values were
Bonferroni-corrected to account for the multiplicity of changepoints
in the model selected by the BIC rule, and a power of curve was
computed, as a function of $\delta$, by recording the
fraction of p-values below 0.05.  In the middle panel of  
Figure \ref{fig:smuce}, we can see that the TG test provides better
power until $\delta$ is about 2.5, after which both methods provides
strong power.  The right panel investigates the empirical type I
error of each method, as $n$ varies.  The SMUCE bands are
asymptotically valid, and recall, the TG p-values and intervals are
exact in finite samples (assuming Gaussian errors).  We can see that
SMUCE begins anti-conservative, before the asymptotics have ``kicked
in'', and then as $n$ grows, becomes overly conservative as a means of
testing post-selection targets, because these tests are derived from
a much more stringent simultaneous coverage property.  

\subsection{Trend filtering example}
\label{sec:tfexample}

\begin{figure}[htb]
\centering
\includegraphics[width=.475\linewidth]{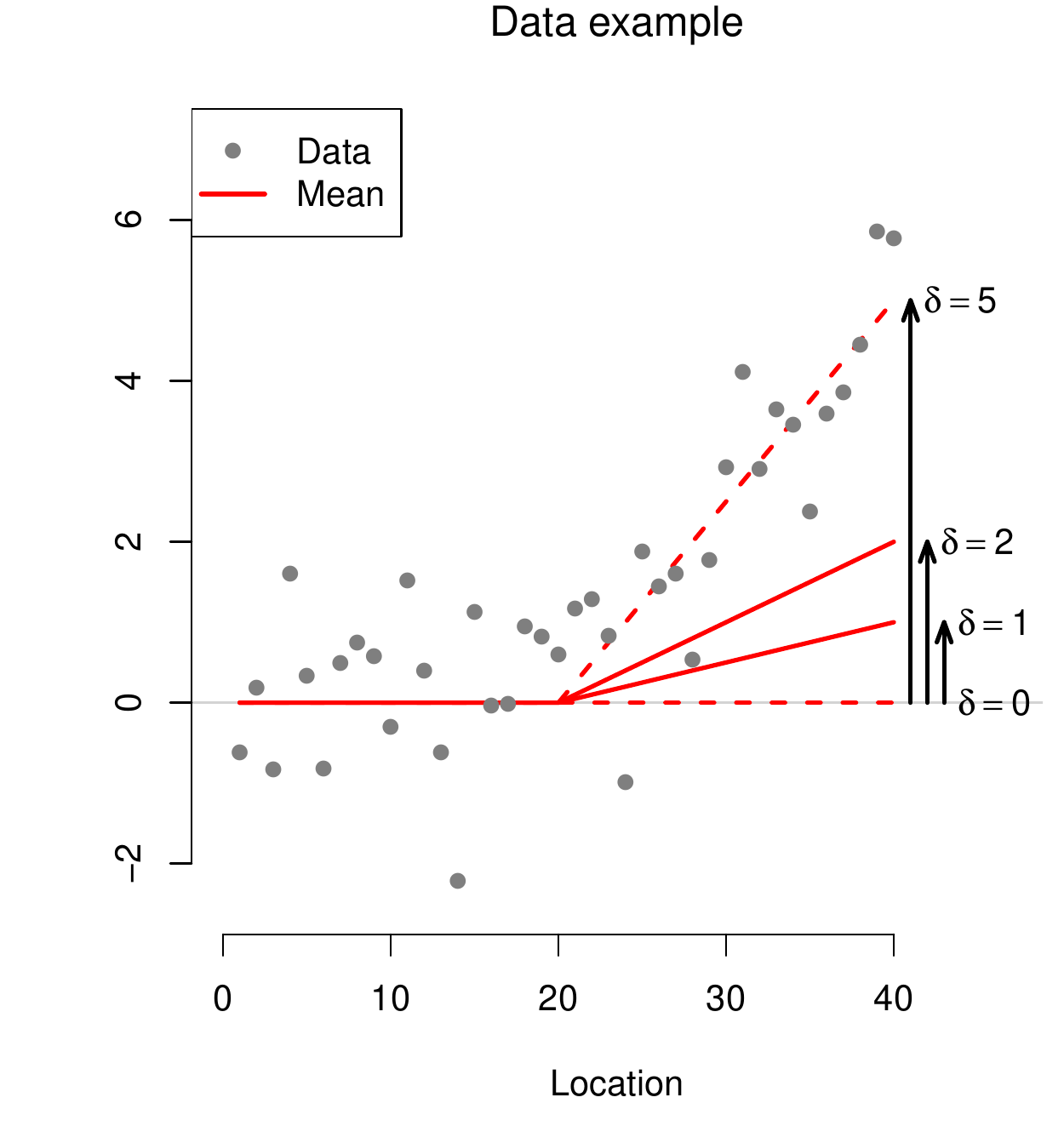}
\hspace{2pt}
\includegraphics[width=.475\linewidth]{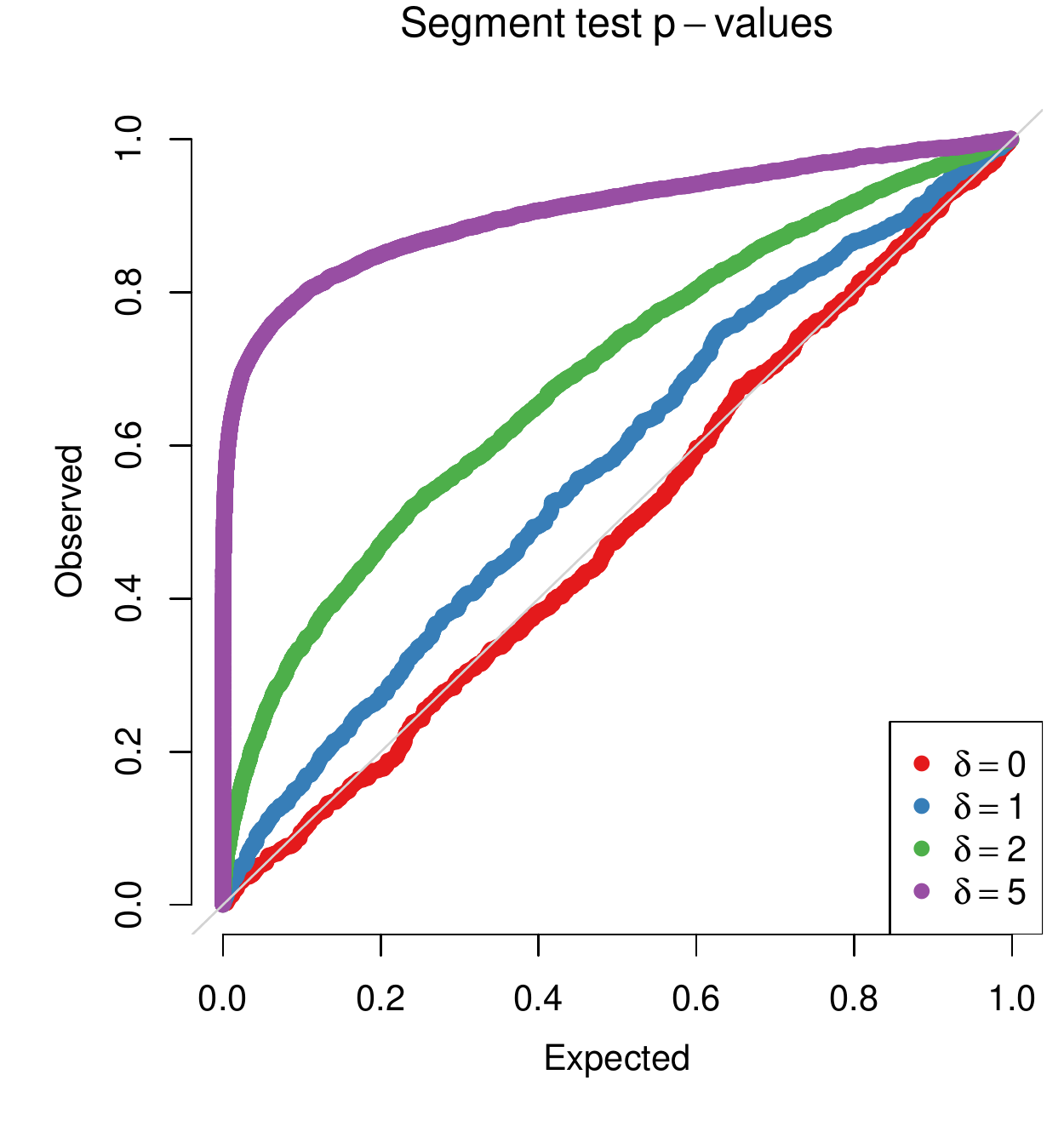}
\caption{\it\small Inferences from the segment
  test, in a setup with $n=40$ points, and one knot
  in the underlying piecewise linear mean at location 20, with the 
  change in slope is $\delta/20$.  We considered the settings
  $\delta=0,1,2,5$.  The left panel
  displays an example simulated data set from this setup, for 
  $\delta=5$. The right panel shows QQ plots of segment test
  p-values at location 20, computed from the trend filtering path,
  stopped by the 2-rise BIC rule.  The p-values  
  were restricted to repetitions in which location 20 appeared in the
  BIC-selected model. When $\delta=0$, we see
  uniform p-values, as appropriate; when $\delta=5$,
  we see nearly full power.}
\label{fig:tf-example}
\end{figure}

We examine a problem with $n=40$, and where $\theta \in \R^{40}$
has its first 20 components equal to zero, and its next 20 components
exhbiting a linear trend of slope $\delta/20$. Data $y \in \R^{60}$
were generated by adding i.i.d.\ $\cN(0,1)$ noise to $\theta$.  We
considered the four settings: $\delta=0$ 
(no signal), $\delta=1$ (weak signal), $\delta=2$ (moderate
signal), and $\delta=5$ (strong signal). See the left panel of Figure 
\ref{fig:tf-example} for an example.  We computed the trend filtering
path, stopped using the 2-rise BIC rule, and considered the segment
test at location 20.
The right panel of Figure
\ref{fig:tf-example} shows the resulting p-values, restricted to
repetitions in which location 20 appeared in the eventual
model.   We can see that when $\delta=0$, the p-values are uniformly
distributed, as we should expect them to be.  As $\delta$ increases,
we can also see the increase in power, with the jump from $\delta=2$
to $\delta=5$ providing the segment test with nearly full power.

\begin{figure}[h!]
\smallskip\smallskip
\centering
\makebox[\linewidth]{
\includegraphics[height=.35\textwidth,valign=c]{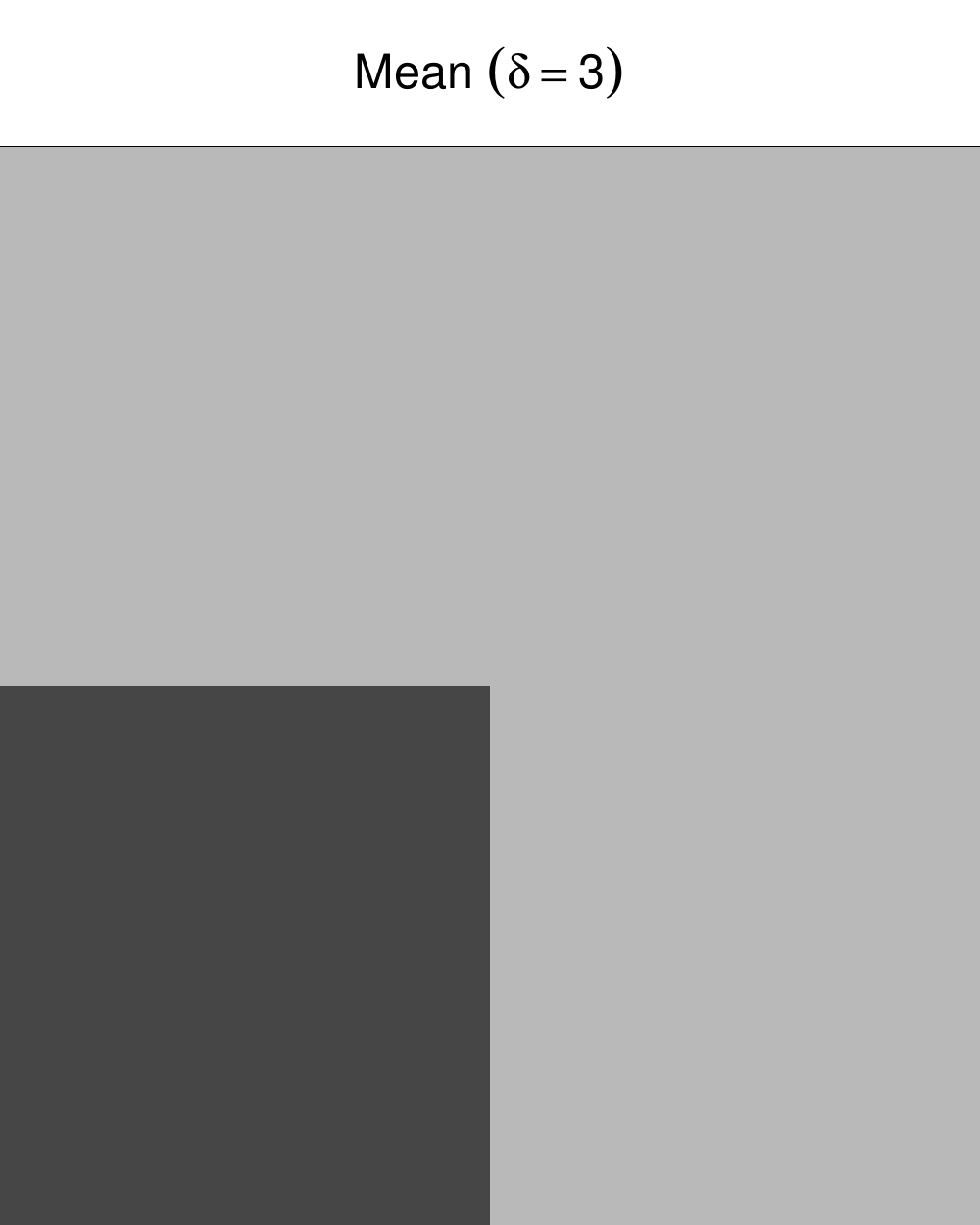}
\includegraphics[height=.35\textwidth,valign=c]{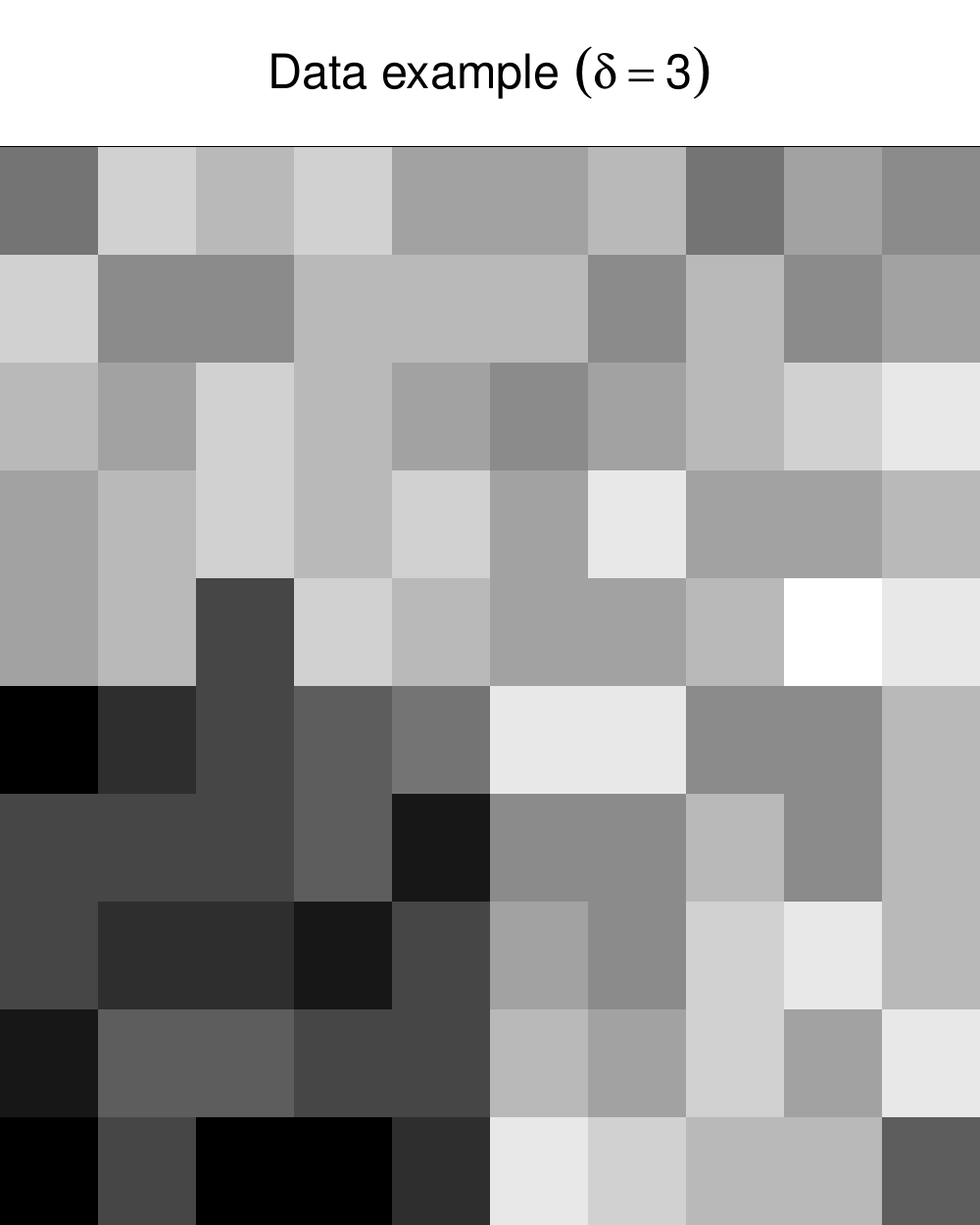}
\includegraphics[height=.35\textwidth,valign=c]{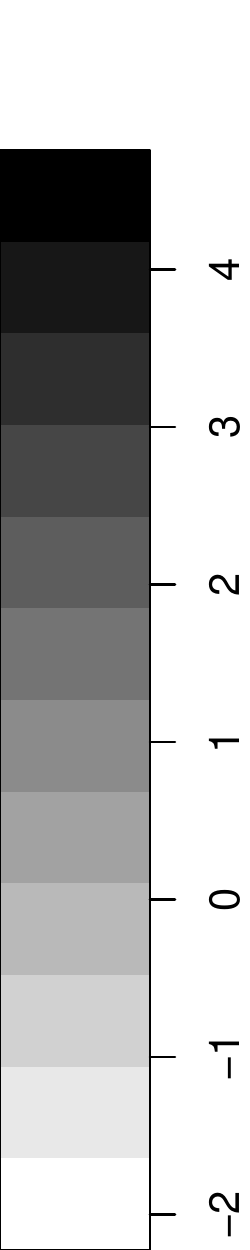}
\includegraphics[width=.35\textwidth,valign=c]{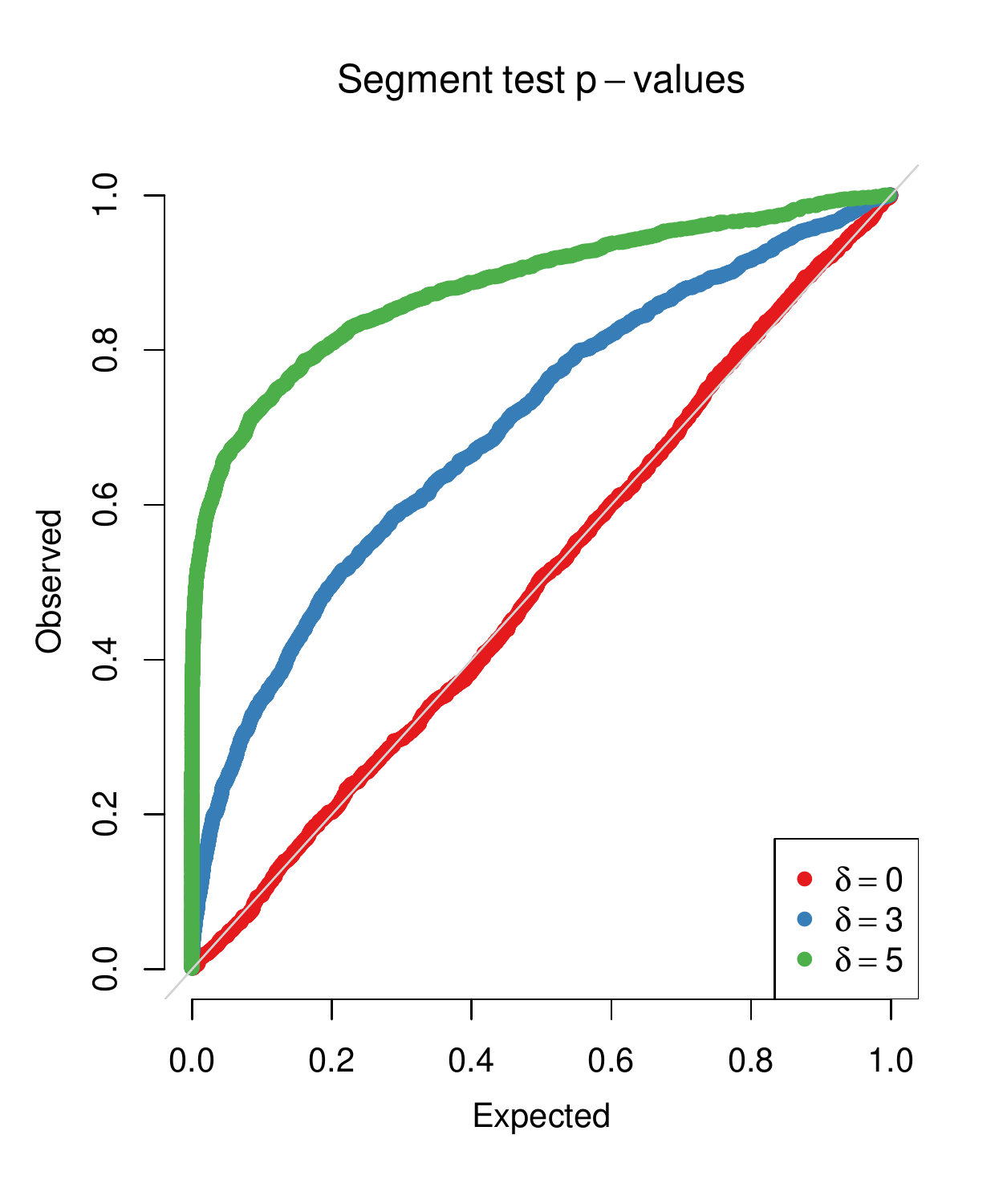}
}
\caption{\it\small Inferences from the segment test, in a 2d problem
  setup with $n=100$, and a mean parameter $\theta$ shaped into a
  piecewise constant $10\times 10$ image.  The bottom $5 \times 5$ 
  block of the mean is assigned a height of $\delta$, and the rest of
  its components 0.  We considered the settings $\delta=0,3,5$. The
  left panel visualizes the mean $\theta$, when $\delta=3$; the middle
  panel shows an example noisy realization $y$, again for $\delta=3$.
  The right panel shows QQ plots of the segment test, with respect to
  two fused components appearing in the 2d fused lasso estimate,
  stopped by the 1-rise BIC stopping rule. When $\delta=3,5$ these
  p-values are restricted to data instances in which the components
  being 
  tested are the lower left $5\times 5$ block and its complement; when
  $\delta=0$, all p-values are shown.  The p-values behave as we would
  expect: uniform for $\delta=0$, and increasing power for
  $\delta=3,5$.}
\label{fig:2d-example}
\end{figure}

\subsection{2d fused lasso example}
\label{sec:gflexample}

We examine a problem setup where the mean $\theta$ is
defined over a 2d grid of dimension $10 \times 10$ (so that $n=100$),
having all components set to zero, except for a $5 \times 5$ patch in
the lower left corner where all components are equal to $\delta$.
Data $y \in \R^{100}$ were generated by adding i.i.d.\ $\cN(0,1)$
noise to 
$\theta$. We considered the following settings: $\delta=0$ (no
signal), $\delta=3$ (medium signal), and $\delta=5$ (strong
signal). See the left panel of Figure \ref{fig:2d-example} for a
visualization of the mean $\theta$, and the middle panel for example
data $y$, both when $\delta=3$. 

Over many draws of data from the described simulation setup,   
we computed the 2d fused lasso solution path, and used the 1-rise BIC 
stopping rule.  For $\delta=3,5$, we retained only the repetitions 
in which the BIC-chosen 2d fused lasso estimate had exactly two
separate fused components---the bottom left $5 \times 5$ patch, and
its complement---and computed segment test p-values with respect to
these two components.  For $\delta=0$, we collected the segment test 
p-values over all repetitions, which were computed with respect to two 
arbitrary components appearing in the BIC-chosen estimate, in each
data instance. The right panel of Figure \ref{fig:2d-example} shows  
QQ plots for each value of $\delta$ in consideration.  
When $\delta=0$, we see uniform p-values, as expected; when
$\delta=3,5$, we see clear power.

\subsection{Regression example} 
\label{sec:regressionexample}

\begin{figure}
\begin{minipage}[c][11cm][t]{.45\textwidth}
  \vspace*{\fill}
  \centering
  \includegraphics[width=\linewidth]{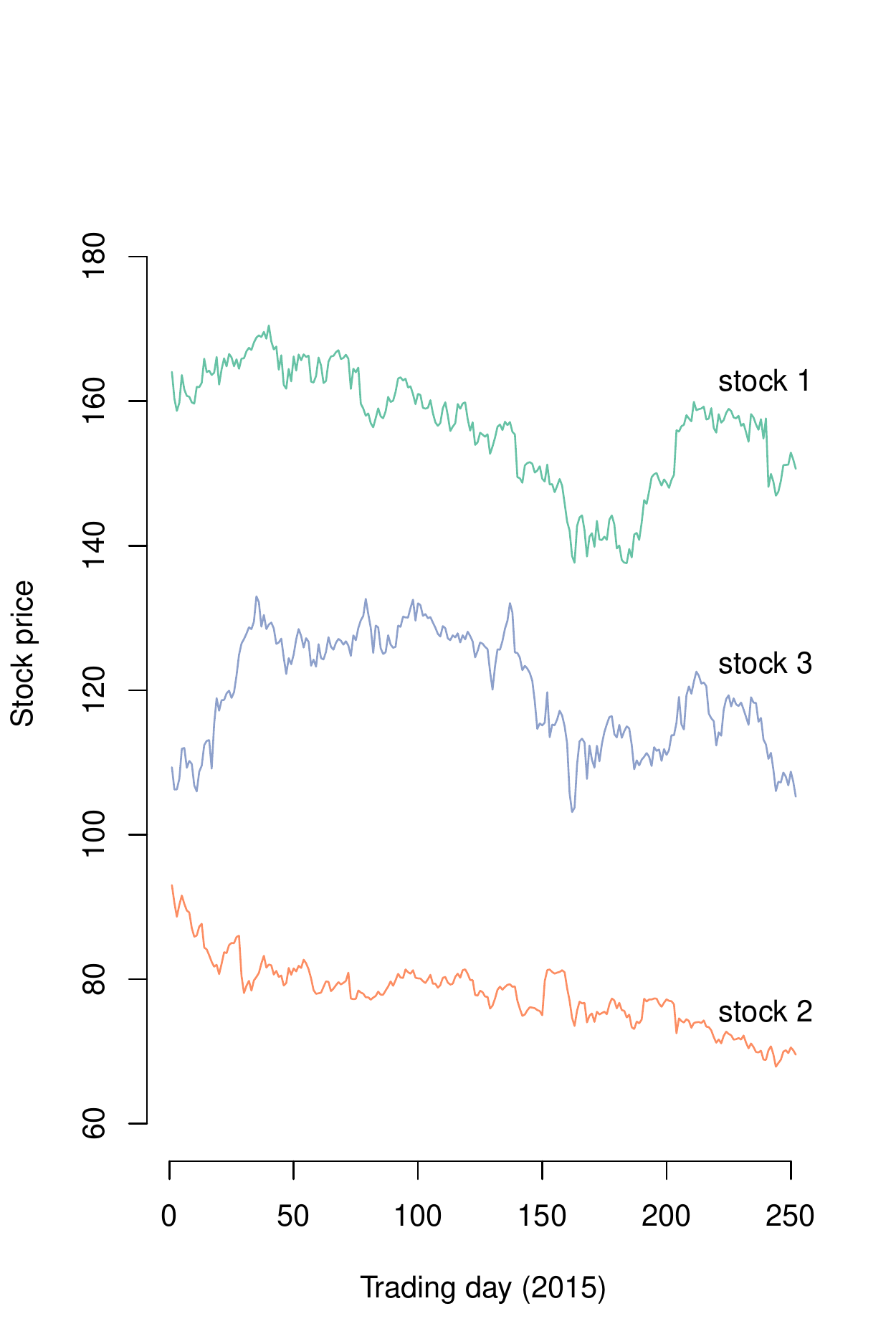}
\end{minipage}%
\begin{minipage}[c][11cm][t]{.5\textwidth}
  \vspace*{\fill}
  \centering
  \includegraphics[width=\linewidth]{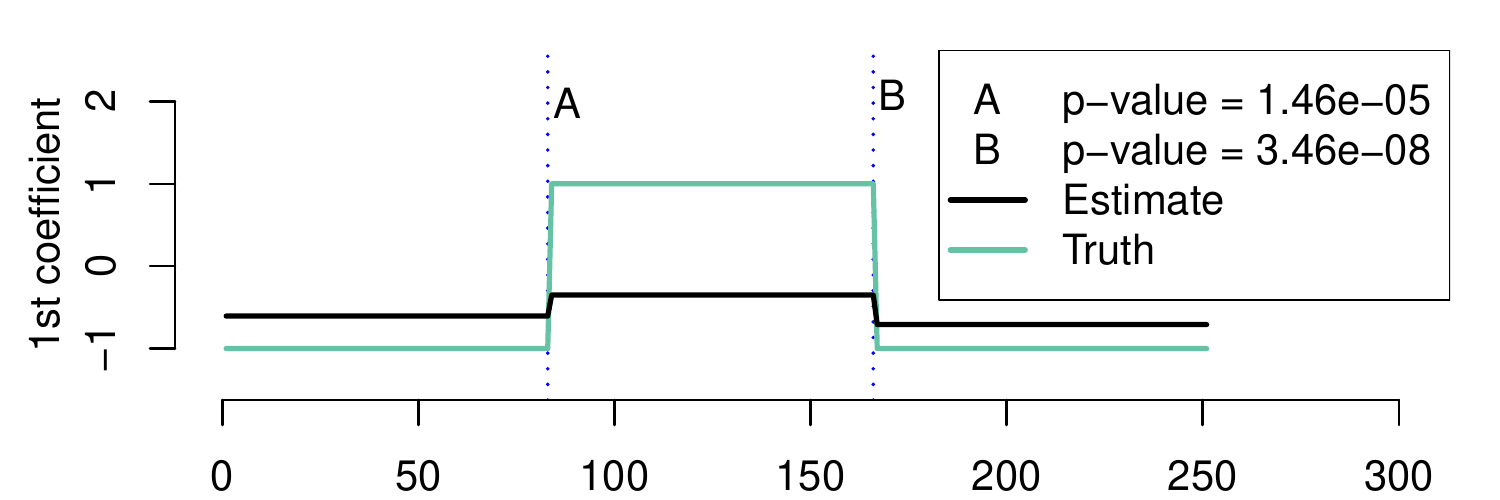}
  \includegraphics[width=\linewidth]{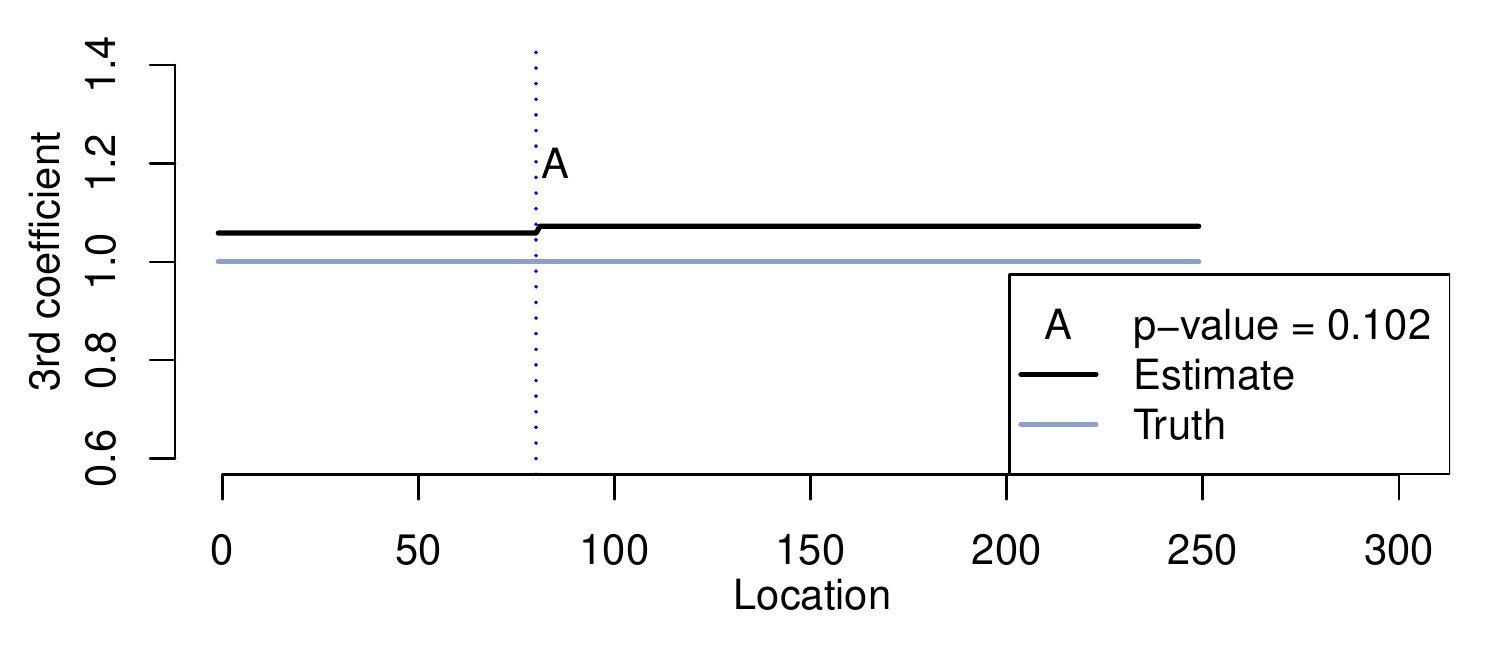}
  \includegraphics[width=\linewidth]{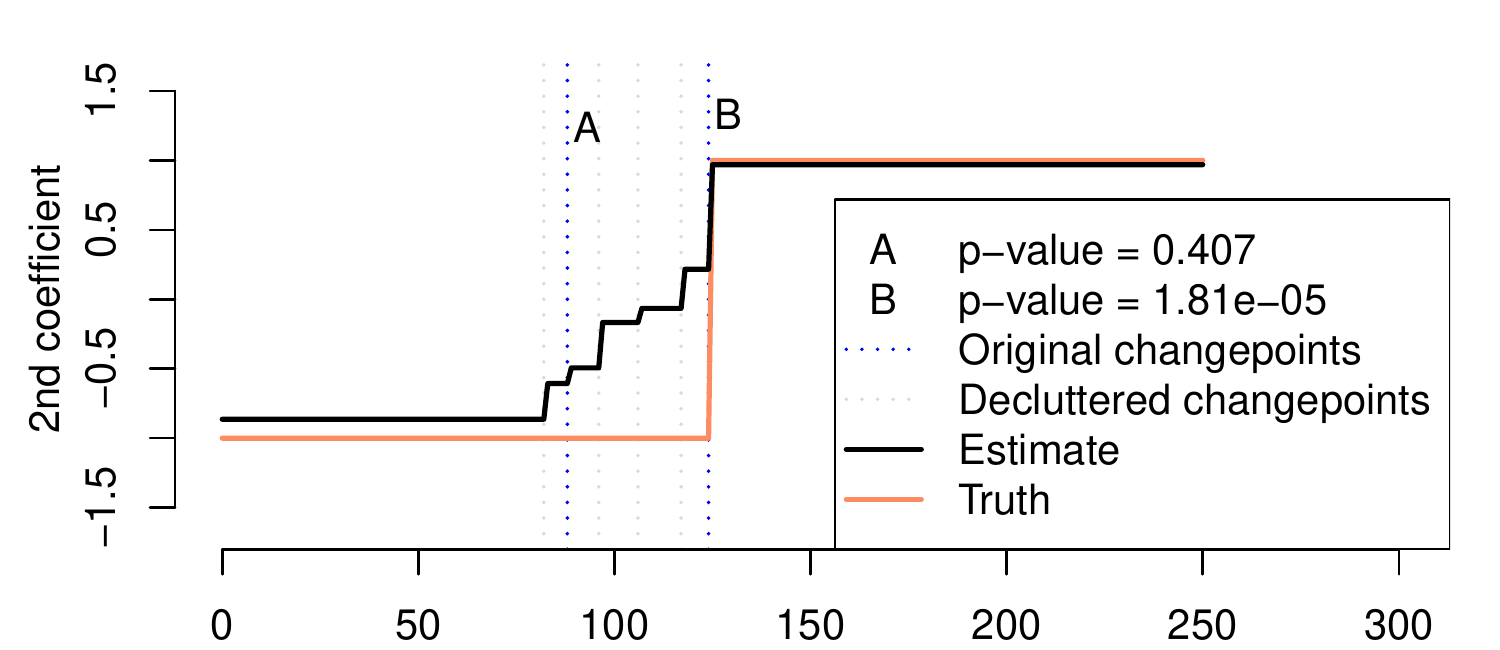}
\end{minipage}
\caption{\it\small A semi-synthetic stock example, with $n=251$
  timepoints or trading days. The left panel shows raw stock
  prices from three DJIA stocks; response data were generated
  according to a linear model with the log daily returns of these 
  stocks as predictors, and time-varying coefficients. The true
  piecewise constant time-varying coefficients are  
  displayed in the right panel. The fused lasso regression path was
  run, and stopped by the 2-rise BIC rule, delivering the estimated 
  coefficients also displayed in the right panel. After decluttering,
  segment tests were applied to the detected changepoints and 3
  approximately correct locations are deemed significant, with the
  other 2 spurious detections deemed insignificant.}
\label{fig:stock-example}
\end{figure}

We consider a semi-synthetic stock example, with $n=251$ timepoints, and data
$y \in \R^{251}$ simulated from a linear model of log daily returns of 3 real
Dow Jones Industrial Average (DJIA) stocks, from the year 2015. Data was
obtained from {\tt quantmod} R package. See the left panel of Figure
\ref{fig:stock-example} for a visualization of these stocks (note that what is
displayed is {\it not} the log daily returns of the stocks, but the raw stock
prices themselves).

Denoting the log daily returns as $X_j \in \R^{251}$, $j=1,2,3$, our
model for the data was
\begin{equation}
\label{eq:var_coef_model} 
y_t = \sum_{j=1}^3 X_{tj} \beta^*_{jt} + \epsilon_{jt}, 
\quad \epsilon_{jt} \sim \cN(0,\sigma^2),
\quad \text{i.i.d., for $t=1,\ldots,T$}.
\end{equation}
The coefficient vectors \smash{$\beta^*_j \in \R^{251}$}, $j=1,2,3$
were taken to be piecewise constant; the first coefficient vector
\smash{$\beta^*_1$} had 
two changepoints at locations 83 and 166, and had constant levels -1,
1, -1 from left to right; the second coefficient vector
\smash{$\beta^*_2$} had one changepoint at location 125, switching
from levels -1 to 1; the third coefficient vector \smash{$\beta^*_3$}
had no changepoints, and was set to have a constant level of 1.  
We generated data once from the
model in \eqref{eq:var_coef_model}, with $\sigma=0.002$ (this is a
reasonable noise level, as the log daily returns are
on a comparable scale).  We then computed the fused lasso
regression path, where 1d fused lasso penalties were placed on the
coefficient vectors for each of the 3 stocks, in order to enforce
piecewise constant behavior in the estimates \smash{$\hbeta_j \in
  \R^{251}$}, $j=1,2,3$.  The path was terminated using the 2-rise BIC
stopping rule, which gave a final model with 9 changepoints among 
the coefficient estimates.  After post-processing (``decluttering'')
changepoints that occurred within 10 locations of each other, we
retained 5 changepoints: 2 in the first estimated coefficient vector,
2 in the second, and 1 in the third.  Segment test p-values were
computed at each of the decluttered changepoints, and 3
changepoints that approximately coincided with true changepoints
were found to be 
significant, while the other 2 were found insignificant. See the
right panel of Figure \ref{fig:stock-example}.  For more
details on the fused lasso optimization problem, and the contrasts
used to define the segment tests, see Appendix
\ref{supp:regression}.

\subsection{Application to CGH data}
\label{sec:cgh}

We examine the use of our fused lasso selective inference tools 
on a data set of array comparative genomic hybridization (CGH) 
measurements from two glioblas-toma multiforme (GBM) tumors, from
the {\tt cghFLasso} R package.
CGH is a molecular cytogenetic method for determining DNA copy numbers 
of selected genes in a genome, and array CGH is an improved method which
provides higher resolutions measurement. Each CGH measurement is 
a log ratio of the number of DNA copies of a gene compared to a
reference measurement---aberrations give nonzero log
ratios. \citet{fusedLassoCGH} considered the sparse 1d
fused lasso as a method for identifying regions of DNA copy number
aberrations from CGH data, and analyzed the GBM tumor data set as a
specific example, with $n=990$ data points.

\begin{figure}[htbp]
\centering
\includegraphics[width=\textwidth]{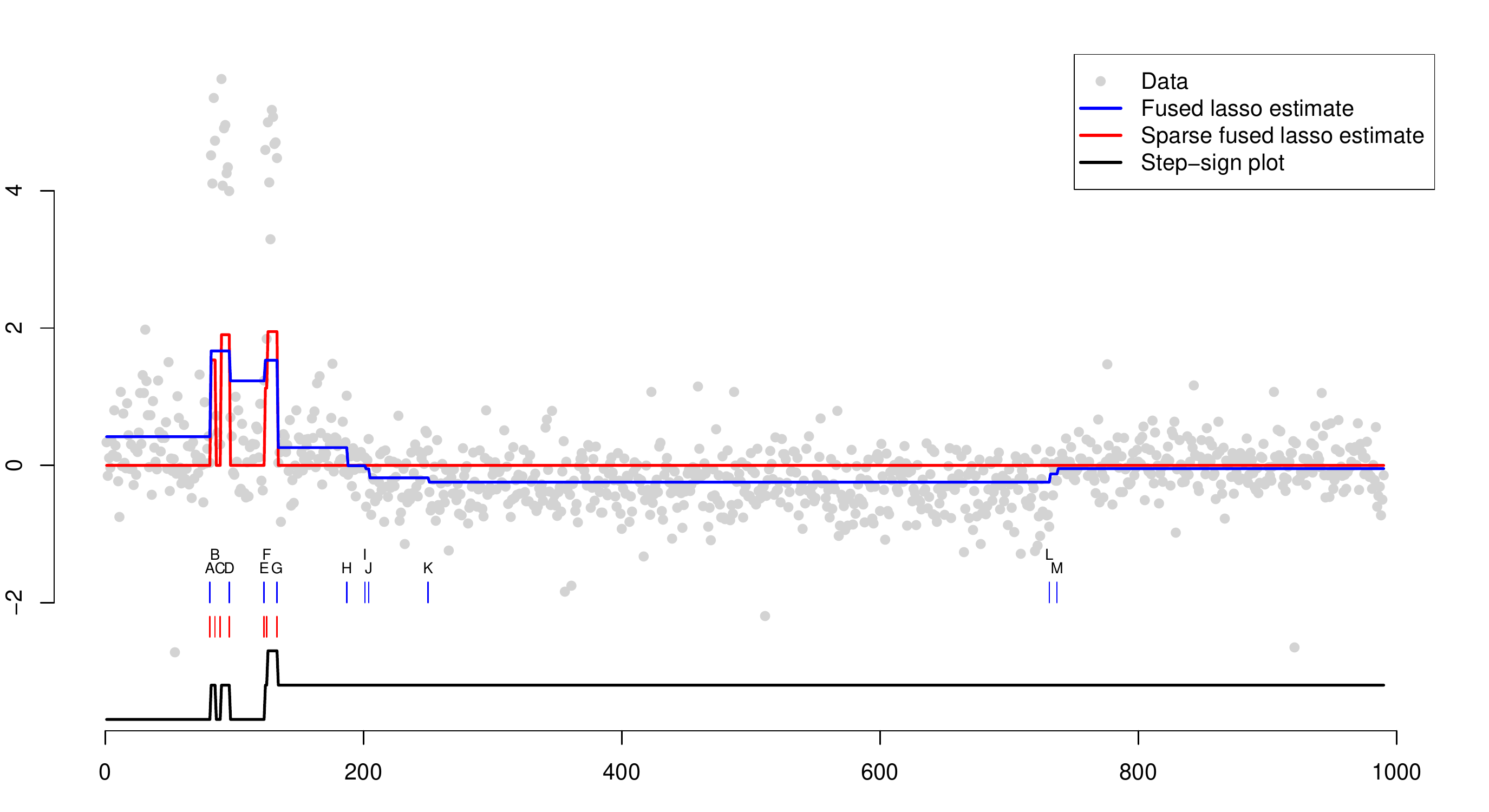}
\\ \bigskip
{\bf Post-selection} 
\\ \smallskip
{\scriptsize
\begin{tabular}{|l|c|c|c|c|c|c|c|c|c|c|c|c|c|}
\hline
 & A & B & C & D & E & F & G & H & I & J & K & L & M \\ 
  \hline
Test location & 81 & 85 & 89 & 96 & 123 & 125 & 133 & 
187 & 201 & 204 & 250 & 731 & 737 \\  
  \hline
  P-value  (non-sparse) & 0.00 &  &  & 0.00 &
  0.00 &  & 0.00 & 0.77 & 0.10 & 0.47 & 0.90 & 0.42
  & 0.58 \\ 
    \hline
  P-value (sparse) & 0.22 & 0.12 & 0.89 & 0.88 & 0.00 & 
  & 0.13 &  &  &  &  &  &  \\  
  \hline
\end{tabular}}
\\ \bigskip
{\bf Sample splitting}
\\ \smallskip
{\scriptsize
\begin{tabular}{|l|c|c|c|c|c|c|c|c|c|c|c|c|c|}
\hline
Test location & 80 & \hphantom{85} & \hphantom{89} & 96 & 122 &  
\hphantom{125} & 132 & \hphantom{187} & \hphantom{201} &
\hphantom{204} & \hphantom{250} & \hphantom{731} & 
\hphantom{737} \\ 
\hline
P-value \hphantom{(non-sparse)} & 0.00 & \hphantom{0.00} & 0.00 & 0.00  
& \hphantom{0.00} & 0.00 & \hphantom{0.00} & \hphantom{0.00} &  
\hphantom{0.00} & \hphantom{0.00}  & \hphantom{0.00} &
\hphantom{0.00} & \\   
\hline 
\end{tabular}}
\bigskip
\caption{\it\small A CGH data set of two GBM tumors, from
  \citet{fusedLassoCGH}, with $n=990$ points.  The plot displays the
  1d fused lasso and sparse 1d fused lasso estimates, in blue and red,
  respectively, each chosen using an appropriate BIC-based stopping
  rule (after 2-rises for the non-sparse estimate, and 1-rise for the
  sparse estimate).  The detected changepoints in each of the two
  models are also labeled. Shown at the bottom of the plot is
  a step-sign plot of the sparse 1d fused lasso solution.  Below the 
  plot are two tables, the first filled with segment test p-values of
  the changepoints in the 1d fused lasso and sparse 1d fused lasso
  models.  Post-processing of changepoints was applied to rule out
  changepoints within 2 locations of each other; this only affected
  the location labeled F in the sparse 1d fused lasso model (hence
  location F was not tested).   The second table shows p-values from a 
  simple sample splitting scheme, where the odd numbered locations
  were used for fused lasso model fitting and the even numbered
  locations for changepoint testing.  We can see that all three tests
  mostly agree on the significance of common locations labeled A, D,
  E, and G, though the sparse 1d fused lasso p-values appear to be 
  under-powered.} 
\label{fig:cgh-example}
\end{figure}

Using the same GBM tumor data set, we examine the significance 
of changepoints that appear in the 10th step of 
the 1d fused lasso path, and separately, changepoints that
appear in the 28th step of the sparse 1d fused lasso path (in general,
unlike the 1d fused lasso, the sparse 1d fused lasso can add and
delete changepoints at each step of the path; the estimate at the
28th step here only had 7 changepoints). These steps were 
chosen by the 2-rise and 1-rise BIC rules,
respectively.\footnote{Anecdotally, for generalized lasso problems 
  in which the penalty matrix 
  $D$ is full row rank (like the 1d fused lasso or trend filtering) we
  have found the 2-rise BIC stopping rule to work well; for problems
  in which $D$ is not full row rank (like the sparse 1d fused lasso,
  sparse trend filtering, or the graph fused lasso over a graph
  with more edges than nodes), we have found the 1-rise BIC stopping
  rule  to work well.}  The resulting estimates are plotted along with
the GBM tumor data, in Figure \ref{fig:cgh-example}.  Displayed below
this is a step-sign plot of the sparse 1d fused lasso
estimate, serving as example of what 
might be shown to the scientist to allow him/her to hand-design
interesting contrasts to be tested. 

Below the plot is a table containing the p-values from segment tests
of the changepoints in the two models, i.e., from the 1d fused lasso
and sparse 1d fused lasso.
The segment test contrasts were
post-processed (i.e., ``decluttered'') so as to exclude changepoints
that occurred within 2 locations of each other---this only affected
the locations labeled E and F in the sparse 1d fused lasso model
(and as a result, the significance of changepoint at location F was
not tested). Commonly detected changepoints occur at locations labeled 
A, D, E, and G; the segment tests from the 1d fused lasso model yield
significant p-values at each of these locations, but those from the 
sparse 1d fused lasso model only yield a significant p-value at
location E. This apparent loss of power with the sparse 1d fused lasso
may be due to the larger amount of conditioning involved.

We also compare the above to results from simple changepoint tests 
carried out using sample splitting. This is possible in a structured
problem like 
ours, where there is a sensible way to split the data (note
that in a less structured setting, like a generic graph fused
lasso problem, 
there would be no obvious splitting scheme). We divided the GBM data
set into two halves, based on odd and even numbered locations.  On the 
first half, the ``estimation set'', we fit the 1d fused lasso path and
chose the stopping point using 5-fold cross-validation (CV), where
the folds were defined to include every 5th data point in the
estimation set. After determining the path step that minimized the CV
error, we moved back towards the start of the path (back towards step
1) until a further move would yield a 
CV error greater than one standard error away from the minimum (this
is often called the ``one standard error rule'', see, e.g., Chapter 7.10
of \citet{esl}).  This gave a path step of 18, and hence 18
changepoints in the final 1d fused lasso model.  Using the second half
of the data set, the ``testing set'', we then ran simple Z-tests to
test for the equality of means between every pair of adjacent segments 
partitioned by the 18 derived changepoints from the estimation set.
For simplicity, in the table in Figure \ref{fig:cgh-example}, we only
show p-values at locations that are comparable to the common locations
labeled A, D, E, G from the fused lasso estimation procedures run
on the full data set.  All are significant.

Lastly, we note that to apply all tests in this subsection,
it was necessary to estimate the noise variance $\sigma^2$.
To do so, we ran 5-fold CV on the full data set, chose the
stopping point using the one standard error rule, and estimated
$\sigma^2$ appropriately based on the residuals.  This gave
\smash{$\hat\sigma = 0.46$}. 

\section{Discussion}
\label{sec:discussion}

We have extended the post-selection inference framework of
\citet{exactlasso,exactlar} to the model selection events along the
generalized lasso path, as studied by \citet{genlasso}.  The
generalized lasso framework covers a fairly wide range of problem 
settings, such as the 1d fused lasso, trend filtering, the graph fused
lasso, and regression problems in which fused lasso or trend filtering 
penalties are applied to the coefficients.  In this work, we 
developed a set of tools for conduting formal inferences on
components of the adaptively fitted generalized lasso model---these
are, e.g., adaptively fitted changepoints in the 1d fused lasso, knots
in trend filtering, and clusters in the graph fused lasso.  Our
methods allow for inferences to be conducted at any fixed step of the
generalized lasso solution path, or alternatively, at a step chosen by
a rule that tracks AIC or BIC until a given number of rises in the
criterion is encountered.

It is worth noting the following important point.  In the language of 
\citet{optimalinf,fithian2015selective}, the development of
post-selection tests in  
this paper was done under a ``saturated model'' for the mean 
parameter $\theta$---this treats $\theta$ as an arbitrary vector in
$\R^n$, and the hypotheses being tested are all phrased in terms of
certain linear contrasts of the mean parameter begin zero, as in $v^T
\theta = 0$. \citet{optimalinf,fithian2015selective} show how to also
conduct tests under the ``selected model''---to use the 1d fused lasso
as an example, this would model the mean as a vector that is piecewise
constant with breaks at the selected changepoints.  The techniques
developed in 
\citet{fithian2015selective}, allow us to perform sequential tests of
the selected model---again to use the 1d fused lasso as an example,
this would allow us to test, at each step of the 1d fused lasso path,
that the mean is piecewise constant in the changepoints detected over
all previous steps, and thus a failure to reject would mean that all
relevant changepoints have already been found.  The selected model
tests of \citet{fithian2015selective} have the following desirable
properties: (i) they do not require the marginal error variance
$\sigma^2$ to be known; (ii) they often display better power (compared
to the tests from this paper) when the selected model is false; (iii)
they yield independent p-values across steps in 
the path for which the selected model is true.  The latter property
allows us to apply p-value aggregation rules, like the
``ForwardStop'' rule of \citet{fdrlasso}, to choose a stopping point
in the path, with a guarantee on the FDR.  This is an appealing
alternative to the AIC- or BIC-based stopping rules described in
Section \ref{sec:kchosen}. The downside of the 
selected model tests is that they are computationally expensive
(compared to those described in this paper), and require sampling  
(rather than analytic computation, using a truncated Gaussian pivot)
to compute p-values.  Furthermore, once we use
the selected model p-values to choose a stopping point in the path, it
is not clear how to carry out valid post-selection tests in the
resulting model (due to the corresponding conditioning region being
very complicated).   
Investigation of selected model inference along
the generalized lasso path will be the topic of future work.

There are several other possible follow-up ideas for future work.  One 
that we are particularly keen on is the attachment of post-selection
inference tools to existing, commonly-used methods for 1d changepoint  
detection.  It is not hard to show that the selection
events associated with 
many such methods---like binary 
segmentation, wild binary segmentation of  
\citet{fryzlewicz2014}, and all wavelet thresholding procedures
(provided that soft- or hard-thresholding is used)---can be
characterized as polyhedral sets in the data $y$.  The ideas in
this paper can therefore be used to conduct significance tests for the
detected changepoints after any number of steps of binary
segmentation, wild binary segmentation, or wavelet thresholding,
this number of steps either being fixed or chosen by an AIC- or
BIC-type rule.  Because other 1d changepoint detection methods can
often outperform the 1d fused lasso in terms of their accuracy in
selected relevant changepoints (wild binary segmentation,
specifically, has this property), pairing them with formal tools
for inference could have important practical implications.



\newpage
\appendix

\section{QQ plots for the 1d fused lasso at one-off detections}
\label{supp:oneoff}

\begin{figure}[h!]
\centering
\includegraphics[width=.4\textwidth]{figures/onejump-example-data-and-contrast.pdf} 
\hspace{2pt}
\includegraphics[width=.4\textwidth]{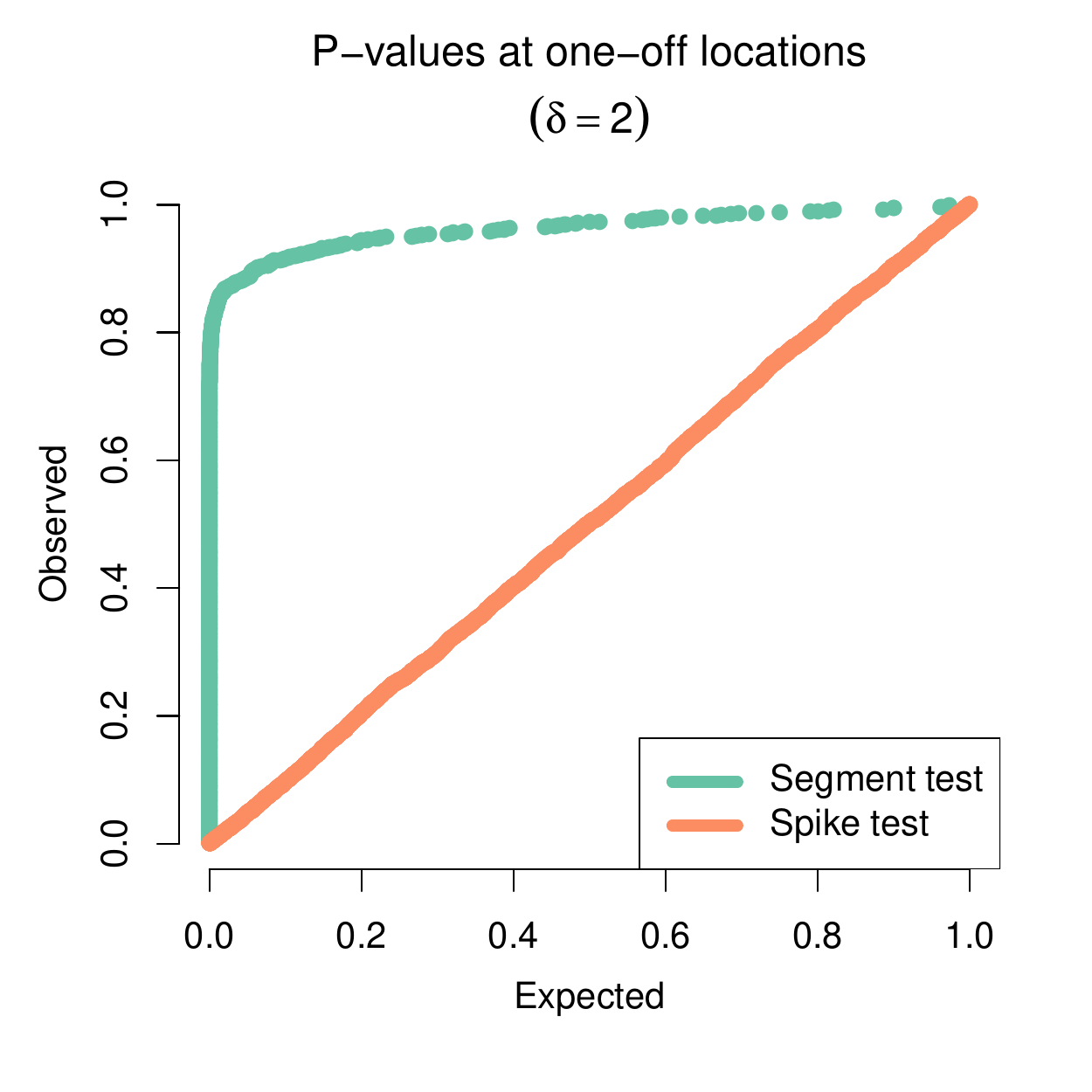}
\caption{ \it\small The left panel is copied from Figure
  \ref{fig:fl-example} in the main text, and shows an example data set
  with $n=60$ and a piecewise constant mean $\theta$ with one jump at
  location 30.  The right panel shows QQ plots from the spike and
  segment tests run at the detected changepoint from the 1-step fused
  lasso, over data instances in which the detected changepoint
  occurred at location 29 or 30, i.e., off by one from the true location
  30.  We can see that the spike test p-values are uniform, and the
  segment test p-values are highly sub-uniform.} 
\label{fig:fl-oneoff}
\end{figure}

We consider the same simulation setup as in the top row of Figure 
\ref{fig:fl-example}, where, recall, the sample size was $n=60$ and
the mean $\theta \in \R^{60}$ had a single changepoint at location
30.  Here we consider the changepoint to have height $\delta=2$, draw
data $y \in \R^{60}$ around $\theta$ using i.i.d.\ $\cN(0,1)$ errors,
and retain instances in which 1 step of the fused lasso path detects
a changepoint at location 29 or 31, i.e., off by one from the true
location 30.  Figure \ref{fig:fl-oneoff} (right panel) shows QQ plots
for the spike and segment tests, applied to test the significance of 
the detected changepoint, in these instances.  We can see that the
spike test p-values are uniformly distributed, which is appropriate,
because when the detected changepoint is off by one, the
spike test null hypothesis is true.  The segment test, on the
other hand, delivers very small p-values, giving power against its own
null hypothesis, which is false in the case of a one-off detection.

\section{Regression example details}
\label{supp:regression}

Recall the notation of Section \ref{sec:regressionexample}, where $X_j
\in \R^{251}$, $j=1,2,3$ denote the log daily returns of 3 real DJIA
stocks, and \smash{$\beta^*_j \in \R^{251}$}, $j=1,2,3$ were synthetic 
piecewise constant coefficient vectors.  Denote by $\theta \in
\R^{251}$ the mean vector, having components
\begin{equation*}
\theta_t = \sum_{j=1}^3 X_{tj} \beta^*_{jt}, \quad
t=1,\ldots,251.
\end{equation*}
Denote by $X \in \R^{251 \times 753}$ the predictor matrix  
\begin{equation*}
X = \left[\begin{array}{ccc}
\diag(X_1) & \diag(X_2) & \diag(X_3) 
\end{array}\right],
\end{equation*}
where $\diag(X_j) \in \R^{251 \times 251}$ is the diagonal matrix in
the entries \smash{$X_{j1},\ldots,X_{j,253}$}, $j=1,2,3$.  Also, let
\smash{$\beta^*=(\beta_1^*,\beta_2^*,\beta_3^*) \in \R^{753}$}. Then, 
in this abbreviated notation, the mean is simply \smash{$\theta =
  X\beta^*$}, and data is generated according to the model 
regression model $y \sim N(\theta, \sigma^2 I)$. 

\paragraph{Optimization problem.}
The fused lasso regression problem that we consider is
\begin{equation}
\label{eq:genlasso_again}
\hbeta = \argmin_{\beta \in \R^{751}} \; \half \|y-X\beta\|_2^2 + 
\lambda \|D\beta\|_1 + \rho \|\beta\|_2^2,
\end{equation}
where $X \in \R^{251 \times 753}$ is as defined above, and 
using a block decomposition $\beta=(\beta_1,\beta_2,\beta_3) \in
\R^{751}$, with each $\beta_j \in \R^{251}$, we may write the penalty
matrix $D \in \R^{750 \times 251}$ as
\begin{equation*}
D = \left[\begin{array}{c}
D^{(1)} \\ D^{(1)} \\ D^{(1)}
\end{array}\right], \quad
\text{where} \quad
D^{(1)} = \left[\begin{array}{ccccc}
-1 & 1 & 0 & \ldots&  0\\
0 & -1 & 1 & \ldots&  0\\
\vdots &  & \ddots & \ddots & \\ 
0 & 0 & \ldots & -1 & 1 
\end{array}\right] \in \R^{250 \times 251}, 
\end{equation*} 
so that
\smash{$\|D\beta\|_1 = \sum_{j=}^3 \|D^{(1)} \beta_j\|_1$}.  Note that
small ridge penalty has been added to the criterion in
\eqref{eq:genlasso_again} (i.e., $\rho>0$ is taken to be a small
fixed constant), making the problem strictly convex, thus 
ensuring it has a unique solution, and also ensuring that we can run
the dual path algorithm of \citet{genlasso}.  Of
course, the blocks of the solution 
$\hbeta=(\hbeta_1,\hbeta_2,\hbeta_3)$ in \eqref{eq:genlasso_again} 
serve as estimates of the underlying coefficient vectors
\smash{$\beta_1^*,\beta_2^*,\beta_3^*$}.  

\paragraph{Segment test contrasts.}
Having specified the details of the generalized lasso regression
problem solved in Section \ref{sec:regressionexample}, it
remains to specify the contrasts that were used to form the segment
tests.  Let $\cB \subseteq \{1,\ldots,753\}$ be the indices of
changepoints in the solution \smash{$\hbeta$}, assumed to be in sorted
order.  We can decompose $\cB
= \cB_1 \cup (251 + \cB_2) \cup (452 + \cB_3)$, where each $\cB_j
\subseteq \{1,\ldots,251\}$, $j=1,2,3$. Write $X_\cB \in \R^{251  
  \times {|\cB|+3}}$ for the ``effective'' design matrix when 
changepoints occur in $\cB$, whose columns are defined by splitting
each $X_j$ into segments that correspond to breakpoints in $\cB_j$,
and collecting these across $j=1,2,3$.  For example, if 
$\cB_1=\{60,125\}$, then $X_1$ gets split into $|\cB_1|+2=3$ columns: 
\begin{equation*}
\left[\begin{array}{c}
X_{j1} \\ \vdots \\ X_{j,60} \\ 0 \\ \vdots \\ 0 \\ 0 \\ \vdots \\ 0   
\end{array}\right], \;\;
\left[\begin{array}{c}
0 \\ \vdots \\ 0 \\ X_{j,61}  \\ \vdots \\ X_{j,125} \\ 0 \\\vdots \\ 0    
\end{array}\right], \;\;
\left[\begin{array}{c}
0 \\ \vdots \\ 0 \\ 0 \\ \vdots \\ 0 \\  X_{j,126} \\ \vdots \\
X_{j,251}    
\end{array}\right].
\end{equation*}

For each detected changepoint $I_j \in \cB$, we now
define a segment test contrast vector by 
\begin{equation}
\label{eq:vsegment_reg}
v_{\mathrm{seg}} = s_{I_j} (X_\cB^+)^T (0, \ldots, 0,
\underset{\underset{r_{I_j}}{\uparrow}}{-1},  
\underset{\underset{r_{I_j}+1}{\uparrow}}{1}, 0,\ldots, 0),  
\end{equation}
where \smash{$s_{I_j}$} is the observed sign of the difference between
coordinates $I_j$ and $I_j+1$ of the fused lasso solution 
\smash{$\hbeta$}, and \smash{$r_{I_J}$} is the rank of $I_j$ in $\cB$.
Then the TG statistic in \eqref{eq:tgstatistic}, with
\smash{$v=v_{\mathrm{seg}}$}, tests
\begin{equation}
\label{eq:hsegment_reg}
H_0 : (0, \ldots, 0,
\underset{\underset{r_{I_j}}{\uparrow}}{-1}, 
\underset{\underset{r_{I_j}+1}{\uparrow}}{1}, 0,\ldots, 0)^T X_\cB^+ 
\theta = 0 \quad \text{versus} \quad
H_1: s_{I_J} (0, \ldots, 0,
\underset{\underset{r_{I_j}}{\uparrow}}{-1}, 
\underset{\underset{r_{I_j}+1}{\uparrow}}{1}, 0,\ldots, 0)^T X_\cB^+ 
\theta > 0.
\end{equation}
In words, this tests whether the best linear model fit to the mean
$\theta$, using the effective design $X_\cB$, yields coefficents that 
match on either side of the changepoint $I_j$.   The alternative
hypothesis 
is that they are different and the sign of the difference is the same
as the sign in the fused lasso solution. 

\paragraph{Alternative motivation for the contrasts.}
An alternative motivation for the above definition of contrast at a
changepoint $I_j \in \cB$ stems from consideration of the hypotheses 
\begin{equation*}
H_0 :  \theta \in \col(X_{\cB\setminus\{I_j\}})
\quad \text{versus} \quad 
H_1 :  \theta \in \col(X_{\cB}).
\end{equation*}
When $\cB$ is considered to be fixed (and hence so are these
hypotheses), the corresponding likelihood ratio test is
is \smash{$v_{\mathrm{lik}}^T y$}, where \smash{$v_{\mathrm{lik}}$} is
a unit vector that spans the rank 1 subspace
\smash{$\col(X_{\cB\setminus\{I_j\}})^\perp \col(X_{\cB})$}, i.e., 
\begin{equation}
\label{eq:vlikelihood_reg}
v_{\mathrm{lik}} v_{\mathrm{lik}}^T =
P_{\col(X_{\cB})} - P_{\col(X_{\cB\setminus\{I_j\}})}.
\end{equation}
We now prove that indeed, \smash{$v_{\mathrm{lik}}$} in
\eqref{eq:vlikelihood_reg} and \smash{$v_{\mathrm{seg}}$} in
\eqref{eq:vsegment_reg} are equal up to normalization. 
Abbreviate
\begin{equation*}
w = (0, \ldots, 0,
\underset{\underset{r_{I_j}}{\uparrow}}{-1}, 
\underset{\underset{r_{I_j}+1}{\uparrow}}{1}, 0,\ldots, 0), 
\end{equation*}
and $M=\cB$, \smash{$m=X_{\cB\setminus\{I_j\}}$}.  It suffices to show
that 
\begin{equation*}
(X_M^+)^T ww^T X_M^+ \;\propto\; X_M X_M^+ - X_mX_m^+.  
\end{equation*}
To verify the above, multiply from the left by $X_M^+$ and from the
right by $X_M$, and assuming with a loss of generality that
$X_M$ has full column rank, we get 
\begin{align*}
ww^T &\;\propto\; X_M^TX_M - X_M^T X_mX_m^+X_M \\
&= X_M^T(I-X_mX_m^+)X_M\\
&= X_M^T P_{\col(X_m)}^\perp X_M \\
&= X_M^T P_{\col(X_m)}^\perp P_{\col(X_m)}^\perp X_M.
\end{align*}
But it is easy to see that \smash{$X_M^T P_{\col(X_m)}^\perp
  P_{\col(X_m)}^\perp X_M$} is proportional to $ww^T$, because  
if $a$ is any vector that has identical entries across 
coordinates \smash{$r_{I_j}$} and \smash{$r_{I_j}+1$}, then   
\begin{equation*}
P_{\col(X_m)}^\perp X_M a = P_{\col(X_m)}^\perp X_m a' = 0,
\end{equation*}
where $a'$ is simply $a$ with its \smash{$(r_{I_j})$}th coordinate   
removed. This completes the proof.

\bibliographystyle{agsm}
\bibliography{paper}

\end{document}